\PassOptionsToPackage{usenames,dvipsnames}{xcolor}
\documentclass[10pt]{article}
\usepackage{geometry}
\usepackage[english]{babel}\usepackage[T1]{fontenc}
\usepackage[latin1]{inputenc}
\usepackage[dvipsname]{xcolor}
\usepackage{enumerate}
\usepackage{pstricks}
\usepackage{graphicx}
\usepackage{amsmath}
\usepackage{amsfonts}
\usepackage{amssymb}
\usepackage{float}
\usepackage{newtxtext}
\usepackage{newtxmath}
\usepackage{natbib}
\title{Collapse of transitional wall turbulence captured using a rare events algorithm}
\author{Joran Rolland
\footnote{ joran.rolland@centralelille.fr }
\footnote{Laboratoire de Physique \`a l'ENS de Lyon, UMR CNRS 5672, Univ. Lyon, Univ. Claude Bernard Lyon 1, France.} \footnote{ UMR CNRS 9014 - LMFL - Laboratoire de M\'ecanique des Fluides de Lille - Kamp\'e de F\'eriet, Univ. Lille, Centrale Lille, ENSMA, ONERA, France.}
}
\date{\today}
\begin{document}

\maketitle

\begin{abstract}
This text presents one of the first successful applications of a rare events sampling method for the study of multistability in a turbulent flow without stochastic energy injection.
The trajectories of collapse of turbulence in plane Couette flow, and their probability and rate of occurrence are systematically computed using Adaptive Multilevel Splitting (AMS).
The AMS computations are performed in a system of size $L_x\times L_z=24\times 18$ at Reynolds number $R=370$ with an acceleration by a factor $\mathcal{O}(10)$ with respect to DNS and in a system of size $L_x\times L_z=36\times 27$ at Reynolds number $R=377$ with an acceleration by a factor $\mathcal{O}(10^3)$.
The AMS results are validated with a comparison to DNS in the smaller system.
Visualisations indicate that turbulence collapses because the self sustaining process of turbulence fails locally.
The streamwise vortices decay first in streamwise elongated holes, leaving streamwise invariant streamwise velocity tubes that experience viscous decay.
These holes then extend in the spanwise direction.
The examination of more than a thousand trajectories in the $(E_{k,x}=\int u_x^2/2\,{\rm d}^3\mathbf{x},E_{k,y-z}=\int (u_y^2/2+u_z^2/2)\,{\rm d}^3\mathbf{x})$ plane in the smaller system confirms the faster decay of streamwise vortices and shows concentration of trajectories.
This hints at an instanton phenomenology in the large size limit.
The computation of turning point states, beyond which laminarisation is certain, confirms the hole formation scenario and shows that it is more pronounced in larger systems.
Finally, the examination of non-reactive trajectories indicates that both the vortices and the streaks reform concomitantly when the laminar holes close.

\end{abstract}



\section{Introduction}

  Many turbulent flows of aerodynamical or geophysical interest are not homogeneous nor isotropic. They often display several possible turbulent flow configurations and rare switches between these flow configurations. For these reasons these flows are often termed multistable. Examples of multistability in turbulent flows include turbulent dynamos \citep{Berhanu2007}, turbulent convection \citep{podvin2017precursor}, bluff body wakes \citep{grandemange2013turbulent}, jets in the wake of a pair of cylinders \citep{kim1988investigation} and barotropic atmospheric-type jets \citep{prl_jet,simonnet2020multistability}. In order to understand these turbulent flows, uncovering what drives the switches is as important as explaining the mechanisms maintaining each metastable configuration. These switches are characterised by the mean first passage time before a change of configuration occurs. The switches often take place through the same chain of events, termed a \emph{transition path} or a \emph{reactive trajectory}, a notion originating from kinetic chemistry \citep{metzner2006illustration}. It is fairly difficult to compute mean first passage times and transition paths, either experimentally or numerically for two main reasons.
   \begin{enumerate}[a)]
   \item ~One firstly has to deal with the very large number of degrees of freedom of turbulent flows, particularly in numerical simulation. These complex flows are often  difficult to simulate even by means of Large Eddy Simulations.
    \item ~One secondly has to deal with the extremely long waiting times between each event. These waiting times are several orders of magnitude larger than the duration of a realisation of a switch and even longer than the typical eddy turnover time \citep{kim1988investigation}. This means that the cost of sampling more than a few events is prohibitive by classical means  (see table in \citep{prl_jet}).
\end{enumerate}

\begin{figure}
\begin{center}
\includegraphics[width=0.8\textwidth]{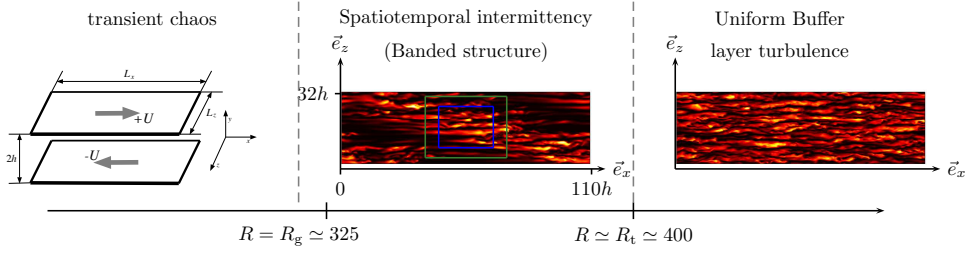}
\end{center}
\caption{Overview of transition to turbulence: the bottom axis indicates the Reynolds number and the text on the top indicating typical states. The left panel displays a sketch of plane Couette flow. The middle panel displays colour levels of the kinetic energy density in a $y=0$ plane at $R=370$, showing banded laminar turbulent coexistence. The green box indicates a domain of size $L_x\times L_z=36\times 27$, the blue box indicates a domain of size $L_x\times L_z= 24\times 18$. The right panel displays colour levels of the kinetic energy in a $y=0$ plane at $R=500$.}
\label{rolland_figint}
\end{figure}

  In order to propose a method to systematically study multistability, one first needs a turbulent system which has fewer effective degrees of freedom but is still complex enough to display multistability, with a moderate need for extrinsic stochastic forcing.
  One can thus temporarily bypass problem a).
  A transitional wall flow such as plane Couette flow can provide such a situation (see Fig.~\ref{rolland_figint}).
  Unlike thermal convection, for instance, a wall flow like plane Couette flow is linearly stable for all Reynolds numbers \citep{romanov1973stability}.
  Meanwhile transitional turbulence can exist at moderate Reynolds numbers, albeit transiently \citep{eckhardt2008dynamical,schmiegel1997fractal,bottin1998statistical,eckhardt2007turbulence}.
  Transitional wall turbulence takes the form of velocity streaks: long wavy streamwise tubes flanked by short streamwise vortices.
  The streaks and vortices sustain one another in a cycle termed the Self-Sustaining Process (SSP) of wall turbulence \citep{waleffe1997self,hamilton1995regeneration}.
  The streamwise vortices extract energy from the mean flow to regenerate the streaks, a process termed lift-up.
  Meanwhile the streaks regenerate the streamwise vortices through a Kelvin-Helmholtz instability and vorticity tilting by the base flow.
  How does wall turbulence collapse, that is to say whether the SSP fails because vortices disappear first or streaks disappear first or both collapse at the same time, is a question that could only be address very recently \citep{gome2020statistical,liu2020anisotropic,rolland_pre18}.
  Conversely, turbulence can build up from laminar flow under a finite amplitude forcing \citep{wan2017dynamic,rolland_pre18,liu2020anisotropic}, or somewhat equivalently from a finite amplitude initial condition \citep{faisst2003sensitive}.
  This gives a first type of multistability, where the transitional turbulence of plane Couette flow can collapse down to laminar flow under its own fluctuations, and go back to turbulence if it is forced.
  In a way this is a purely temporal view. This image should be completed by noting that transitional wall turbulence tends to be localised (the type of organisation depends on the flow geometry).
  At moderate Reynolds numbers, wall flows where transitional turbulence extends in one dimension (such as Hagen--Poiseuille flow \citep{moxey2010distinct}, tilted plane Couette \citep{shi2013scale} flow or tilted plane Poiseuille flow \citep{gome2020statistical}) can display splitting: the turbulent puff elongates and splits in half.
  This can lead to an effective extension of the area occupied by turbulence.
  This can provide a third type of multistability events in some flow configurations.
When all these processes are taken into account, one can find laminar-turbulent coexistence in large flow domains containing several puffs.
In models, it can be shown that the laminar-turbulent coexistence is also transient, and that its lifetime grows exponentially with the system size \citep{rolland_pre18}.
The laminar-turbulent coexistence and the possibility of reinvasion of laminar holes through puff splitting are actually preeminent ingredients in the collapse scenario and size scaling of lifetimes in the model.
Transitional wall flows are relevant for high Reynolds number wall turbulence, since the processes described here still take place very close to the wall (up to thirty wall units \cite{pope2001turbulent}), in a layer termed the buffer layer, or the viscous layer when a slightly thicker layer is taken into account.
Moreover, canonical flows such as plane Couette can thus provide a good laboratory to test rare events methods on turbulent flows.
These multistable turbulent flows are not so complex that they cannot be simulated by Direct Numerical Simulations (DNS).

  Even if the number of degrees of freedom is reduced, as is the case for transitional wall flows, one still has to deal with problem b).
  Transitional turbulence, like other multistable systems, displays very long waiting times before a multistability event occurs \citep{eckhardt2007turbulence,bottin1998statistical,shi2013scale,gome2020statistical}.
  This means that even for the study of turbulence collapse in plane Couette flow, the use of DNS to sample many events comes with a prohibitive computational cost.
  There are of course alternative methods to study rare events. Many of them originate from the study of kinetic chemistry and borrow much of its vocabulary. All these methods aim at performing the computation of the reactive trajectories, the probability that a reactive trajectory occurs, and the mean waiting time before they occur.
  These methods can be divided in three main families \citep{bouchet2019rare}, two of them (action minimisation and cloning methods) will be invoked in this text.

  \begin{itemize}
  \item Firstly, one finds mostly theoretical, optimisation methods, which are applied to systems where a stochastic forcing is clearly identified.
  In that case, in the limit where the variance of said forcing goes to zero, all the transition paths concentrate around an \emph{instanton} \citep{touchette2009large} \S~6.
  The instanton represents the most probable transition path and can be computed using action minimisation \citep{grafke2019numerical,wan2017dynamic}.
  One fundamental property of said instantons is that they evolve from a first multistable state toward a saddle point of the deterministic dynamics under the action of the noise. The instanton then evolves freely from the saddle point toward the second multistable state.
  The mean first passage time can be estimated using the result of said action minimisation and can loosely be thought of as depending on the distance between the first metastable state and the saddle point.
  In gradient stochastic systems, this yields the celebrated Eyring--Kramers formula, also known as the Arrhenius law \citep{hanggi1990reaction}. Such results can be extended to non gradient systems \citep{bouchet2016generalisation}.
  While these methods give a lot of qualitative results on the structure of transition paths, they can prove tricky to implement for fluid flows \citep{wan2017dynamic}. A key property of these formulae and methods is that they are formulated as \emph{large deviations} \citep{touchette2009large}.
This means that probabilities $\alpha$, probability density functions $\rho$, rates of transitions $1/T$ are considered in the limit of a vanishing small parameter $\epsilon \rightarrow 0$. In that case, we have that $\lim_{\epsilon \rightarrow 0} -\epsilon \log(\alpha)=I_\alpha$, $\lim_{\epsilon \rightarrow 0} -\epsilon \log(\rho)=I_\rho$, $\lim_{\epsilon \rightarrow 0} \epsilon \log(T)=I_T$, with $I_\alpha$, $I_\rho$ and $I_T$ independent on $\epsilon$. In other words, the leading dependence is exponential, and noted as $\alpha\underset{\epsilon\rightarrow 0}{\asymp}e^{-\frac{I_\alpha}{\epsilon}}$.

\item Secondly, one finds mostly numerical cloning methods, which use the actual fluctuating dynamics of the system and push them toward realisations of the reactive trajectories.
  These methods compute the transition paths using $N$ clone dynamics of the system and apply a mutation selection procedure to compute the reactive trajectories. One such method is termed \emph{Adaptive Multilevel Splitting} (AMS) \citep{rolland_CG07,cerou2019adaptive}. AMS and its variants have been successfully used to compute reactive trajectories and extreme events in kinetic chemistry \citep{lopes2019analysis} theoretical physics models \citep{rolland2016computing}, models of transitional flows \citep{rolland_pre18}, idealised atmospheric flows \citep{prl_jet,simonnet2020multistability}. Some variants have been applied to the study of extreme two dimensional turbulent wakes \citep{rolland_Les,lestang2020numerical} and oceanic flow reversals \citep{baars2021application}.
Thirdly, one can use importance sampling methods \citep{l2009importance,hartmann2019variational}.
These methods  modify the dynamics so that the events of interest can then be sampled according to a new probability distribution function (see \citep{rolland_Les}, Fig.~8 \citep{ragone2019computation} Fig.~3 or \citep{ragone2020rare}, Fig~1 (a,b)).
 Under this new distribution, the event of interest is much more probable and is therefore sampled much more precisely.
Conversion factors are then used to rescale the estimated mean first passage times and probabilities.
Importance sampling is often performed in stochastic systems, where the modifications are applied on the noise.
As a consequence, this often makes importance sampling hard to apply to deterministic systems.
The key question is then what modification to apply and how to compute the rescaling factors.
When answering these questions, one can note that the boundaries between optimisation, cloning and importance sampling methods are porous.
For instance, results  of action minimisation can be used to design importance sampling methods, when studying extreme events on a given time interval \citep{grafke2019numerical,ebener2019instanton}.
Similarly, some cloning methods lead to a situation comparable to importance sampling \citep{ragone2019computation,ragone2020rare}, where the rescaling of probabilities is based on Large Deviations for time averaged variables.
\end{itemize}

Action minimisation and cloning methods are often used hand in hand.
Once a small parameter that controls the effective noise variance has been identified, the qualitative insight from theory is used to guide numerical studies and propose manners in which results can be presented, as it has been done in models of pipe flow \citep{rolland_pre18}.
Applying such a program to the collapse of turbulence in plane Couette flow is actually not that straightforward.
Because energy is not injected in the flow through a stochastic forcing whose variance decays, the small parameter is not readily identified. For similar reasons, cloning methods cannot be applied as such \citep{lestang2020numerical}.
Applying basic cloning rules leads to so called extinction: the flow does not separate trajectories from one another and the method does not manage to create a reactive trajectory which contains an excursion far enough from the starting metastable state.
A first goal is therefore to propose a modified cloning method that can bypass this problem and can at least succeed in computing reactive trajectories faster than a DNS would. This is the purpose of anticipated AMS, which is presented and used in this text.
Once the reactive trajectories are computed, if they display concentration around a typical transition path, one needs to make sense of this concentration. For this purpose, one can use both AMS and DNS to identify the right small parameter from the study of probability density functions, probabilities of transition, mean first passage times \emph{etc}. We will keep this in mind in our study. We also note that there are examples of very relevant reactive trajectories (see \citep{rolland2016computing} for instance) that are not necessarily of the instanton type. As a consequence, one does not always have to try to force the results into a large deviations framework.

  We present the study of the collapse of transitional turbulence of plane Couette flow in the following manner. We describe plane Couette flow in section~\ref{pcf}.
  We then present Anticipated Adaptive Multilevel Splitting in section~\ref{AMS}.
  The generation of initial conditions is presented in section~\ref{init}. The reaction coordinate used to compute reactive trajectories is presented in section~\ref{scoord}.
  The DNS which are used as reference are presented in the next section (\S~\ref{dns_couette}).
  We then present the systematic comparison of reactive trajectories computed by AMS and DNS in a system of size $L_x\times L_z=24\times18$ (\S~\ref{traj}). We perform the validation of the computation of the probability of crossing and mean first passage time in this system in section~\ref{mfpt}.
  AMS is then applied to the computation of very rare trajectories and laminar hole formation in section~\ref{hole}. These results are finally discussed together in the conclusion (\S~\ref{concl}).

\section{Flow configuration and numerical procedures}\label{met}

\subsection{Plane Couette flow}\label{pcf}

We will perform the study of collapse in plane Couette flow, the flow between two parallel walls located at $y=h$ and $y=-h$, respectively moving at velocities $U\textbf{e}_x$ and $-U\textbf{e}_x$ (Fig.~\ref{rolland_figint}, left). We term $\textbf{e}_x$ the streamwise direction, $\textbf{e}_y$ the wall normal direction and $\textbf{e}_z$ the spanwise direction. Lengths are nondimensionalised by $h$, velocities are nondimensionalised by $U$ and times by $h/U$. The first and foremost control parameter is the Reynolds number $R=hU/\nu$, with $\nu$ the kinematic viscosity. The nondimensional streamwise and spanwise sizes $L_x$ and $L_z$ are two other control parameters of the system. The full velocity field is written $\textbf{v}=y\textbf{e}_x+\textbf{u}$, where $y\textbf{e}_x$ is the laminar baseflow.

The forced incompressible Navier--Stokes equations for the field $\textbf{u}$, the departure to the laminar base flow $y \textbf{e}_x$ and the pressure $q$, read
\begin{equation}
\frac{\partial u_l}{\partial t}+u_m\frac{\partial u_l}{\partial x_m}+y\frac{\partial u_l}{\partial x}+\delta_{l,x}u_y=-\frac{\partial q}{\partial x_l}+\frac{1}{R}\left( \frac{\partial^2u_l}{\partial x^2}+\frac{\partial^2u_l}{\partial y^2}+\frac{\partial^2u_l}{\partial z^2}\right)
+f_l(x,t)
\,,\,
 \frac{\partial u_m}{\partial x_m}=0\,,\label{nsf}
\end{equation}
using tensorial notations.
We include the term $\mathbf{f}$. It is a very general forcing which can be switched on or off. When it is on, it is white in time and in $y$. It can be red or white in $x-z$: in AMS simulations, we will use a temporally localised fully white perturbation (see \S~\ref{princams}), while in perturbed DNS, we use a red forcing (see \S~\ref{init},~\ref{dns_couette} for details).
These equations are discretised in space on $N_x$ and $N_z$ de-aliased Fourier modes (so that  $\frac32 N_x$ and $\frac32 N_z$ modes are used in total) in the streamwise $\textbf{e}_x$ and $\textbf{e}_z$ directions and $N_y$ Chebyshev modes in the $\textbf{e}_y$ direction. Time integration is performed using {\sc channelflow}, by J. Gibson \citep{rolland_Gib08}. We investigate in detail systems of two sizes. The smaller system has size $L_x\times L_z=24\times 18$ (see Fig.~ \ref{rolland_figint} for scale), we set $N_y=27$, $\frac32 N_x=128$ and $\frac32 N_z=96$. The larger system has size $L_x\times L_z=36\times 27$, we set $N_y=27$, $\frac32 N_x=196$ and $\frac32 N_z=144$. In both cases we will set a constant timestep $\Delta t$ during the time integrations, $\Delta t=0.05$ (at $R=370$ and $R=377$) and $\Delta t=0.02$ (at $R=600$). Both these values ensure stability of the time integration and respect the CFL criterion. We do not use adaptive time steps in order to have more control on trajectory reconstruction and effect of the time step on convergence.

We define the spatially averaged kinetic energy as
\begin{equation}
E_k=\frac{1}{2L_xL_z}\int_{x=0}^{L_x}\int_{y=-1}^1\int_{z=0}^{L_z}\frac{u_x^2+u_y^2+u_z^2}{2} \,{\rm d}x{\rm d}y{\rm d}z\,.
\end{equation}
We will also distinguish the kinetic energy contained in the streamwise component $E_{k,x}$, on the one hand, and the kinetic energy contained in the spanwise and wall normal components $E_{k,y-z}$, on the other hand
\begin{align}\notag
E_{k,x}=\frac{1}{2L_xL_z}\int_{x=0}^{L_x}\int_{y=-1}^1\int_{z=0}^{L_z}\frac{u_x^2}{2} \,{\rm d}x{\rm d}y{\rm d}z\,, \\   E_{k,y-z}=\frac{1}{2L_xL_z}\int_{x=0}^{L_x}\int_{y=-1}^1\int_{z=0}^{L_z}\frac{u_y^2+u_z^2}{2} \,{\rm d}x{\rm d}y{\rm d}z\,.\label{ekin}
\end{align}
The first kinetic energy $E_{k,x}$ roughly quantifies the energy contained in velocity streaks, while the second $E_{k,y-z}$ roughly quantifies the energy contained in streamwise vortices \citep{jimenez1991minimal,hamilton1995regeneration}. These are the two main flow structures of transitional wall flow turbulence. They regenerate one another in the cyclic self sustaining process of turbulence \citep{waleffe1997self}.

\subsection{Adaptive Multilevel Splitting}\label{AMS}

\subsubsection{Principle of the algorithm}\label{princams}

Before presenting the principle of anticipated AMS, let us first give a formal phase space description of the rare events we will study in this text. Let us sketch the collapse of turbulence in figure~\ref{rolland_figrare} (a) and consider the set $\mathcal{A}$, a neighbourhood of the turbulent flow in phase space, and the set $\mathcal{B}$, a  neighbourhood of the laminar flow.  A realisation of the dynamics which starts in $\mathcal{A}$
fluctuates around it, has several excursions out of $\mathcal{C}$, a hypersurface closely surrounding $\mathcal{A}$, and eventually crosses $\mathcal{C}$ and reaches
$\mathcal{B}$ before coming back to $\mathcal{A}$, is termed a first passage. Its average duration is termed the mean first passage time $T$. The last stage
of the dynamics is termed a reactive trajectory: this is the part of the dynamics that starts in $\mathcal{A}$,
crosses $\mathcal{C}$ and reaches $\mathcal{B}$ before
$\mathcal{A}$. Precise definitions of sets $\mathcal{A}$, $\mathcal{B}$ and hypersurface $\mathcal{C}$ for collapse  will be given in sections~\ref{val} and~\ref{hole}, based on reaction coordinates defined in section~\ref{scoord}.


We then give a brief overview of the variant of AMS, termed \emph{Anticipated Adaptive Multilevel Splitting}, which was used for the computation presented in this text (see \citep{rolland_CG07,rolland2015statistical,brehier2016unbiasedness,rolland_pre18,cerou2019adaptive,lestang2020numerical} for more details on the general methods).
All variants of AMS use a reaction coordinate (or observable) $\phi(\mathbf{u})$, a real-valued function of the velocity field.
The reaction coordinate gives a relative distance in phase space between $\mathbf{u}$ and the sets $\mathcal{A}$ and $\mathcal{B}$. For practical reasons, we will see the reaction coordinate as a function of time $\Phi(t)=\phi(\mathbf{u}(t))$ on the trajectories.
The reaction coordinate is often rescaled such that $\phi(\partial \mathcal{A})=0$, on the boundary of set $\mathcal{A}$, $\phi(\partial \mathcal{B})=1$, on the boundary of set $\mathcal{B}$, and grows monotonously inbetween.
All variants run $N$ clone dynamics of the system to compute iteratively at least $N-N_c>0$ reactive trajectories going from a hypersurface $\mathcal{C}$, close to set $\mathcal{A}$, towards the set $\mathcal{B}$.
At each iteration, $N_c<N$ clones are replaced. The algorithm is sketched in figure~\ref{rolland_figrare} (b)  and proceeds in the following manner:
\begin{itemize}
\item There is a first stage of natural dynamics, where each clone dynamics starts inside set $\mathcal{A}$. As much as possible, these initial conditions should be distributed according to the natural flow, restricted to $\mathcal{A}$ (see \S~\ref{init} for an example of procedure). We let all the initial conditions evolve according to their natural dynamics until they cross $\mathcal{C}$ and we stop them when they reach either $\mathcal{A}$ or $\mathcal{B}$. We set the number of iterations $\kappa=0$.
\item In a second stage, the algorithm iterates the mutation-selections as long as there are less than $N-N_c$ clones that transit to $\mathcal{B}$.

    At each iteration, we set $\kappa=\kappa+1$. Then all the clones are ordered with index $i$, $1\le i\le N $, according to the maximum of the reaction coordinate on the trajectory $\max_t\Phi_i(t)$. The clones $1\le i\le N_c$, that have the shortest excursion out of $\mathcal{A}$ are removed from the set of clones.
In order to keep a constant number of clones, $N_c$ new clones are created by branching on $N_c$  clones, labeled $i'$, drawn uniformly out of the $N-N_c$ non-removed clones at level $\Phi_b$, a function (specified later in the text) of $\Phi_{N_c}$ (the maximal value of reaction coordinate reached by the removed trajectories).
The branched trajectories first share the dynamics of the clone on which they are branched, \emph{i.e.} we set $\{\mathbf{u}(t),q(t)\}_i=\{\mathbf{u}(t),q(t)\}_{i'}$ from $t=0$ to the time where clone $i'$ first reach $\Phi_b$.  Then, the branched clones follow their natural dynamics until they reach either $\mathcal{A}$ or $\mathcal{B}$, with a new realisation of the noise. One may have to repeat this branching operation several times to make sure that the maximum of $\Phi$ on the branched trajectory is strictly larger than $\Phi_{N_c}$.
\end{itemize}
 The algorithm stops when $Nr\ge N-N_c+1$ clones trajectories have reached $\mathcal{B}$. We usually perform $o>1$ independent AMS runs. In each of the runs labeled by the number $a$, $1\le a\le o$, we  obtain the total number of iterations $\kappa_a$, and $r_a$ the proportion of clones that transit to $\mathcal{B}$. This yields an estimator of the probability $\alpha$ of reaching $\mathcal{B}$ before $\mathcal{A}$ \citep{rolland_CG07}, and the corresponding mean first passage time $T$ \citep{cerou2011multiple}
 \begin{equation}\alpha_a=r_a\left(1-\frac{N_c}{N} \right)^{\kappa_a}\,,\,
 \hat{\alpha}=\left\langle \alpha_a\right\rangle_o\,, \,
 \hat{T}=\left\langle\left(\frac{1}{\alpha_a}-1\right)(t_{1,a}+\tilde{\tau}_a)+(t_{1,a}+\tau_a)\right\rangle_o \label{EQT}\,,\end{equation}
where $\tau_a$ is the mean duration of reactive trajectories, $t_{1,a}$ is the mean duration to go from $\mathcal{A}$ to $\mathcal{C}$,  and $\tilde{\tau}_a$ is the mean duration of non-reactive trajectories, computed in each AMS run $1\le a \le o$. The notation $\langle \cdot\rangle_o=\frac{1}{o}\sum_{a=1}^o\cdot_a$
corresponds to an average over the $o$ independent AMS runs \citep{rolland_CG07,rolland_pre18}.

We will often record the velocity field, noted $\mathbf{u}_{\rm last}$ and termed the last state at the last stage, that corresponds to $\max_t\Phi_{N_c}(t)$ during the last stage of the algorithm.
It often gives a precise idea of the turning point in reactive trajectories. Before the flow visits the neighbourhood of that state, returning towards turbulence is more likely, beyond that point, relaminarising becomes more likely.
In systems which correspond to a simple deterministic part forced by noise, that state actually corresponds to the saddle point of the deterministic part of the dynamics crossed by the instanton in the limit of the noise variance going to $0$.
This has been verified for models with few degrees of freedom and the one dimensional Ginzburg--Landau equation (not shown here).
It can be used to educe an effective saddle between two multistable states \citep{simonnet2020multistability}. Dichotomy procedures have been started from states seen during turbulence collapse \citep{de2012edge}. However, we have  \emph{a priori} no certainty that the field $\mathbf{u}_{\rm last}$ corresponds to an actual saddle of the Navier--Stokes equations (as computed by dichotomy or other methods \citep{schneider2007turbulence,willis2009turbulent}).

 The algorithm is naturally parallelised over the $N_c$ suppressed clones.
 This will be done for the calculation presented in this text.
 We usually choose the number of threads $p$ such that $N_c/p$ is an integer larger than or equal to two.
 Since the trajectories have a random duration, we cannot have a perfect load balancing in this parallelisation.
 Note however that as $N_c/p$ increases, it has been observed that the differences in trajectory durations average out and that we can reach a reasonable load balancing between threads.

 Note that in AMS computations in deterministic systems, we  add a small background noise which helps the separation of trajectories after branching.
In practice we will switch on the forcing $\mathbf{f}$ (see Eq.~(\ref{nsf})), white in time and space $\langle f_l(\mathbf{x},t)f_m(\mathbf{x}',t')\rangle=\frac{1}{\beta_f}\delta_{lm}\delta(\mathbf{x}-\mathbf{x}')\delta(t-t')$, with inverse variance $\beta_f=10^{10}$,  at branching timestep and switch it off afterwards.
The realisation of this noise is decorrelated from one branching to another.
The variance is small enough so as not to perturb the laminar flow too much.
The trade-off is that we compute trajectory properties with a small error : $T+\delta T$, $\alpha+\delta \alpha$ and $\tau+\delta \tau$. We will comment on the visible effects of this additional force on the trajectories and their properties in section~\ref{comp_trajs} and~\ref{mfpt}.

\subsubsection{Anticipated AMS and branching level $\Phi_b$}

\begin{figure}
\centerline{\textbf{(a)}
\includegraphics[width=0.3\textwidth]{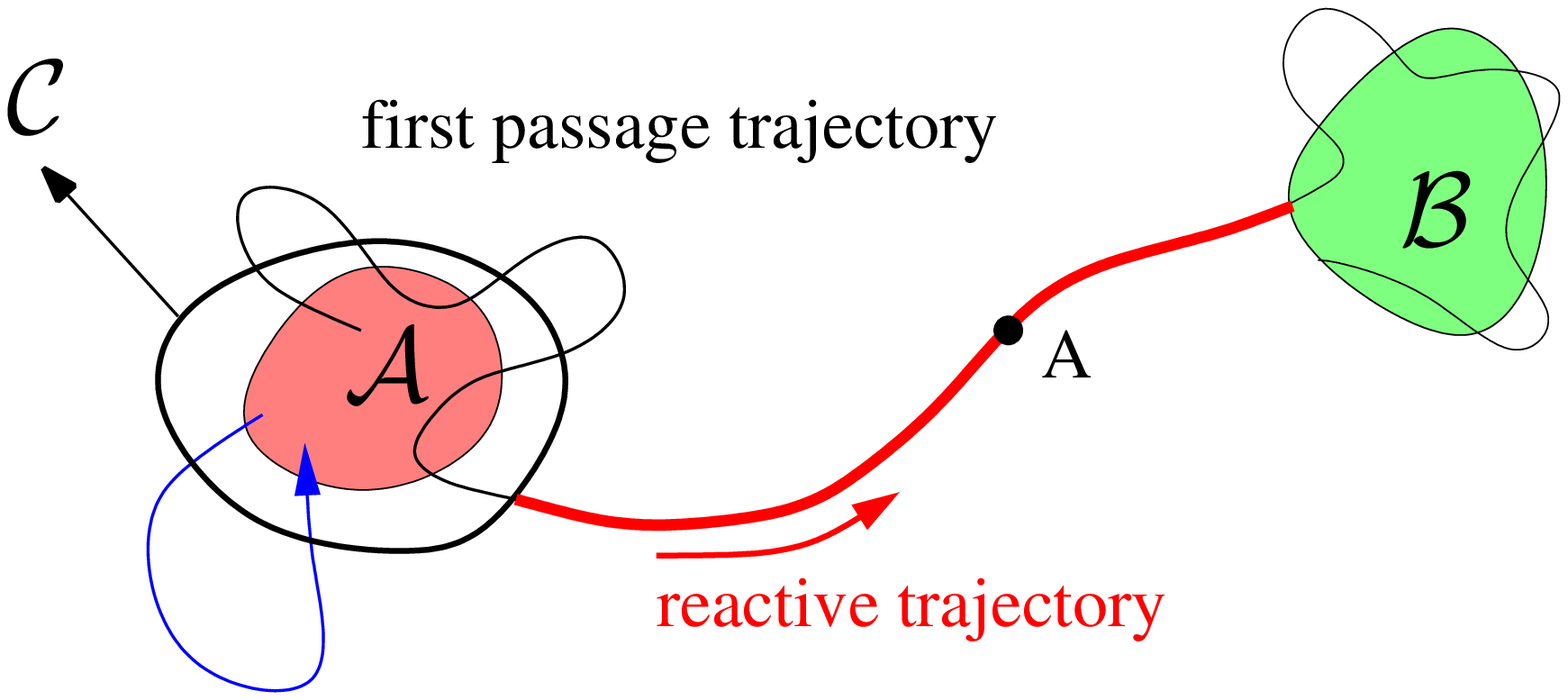}
\textbf{(b)}
\includegraphics[width=0.3\textwidth]{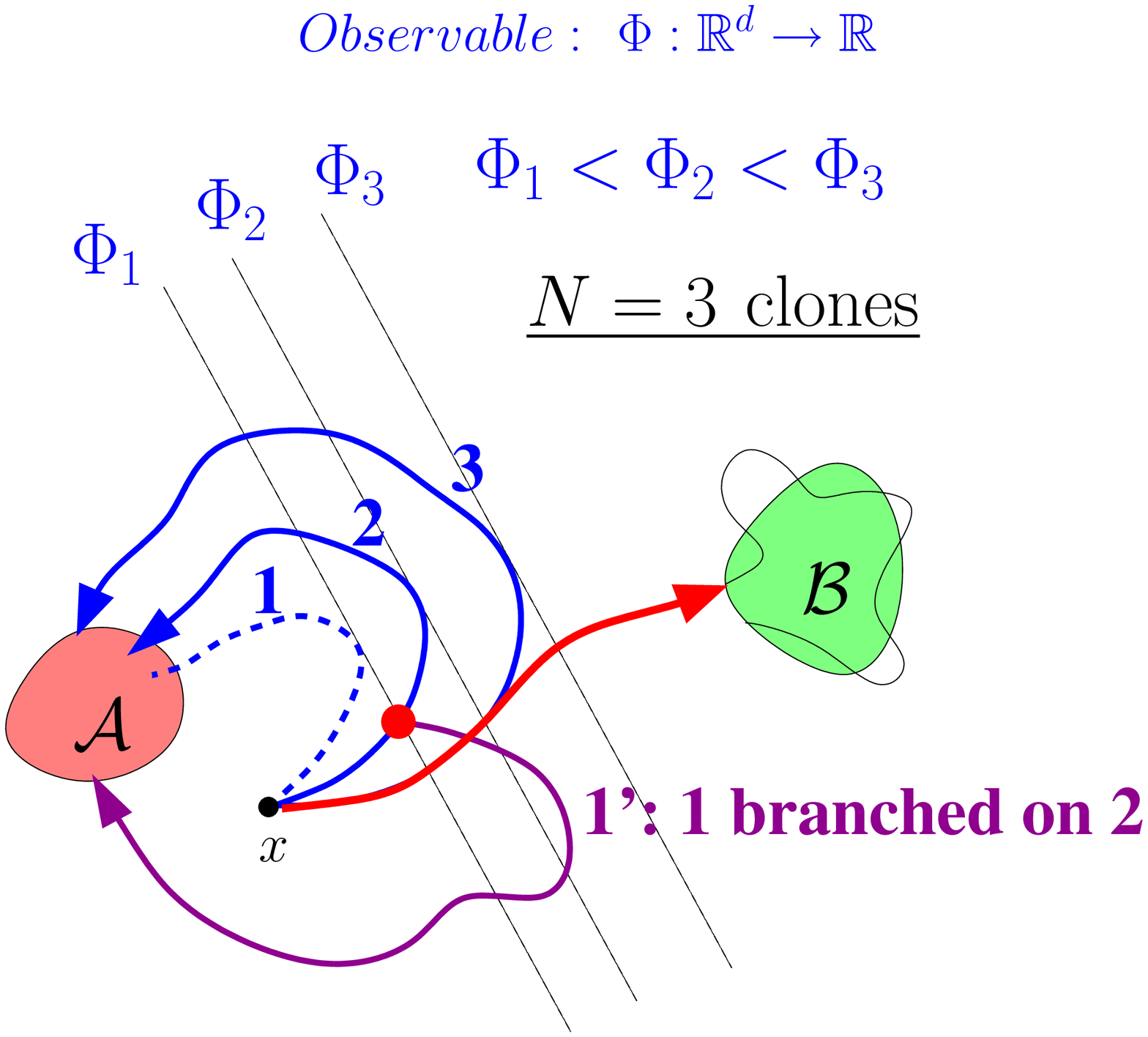}
\hspace{0.5cm}
\includegraphics[width=0.3\textwidth]{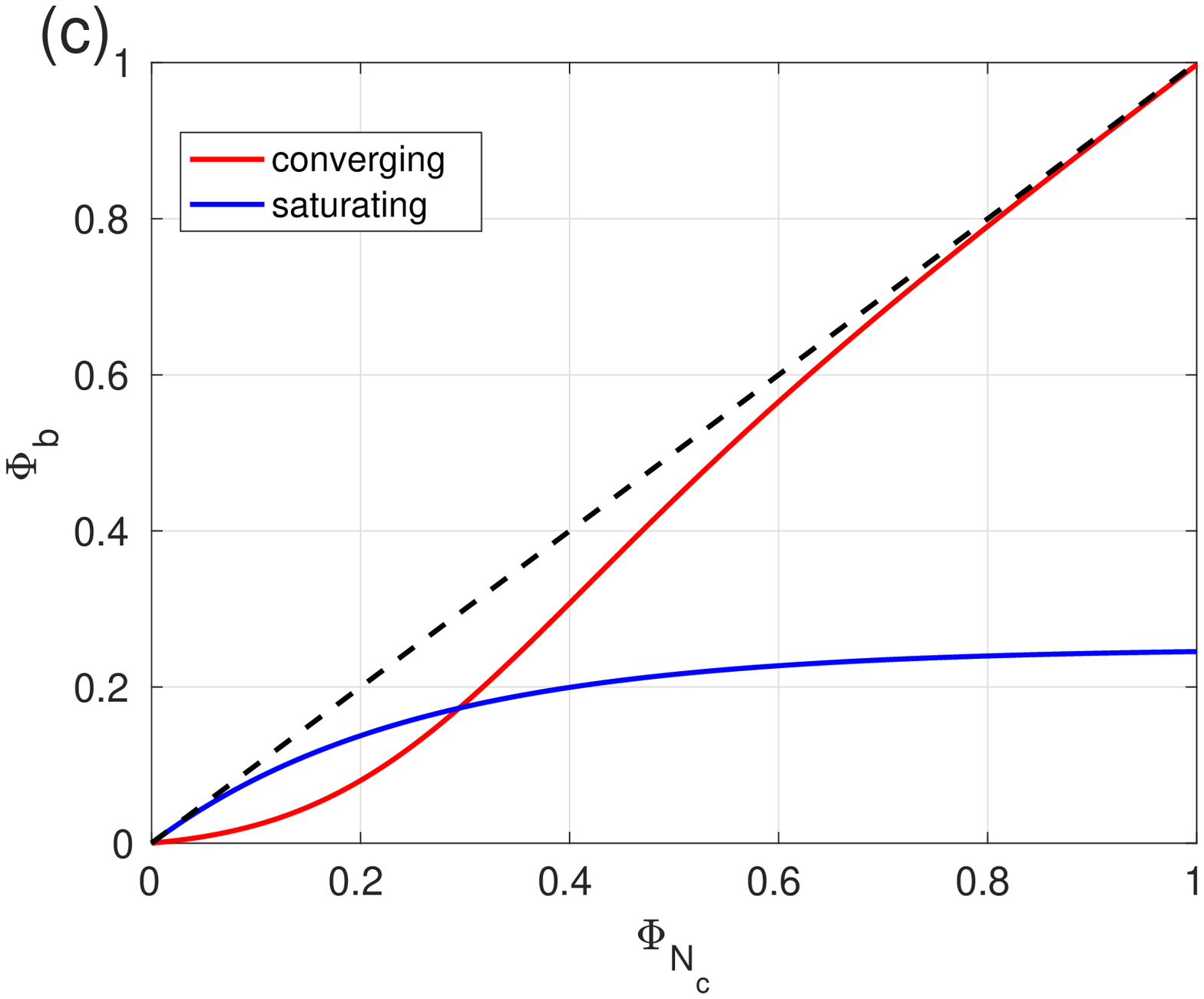}}
\caption{(a) Sketch of two bistable states $\mathcal{A}$ and $\mathcal{B}$ and the hypersurface $\mathcal{C}$ closely surrounding $\mathcal{A}$. two realisations of the dynamics are sketched: a single excursion in blue and a first passage trajectory in black and red. The red part of the first passage trajectory is the reactive trajectory (Figure originally made for \citep{rolland2015statistical}). (b) Sketch of the principle of AMS, showing two iterations of the algorithm, with $N=3$ clones, indicating the starting state $\mathcal{A}$ and its neighbourhood, the arrival state and its neighbourhood $\mathcal{B}$, three trajectories are ordered by their $\max_t\Phi$. Trajectory one (dashed blue line) is suppressed and branched on another trajectory at level $\max_t\Phi_1$ and then ran according to its natural dynamics. Trajectory $2$ is then suppressed and branched on 3 at level $\max_t\Phi_2$ (Figure originally made for \citep{simonnet2016combinatorial}). (c) Two examples of anticipation of branching level of reaction coordinate as a function of maximum reaction coordinate reached by suppressed trajectories $\Phi_{N_c}(\Phi_b)$  tested in anticipated branching (see \S~\ref{rolland_sappa}, Eq.~(\ref{sat}),~(\ref{conv}) for details, both examples use a parameter $\xi=0.25$).}
\label{rolland_figrare}
\end{figure}

Schematically, we can apply AMS to two types of systems. On the one hand, we find systems with a large time scale separation between some fast fluctuating degrees of freedom and slower degrees of freedom which represent the main features of the flow travelling between $\mathcal{A}$ and $\mathcal{B}$.
This is often the case for stochastically forced systems.
In these systems, two slightly different initial conditions will quickly separate and the odds of creating an excursion toward $\mathcal{B}$ instead of $\mathcal{A}$ by slightly changing the noise realisation at a branching are non-negligible.
On the other hand, we find systems with absolutely no clear time scale separation between degrees of freedom.
This is often the case of purely deterministic systems. Two slightly different initial conditions do not separate until it is too late (they both reach $\mathcal{A}$).
We can find situations where, no matter the structure of the small perturbation (typically at a branching), the odds of creating a further excursion toward $\mathcal{B}$ instead of $\mathcal{A}$ can be exceedingly small. This is especially the case if we perturb at the peak of an existing fluctuation.
If we apply basic AMS to this second type of system, where we branch at $\Phi_{N_c}$, we run the risk of a so called extinction \citep{lestang2020numerical}.
 This occurs when all trajectories have the same maximum of reaction coordinate but none of them reach the arrival set, so that $\forall 1\le i\le N, \max_{t}\Phi_i=\Phi_{\rm ext}<\phi (\partial \mathcal{B})$, where $\partial \mathcal{B}$ is the boundary of set $\mathcal{B}$. The algorithm does not manage to proceed any further\footnote{In that case, when extinction is detected, the computation is terminated and the way AMS is used is reassessed.}. In order to bypass this limitation, we can perform anticipated branching, that is to say branch the new trajectories at $\Phi_b<\Phi_{N_c}$.  In that case, it may be necessary to reiterate the branching several times in order to ensure that the branched trajectories have $\max_t\Phi>\Phi_{N_c}$: all the branched trajectories have to go further than the maximum level of reaction coordinate reached by removed trajectories. In figure~\ref{rolland_figrare} (c), we give two examples of relations $\Phi_b(\Phi_{N_c})$ that were tested for plane Couette flow. Each one is adapted to a given situation. The converging anticipation is mostly used in this article. The point is to take advantage of higher mixing and faster separation of trajectories that take place when the flow is closer to the fully turbulent state. For this matter, one has first has a small branching level $\Phi_b\ll \Phi_{N_c}$ for $\Phi_{N_c}<0.5$. We then almost branch at the maximal possible level $\Phi_b\lesssim \Phi_{N_c}$ for $\Phi_{N_c}\ge 0.5$, so as not to lose too much computational time rerunning trajectories when the flow is very close to turbulence collapse. The saturated anticipation takes a very different point of view. It has proved very efficient in very small domains where the collapse of turbulence is very well described by transient chaos. This modification of AMS uses the fact that there is much more mixing when the flow is close to the turbulent state, but that mixing gradually stops during excursions. If a trajectory is on the wrong track, there is no derailing it at large $\Phi$.
 In this type of systems, the task of AMS is really about finding the right exit point.
 More details on the necessity of anticipation are given in appendix~\ref{rolland_sappa}.

\subsubsection{On the goodness of AMS computations}\label{sgood}

We use AMS (or any other rare event simulation method) on top of a numerical discretisation of the Navier--Stokes equations to compute (by order of priority):
\begin{enumerate}[a)]
\item ~a large number of reactive trajectories, in our case from turbulent to laminar flow,
\item ~provide an estimate of the probability that these reactive trajectories occur,
\item ~estimate the mean first passage time before a reactive trajectory occurs.
\end{enumerate}
We wish these computations to be precise, with an emphasis firstly on reactive trajectories, secondly on probabilities and ideally on mean first passage times. Like every other numerical procedure, the use of AMS can lead to errors on estimates which are deemed to large because of an inadapted reaction coordinate or an insufficient number of clones. This is similar to the effect of time or space steps too large that lead to excessive errors in a numerical discretisation. Similarly, an unadapted scheme can lead to a numerical convergence with time and space steps which is too slow.

A successful application of AMS first means that a) the resulting reactive trajectories should faithfully represent the actual reactive trajectories that would be computed in very long DNS (in term of path chosen, of duration of the paths \emph{etc}.).
This can be checked by comparing the paths computed by AMS to some reference, for instance a DNS.
The comparison is usually done in a space with few dimensions (two or three), where we display the most probable paths as computed by AMS and by the reference and show that they go through the same stages.
If several clearly distinct types of paths are possible, one should check that in the set of trajectories computed by AMS, there is the correct proportion of trajectories going through each path (see for instance \citep{rolland2015statistical}, Fig.~8).
In order to validate quantities like the average duration of trajectories $\langle \tau\rangle_o$ (averaged over all the sampled durations over a series of independent AMS runs), one compares the sample mean to the sample mean of the reference.
Since these means are the sum of independent random variables, drawn from identical distributions with a finite variance, one often invokes the Central Limit Theorem to state that the actual average has a 66\% chance of being within $\pm$ the variance of the
 distribution $\sigma_\tau=\sqrt{\frac{1}{o}\sum_{a=1}^o\tau_a^2-\langle \tau\rangle_o^2}$ divided by the square root of the number of samples (or 98\% chances of being $\pm$ twice the variance).
 The interval $\left(\langle \tau\rangle_o-\frac{\sigma_\tau}{\sqrt{o}};\langle \tau\rangle_o+\frac{\sigma_\tau}{\sqrt{o}}\right)$ is termed the confidence interval.
The averaged quantities should be close in that their respective confidence interval should overlap: in that case, we cannot assert that the estimation is biased.
The result of AMS computations should first pass this first test in order to be validated.
Indeed, it may be that incorrect trajectories are selected, usually because of a very poor reaction coordinate \citep{rolland2015statistical,brehier2016unbiasedness}.
If one can avoid this phenomenon, one can ensure that the more clones $N$ are used in AMS computations and the more independent runs $o$ are performed, the more precise the result is going to be.
Furthermore the correctly computed trajectories can be used to improve the computation of more sensitive quantities like the probability $\alpha$ and the mean first passage time $T$

A successful application of AMS secondly means that b) the estimate of the probability that the trajectory occurs is precise.
One can perform an estimate of $\alpha$ by averaging the result over AMS runs.
The AMS runs are independent.
Moreover, one can ensures that the distribution of $\alpha_a$ computed in AMS runs has a finite variance $\sigma_\alpha=\sqrt{\frac{1}{o}\sum_{a=1}^o\alpha_a^2-\langle \alpha\rangle_o}$.
One can provide an estimate with a 66\% interval of confidence $\langle \alpha \rangle_o\pm \frac{\sigma_\alpha}{\sqrt{o}}$ for the probability, using the sample variance and the number of samples, and compare this to a reference.
Of course this estimate is tainted by the finite number of clones $N$, the finite number of AMS realisations $o$, and possibly a poor reaction coordinate.
It is often observed that $\langle\alpha\rangle_o$ underestimates $\alpha$ with a probability close to one.
However,  it is demonstrated that for most version of AMS, this quantity is estimated without bias \citep{brehier2016unbiasedness} or a controlled bias decaying with $N$ \citep{rolland_CG07}.
It is conjectured that this discrepancy is caused by an effect  called the \emph{apparent bias} in multilevel splitting \citep{glasserman1998large}.
A correct estimation would then require an infinite number of realisations.
Note that this also occurs when Importance Sampling is performed \citep{devetsikiotis1993statistical}.
The quality of the estimate of $\alpha$ can further be tested using the sample variance of $\hat{\alpha}$, which should not be too large compared to the ideal variance.
The ideal variance is obtained numerically if one uses the ideal reaction coordinate, termed the committor \citep{cerou2019asymptotic}.
If a problem is suspected, because the confidence intervals of $\alpha$ estimated by AMS and the reference absolutely do not overlap and/or the sample variance of $\alpha$ is too large, one can compute the histograms of $\alpha$.
When the apparent bias phenomenon occurs, heavy powerlaw tails appear, usually toward large $\alpha$: a few very large overestimate compensate the large majority of underestimates.
If the apparent bias phenomenon is avoided, one can ensure that the estimate of $\alpha$ is unbiased: that this to say that there is no additive irreducible error on the $\alpha$ computed after each AMS run (dependent on $N$ or not) and that the average over AMS realisations will converge toward the reference when the number of realisations $o$ goes to infinity.
When errors occur, the number of clones in each realisation can be increased.
One can also rely on the information brought by correctly computed reactive trajectories to construct a better reaction coordinate.

The most successful applications of AMS finally  means that c) the estimated mean first passage time is precise.
Again, one can provide an estimate of $T$ and an interval of confidence.
Note however that unlike the properties of trajectories or the probability of collapse,
one cannot demonstrate that the estimator of equation~(\ref{EQT}) is unbiased.
In practice the relative error on the estimate of $T$ using a small number of clones can be quite larger that the error on the estimate of each of the separate terms involved in equation~(\ref{EQT}).
In order to reduce said bias, it has been observed that increasing the number of clones and improving the reaction coordinate will improve the estimate of $T$.
Biases in the estimate of $T$ generally lead to overestimates.
These biases arise because the estimate of $T$ is not direct and comes from the product of several other random variables.
There are version of AMS that lead to a more direct estimate of $T$.
However, this rewriting comes with additional constraints, such as fixed durations for trajectories \citep{rolland_Les}.

All things considered, we can assert that AMS computations give realiable results when the reactive trajectories and the probability of crossing are computed precisely,
with clear accelerations of computations with respect to DNS.
When these two quantities are correctly estimated, we can use the AMS results to discuss the physics of the multistability of the problem we investigate.
This also means that we can reuse the information on reactive trajectories from these computations in order to improve the reaction coordinate.
Since the probability of crossing follows the same exponential scalings as the PDFs or the mean first passage times,
it can be used as a proxy to investigate the Large Deviations or PDF tails if we are not satisfied with the estimate of the mean first passage time.

\subsection{Initial condition generation}\label{init}

In this section we present the procedure used for the generation of turbulent initial conditions used to study the collapse of turbulence at Reynolds number $R$ (by means of DNS or AMS).
This procedure uses mixing at a higher Reynolds number $R_+=600$ to naturally decorrelate turbulent initial conditions.
  It is  easily parallelised with a minimal load imbalance. If we use $p$ threads, we generate $N_j=\left\lceil \frac{N}{p} \right\rceil$ or $N_j=\left\lfloor \frac{N}{p} \right\rfloor$ initial conditions on each thread $j$. Each thread uses distinct seeds for random number generation.
\begin{enumerate}[a)]
\item ~On each thread $j$, $1\le j\le p$, we first create an artificial velocity field $u_y=0$, $u_z=0$, $q=0$, $u_x=0.4\sin\left(\pi \frac{y+1}{2} \right)\cos\left(4\pi \frac{z}{L_z}M_z \right)$, with spanwise wavenumber $M_z=\max\left( 1,\left\lfloor \frac{L_z}{5} \right\rfloor\right)$.
This corresponds to streamwise velocity tubes which are prone to the streaks instability, a key process in the cyclic SSP  \citep{waleffe1997self} and should thus lead to wall turbulence.
\item\label{iu0} ~On top of the velocity components, we add a noise red in $x$ and $z$ and white in $y$. This yields an initial velocity field $\mathbf{u}_0$.
This red noise is such that the variance of its  Fourier mode for streamwise wavenumber $n_x$ and spanwise wavenumber $n_z$ on component $l$ is $\sigma_l\gamma_{n_x}\gamma_{n_z}$, with shape factors $\gamma_{n_{m}}=1$ for $0\le |n_{m}|\le 6$, $\gamma_{n_{m}}=\frac{6}{|n_{m}|}$ for $n_{m}>6$, with $m=x$ or $z$.
We set $\gamma_{n_{m}}=0$ if $n_{m}>28$  in the $L_x\times L_z=24\times 18$ system and if $n_{m}>36$  in the $L_x\times L_z=36\times 27$ system, with $m=x$ or $z$.
We use the variances $\sigma_x=0.05$, $\sigma_y=0.0025$ and $\sigma_z=0.015$.
\item ~This initial condition is evolved for $T_0=500$ in the $L_x\times L_z=24\times 18$ system and $T_0=200$ in the $L_x\times L_z=36\times 27$ system at mixing Reynolds number $R_+=600$. It has been checked that this duration was long enough so that natural buffer layer turbulence forms \citep{pope2001turbulent}. If the kinetic energy of this velocity field is larger than $0.03$, this yields the fields $\{\mathbf{u}_{R_+,1,0}, q_{R_+,1,0}\}$, otherwise we go back to step~\ref{iu0}) and generate a  new $\mathbf{u}_0$ with a new realisation of the red noise.

\item\label{uiR} ~We then generate the $N_j$ initial conditions $\{ \mathbf{u}_n,q_n\}_{0\le n<N_j}$ in the following manner. We first evolve  the fields $\{\mathbf{u}_{R_+,1,n},q_{R_+,1,n}\}$ at $R_+$ for $T_+=500$ ($L_x\times L_z=24\times 18$) or $T_+= 200$ ($L_x\times L_z=36\times 27$), yielding the fields $\{\mathbf{u}_{R_+,2,n},q_{R_+,2,n}\}=\{\mathbf{u}_{R_+,1,n+1},q_{R_+,1,n+1}\}$, which will be used to generate the initial condition at $R$ and generate a subsequent decorrelated field at $R_+$.
\item ~We then set the Reynolds at $R$ (where we study collapse) and we set $\{\mathbf{u}_{R,1,n},q_{R,1,n}\}=\{\mathbf{u}_{R_+,2,n},q_{R_+,2,n}\}$.
These velocity and pressure fields are evolved at $R$ during $T_-=750$ ($L_x\times L_z=24\times 18$) or $T_-=500$ ($L_x\times L_z=36\times 27$).
 This duration is chosen so that enough mixing has occurred and each initial condition is decorrelated from the others.
 We obtain $\{\mathbf{u}_{R,2,n},q_{R,2,n}\}$. We then let it evolve until the kinetic energy is either within $1.25\%$ of $E_0$, in which case we have our $n^{\rm th}$ initial condition $\{\mathbf{u}_n,q_n \}$ or is below $0.03$.
 In that case we restart at~\ref{uiR}) by setting $\{\mathbf{u}_{R_+,1,n},q_{R_+,1,n}\}=\{\mathbf{ u}_{R_+,2,n},q_{R_+,2,n}\}$.
 We use $E_0=0.055$ in the system of size $L_x\times L_z=24\times 18$ and $E_0=0.052$ in the system of size $L_x\times L_z=36\times 27$.
\end{enumerate}

This approach ensures that we have $N$ decorrelated initial conditions which verify a given constraint on kinetic energy (for instance).

\subsection{Reaction coordinates}\label{scoord}

\begin{figure}
\centerline{
\includegraphics[width=0.3\textwidth]{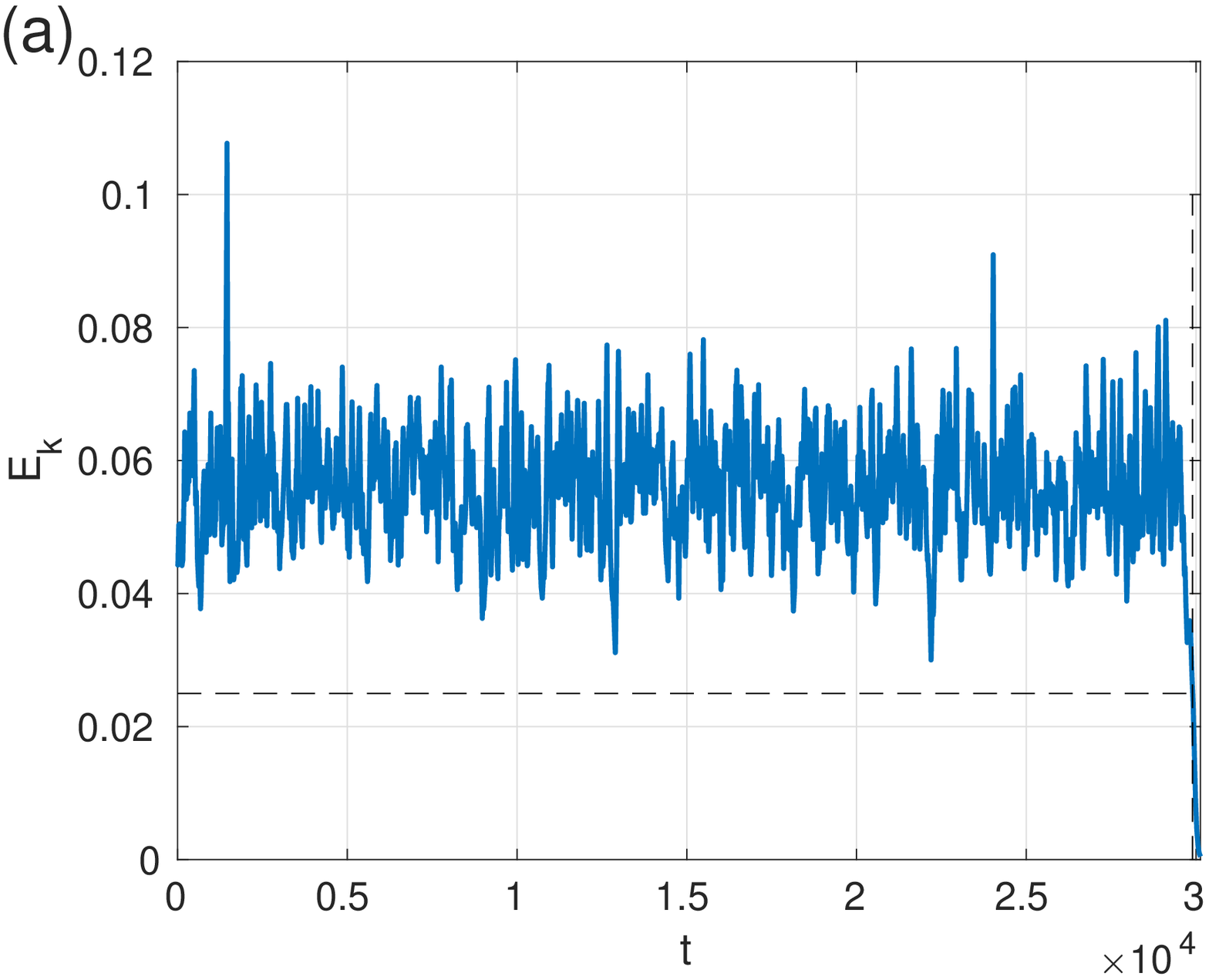}
\includegraphics[width=0.3\textwidth]{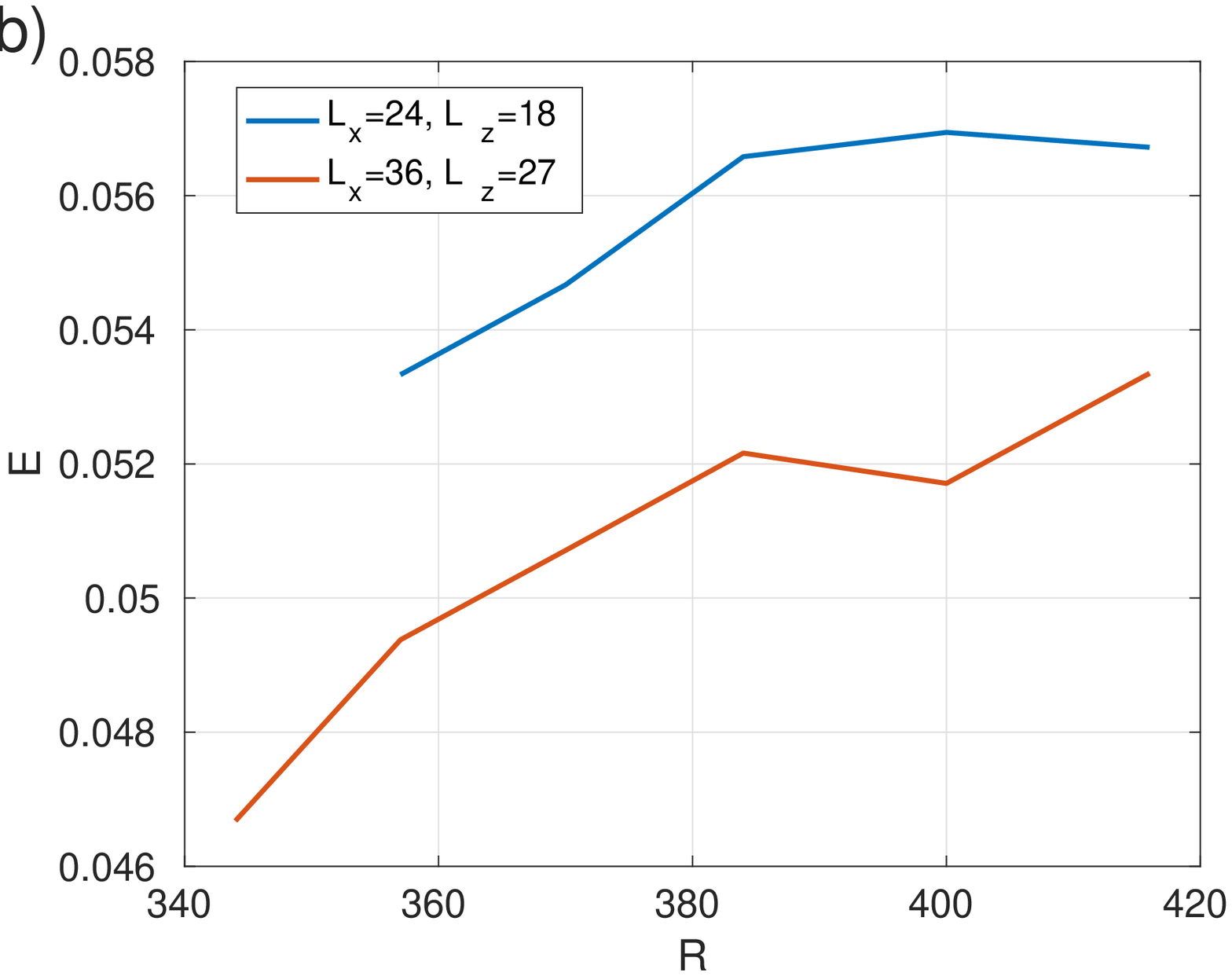}
\includegraphics[width=0.3\textwidth]{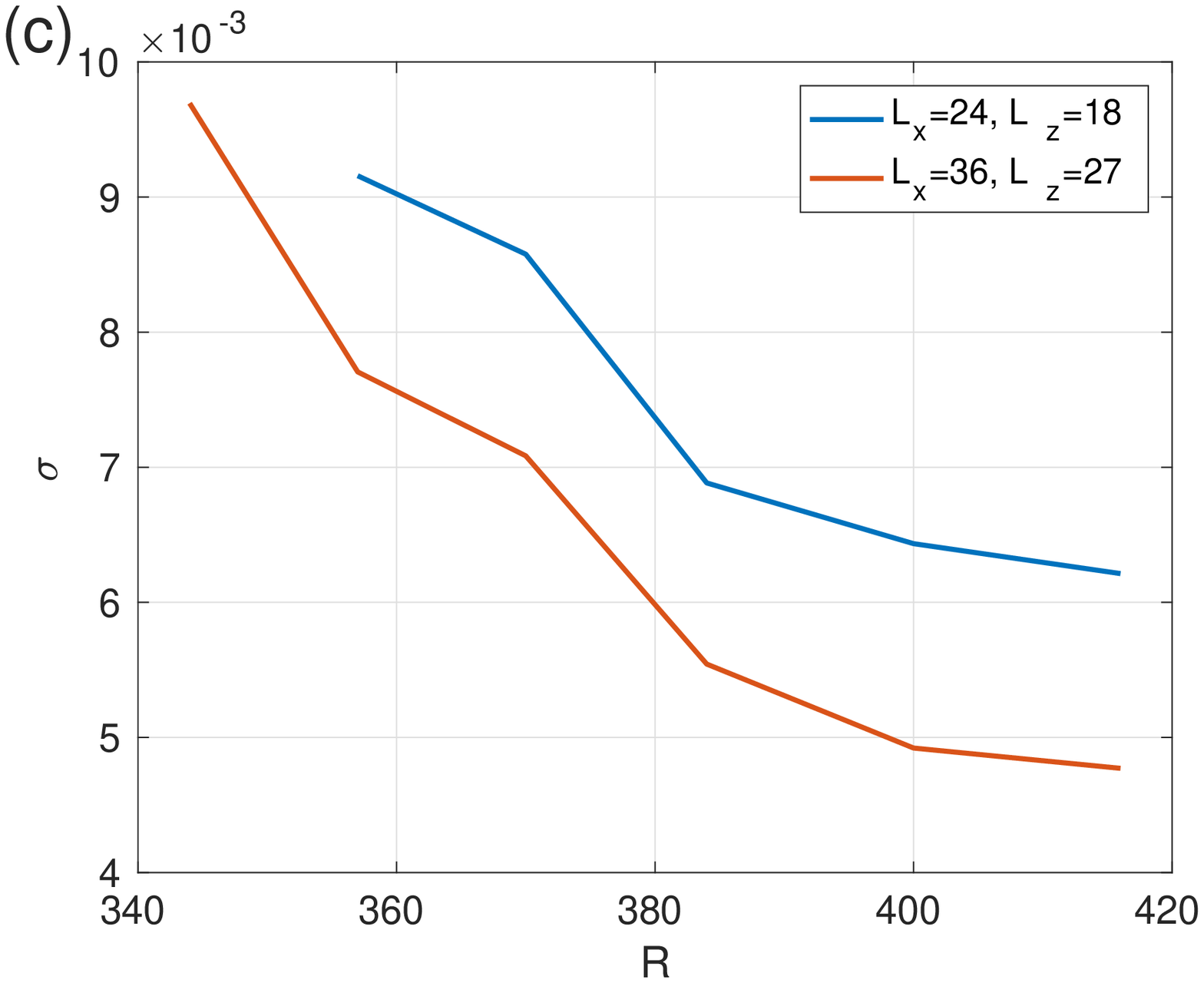}}
\caption{(a) Example of a time series of kinetic energy in a domain of size $L_x\times L_z=24\times 18$ at Reynolds number $R=370$ The black dashed lines indicate were the sampling is stopped for the construction of the empirical probability density function of kinetic energy conditioned to not collapse happening. (b) Conditional average of the kinetic energy as a function of the Reynolds number for domains of size $L_x \times L_z=24\times 18$, $L_x\times L_z=36\times 27$. (c) Conditional variance of the kinetic energy as a function of the Reynolds number for domains of size $L_x \times L_z=24\times 18$, $L_x\times L_z=36\times 27$.}
\label{stat_cond}
\end{figure}

Since the kinetic energy $E_k$ (Eq.~\ref{ekin}) of the turbulent flow is fluctuating around a conditional average, while the kinetic energy of the laminar flow is zero (Fig.~\ref{stat_cond} (a)), a first choice to construct the reaction coordinate is to use $E_k(t)$. The simplest reaction coordinate based on $E_k$ is affine. We can therefore propose the reaction coordinate $\Phi_E$ defined directly as a function of time by
\begin{equation}
\Phi_E(t)=\frac{E_{\rm turb}-E_k(t)}{\Delta E}\,.\label{ephie}
\end{equation}
In this affine function, we use the shift  to the kinetic energy $E_{\rm turb}$ and the normalisation  of the reaction coordinate $\Delta E$.
These two quantities are chosen so that $\Phi_E\simeq 0$ when the flow is turbulent and $\Phi_E\simeq 1$ when the flow is laminar.
It is natural to choose $E_{\rm turb}$ close to some average of the kinetic energy and $\Delta E\lesssim E_{\rm turb}$. In order to estimate $E_{\rm turb}$, one can sample $E_k(t)$ and construct the empirical probability density function conditioned on the flow experiencing no collapse of turbulence, in a system of given size and Reynolds number. One can for instance retain the part of the time series where $E_k(t)\ge 0.025$ (indicated by the black dashed line in figure~\ref{stat_cond} (a)). Using this, one can compute a sample mean of the kinetic energy $E$ (Fig.~\ref{stat_cond} (b)), and then choose close enough $E_{\rm turb}$ and $\Delta E$ accordingly. We will state what values of $E_{\rm turb}$ and $\Delta E$ are chosen in AMS computations in section~\ref{val} and section~\ref{hole}. We also compute the conditional variance $\sigma$ as a function of the Reynolds number for the two system sizes (Fig.~\ref{stat_cond} (c)): we will use this quantity to choose the hypersurface $\mathcal{C}$, so that we typically have $E_{\rm turb}-E_k(t)\simeq \frac{\sigma}{2}$ on $\mathcal{C}$. This corresponds to a short excursion where the kinetic energy is away from its conditional average by typically half a standard deviation. Using the reaction coordinate, we set $\mathcal{A}$ as all velocity fields such that $\phi_{E}\le 0$ and $\mathcal{B}$ as all velocity fields such that $\Phi_E\ge 1$.

One can note that the variance of the kinetic energy decreases with Reynolds number in the range of interest (Fig.~\ref{stat_cond} (c)).
The larger variance for $R\le 380$ are caused by  fluctuations of kinetic energy toward low values at lower Reynolds numbers.
The distributions of kinetic energy are asymmetric for $330\le R \le 380$ with a relative skewness that can be below $-0.5$.
As the Reynolds number is increased, the excursions of $E_k$ toward low values become less and less probable and the PDF of kinetic energy become narrower and more symmetric (see \citep{rolland2015mechanical} Fig.~ 9 (c), in a case where the domain can accommodate laminar-turbulent bands).
In domain large enough to contain laminar-turbulent bands (Fig.~\ref{rolland_figint}, centre),
one can observe a narrow peak of the variance in the range $378\lesssim R\lesssim 382$ (at our numerical resolution),
because temporary closing of laminar holes can be observed on timescales smaller than $\mathcal{O}\left(10^4\right)$.
At higher Reynolds numbers, the variance of kinetic energy increases again because of turbulent fluctuations.

In appendix~\ref{rolland_sappa}, we use an other reaction coordinate $\Phi_{\rm asym}$, based on $\psi(t)$, the asymmetry of the streamwise velocity field with respect to the midplane $y=0$
\begin{equation}
\psi(t)=\frac{1}{V}\int_{x=0}^{L_x}\int_{y=-1}^1\int_{z=0}^{L_z} u_x {\rm sgn}(y)\,{\rm d}x{\rm d}y{\rm d}z\,,\label{eqasym}
\end{equation}
where ${\rm sgn}(y)$ stands for the sign of $y$. The choice of this reaction coordinate is motivated by the fact that we have mostly $u_x<0$ (coming from low speed streaks) for $y>0$ and mostly $u_x>0$ (coming from high speed streaks) for $y<0$ when velocity streaks (and thus wall turbulence) are present in the flow \citep{jimenez1991minimal,hamilton1995regeneration,kawahara2003linear}. Note that with this definition $\psi$ is mostly negative. We proceeded in the same manner as with the kinetic energy in order to construct a reaction coordinate: conditional averages of $\psi$ are performed in order to calculate $\Phi_{\rm asym}(t)=1-\frac{\psi(t)}{\langle \psi\rangle}$.
\subsection{Direct numerical simulations of collapses}\label{dns_couette}

We perform two types of direct numerical simulations of turbulence collapse. The first kind of simulation consists in letting the flow freely evolve from initial conditions computed following the method of section~\ref{init}. These simulations are stopped when the flow has laminarised. This is deemed to have happened when the reaction coordinate has reached one. In these simulations, we only save time series, such as those of $E_k$, $E_{k,x}$, $E_{k,y-z}$ and $\Phi_E$ (an example of such time series in given in figure~\ref{stat_cond} (a)). From the last part of each DNS, we can extract the time series of these quantities in natural collapse trajectories. We will perform two types of such DNS: some free natural DNS, and some forced DNS, in order to check the effect of additional noise on the collapse of turbulence. An additional noise will necessarily be included in AMS computations, and we wish to know what is the minimal error caused  on the probability of collapse, trajectory features and durations and mean first passage times by the addition of this noise. Since we cannot add a comparable temporally localised perturbation in DNS, we will add a noise red in space and white in time. In that case the forcing $\textbf{f}$ (Eq.~(\ref{nsf})) is exerted at all time and it is red in $x$ and $z$. The forcing noise is characterised by its spatial correlation function
\begin{equation}
\langle f_l(\textbf{x},t)f_m(\textbf{x}',t')\rangle=\delta_{lm}\delta(t-t')C_l(x-x',z-z')\tilde{C}_l(y-y')\,,\, \hat{C}_l(n_x,n_z)=\Gamma_{l,n_x,n_z}\,,
\end{equation}
with $C_l$, the correlation function of the noise for variables $x$ and $z$, along component $l$, $\tilde{C}_l$, the correlation function of the noise for variable $y$ along component $l$. The correlation function $C_l$ is  prescribed using its Fourier transform $\hat{C}_l$, with amplitude
 $\Gamma_{l,n_x,n_z}$, where $n_x$ stands for the Streamwise  wavenumber, $n_z$ stands for the spanwise wave number. In this text we always use $\tilde{C}_l(y-y')=\delta(y-y')$ in numerical simulations using a finite number $N_y$ of Chebyshev modes in the wall normal direction. In DNS, we will always set the amplitude $\Gamma_{i,n_x,n_z}=\sigma_l\gamma_{n_x}\gamma_{n_z}$ using
 the previously defined shape factors $\gamma_{n_{m}}$ for $m=x$ or $z$ (see \S~ \ref{init}). We set the variance $\sigma_l=10^{-9}$ for all components.

Another kind of direct numerical simulation consists in repeating the first stage of AMS, where we start the simulation from our initial conditions, let them cross the hypersurface $\mathcal{C}$ and then either laminarise or go back to turbulence as detected by $\Phi_E<0$. From these, we can compute the proportion of trajectory that laminarises and thus have an unbiased estimate of the probability of collapse and trajectory durations and validate AMS estimates of these quantities.

\section{System of size $L_x\times L_z=24\times 18$: reactive trajectories and validation of AMS computations}\label{val}

In this section, we will compare the properties of the reactive trajectories computed by means of AMS and by means of DNS as well as discuss the trajectory properties in a system of size $L_x\times L_z=24\times 18$ at Reynolds number $R=370$.
We choose rectangular shaped domains: a shape  comparable to what is feasible in experiments, albeit at smaller size (see Fig.~\ref{rolland_figint}) (tilted periodical domains being a numerical experiment).
The values of the domain size  and Reynolds number imply that the study of turbulence collapse is affordable by means of direct numerical simulations. Results of DNS and AMS computations will be compared to assert what degree of trust can be placed in general outputs of AMS computations. For estimates (of the probability of collapse $\alpha$, the average duration of trajectories $\tau$,
and the mean first passage time $T$ and paths followed by trajectories) coming from both types of computations, we will provide intervals of confidence and check whether they overlap or if biases are present in the results of AMS computations.
From these we will be able to deduce what should be the error bars \emph{a minima} that should be placed on the results of anticipated AMS computations.
This will be useful when no DNS are available for comparison, for instance in larger domains at larger Reynolds numbers, in section~\ref{hole}.

In these computations, we will use $E_{\rm turb}=0.05$, $\Delta E=0.048$ for the reaction coordinate $\Phi_E$ (Eq.~(\ref{ephie})), in conjunction with the conditional average of the kinetic energy (Fig.~\ref{stat_cond} (b)). The properties of the initial conditions in $\mathcal{A}$ are given in section~\ref{init}.
The hypersurface $\mathcal{C}$ will correspond to the set of velocity fields such that $\Phi_E=0.06$, which roughly corresponds to half a standard deviation from the conditional average (Fig.~\ref{stat_cond} (c)).
We set the parameter $\xi=0.2$ with the converging $\Phi_b$ (Eq.~(\ref{conv}), \S~\ref{rolland_sappa}, Fig.~\ref{rolland_figrare} (c)) for the anticipation of branching.
This corresponds to  a good trade-off between the need for mixing and separation of trajectories, and the minimisation of the number of retries when branching trajectories.
The AMS computations use $N=120$ clones and suppress $N_c=32$ clones at each iteration. They are ran on 16 threads, which leads to reasonable load balancing.

\subsection{Collapse trajectories}\label{traj}

\subsubsection{Visualisation of the collapse trajectories}

We first describe the velocity fields during the collapse, in trajectories computed by AMS.
With the use of AMS, we could produce far more time and space resolved three components velocity fields than was previously shown when using direct numerical simulations.
Reactive and non-reactive trajectories are simply reconstructed and saved  at time intervals $\delta t>1$ from the checkpoints used in AMS computations.

\begin{figure}
\centerline{\includegraphics[width=14cm,clip]{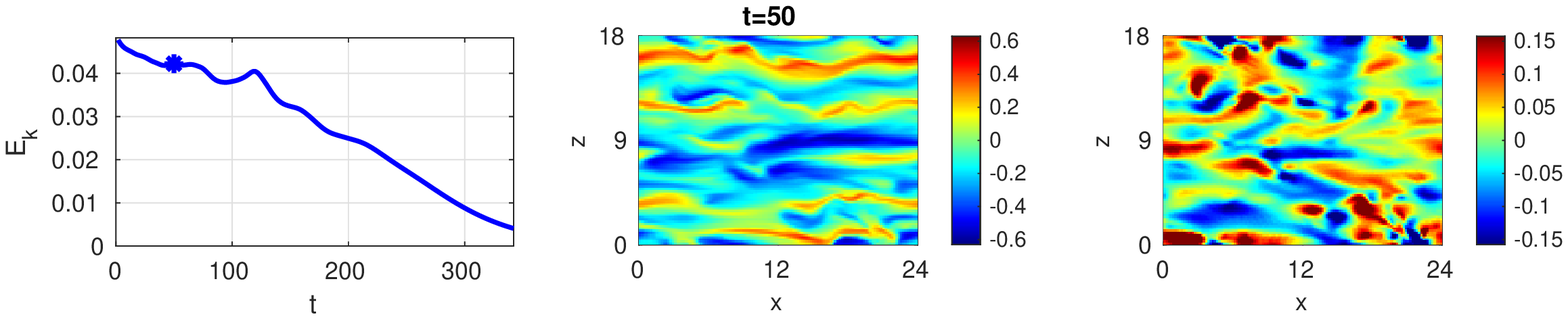}}
\centerline{\includegraphics[width=14cm,clip]{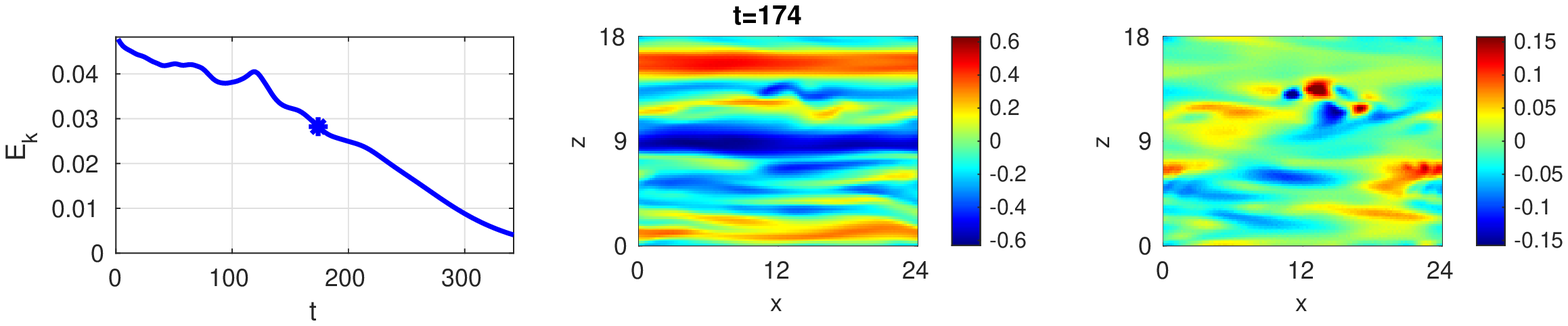}}
\centerline{\includegraphics[width=14cm,clip]{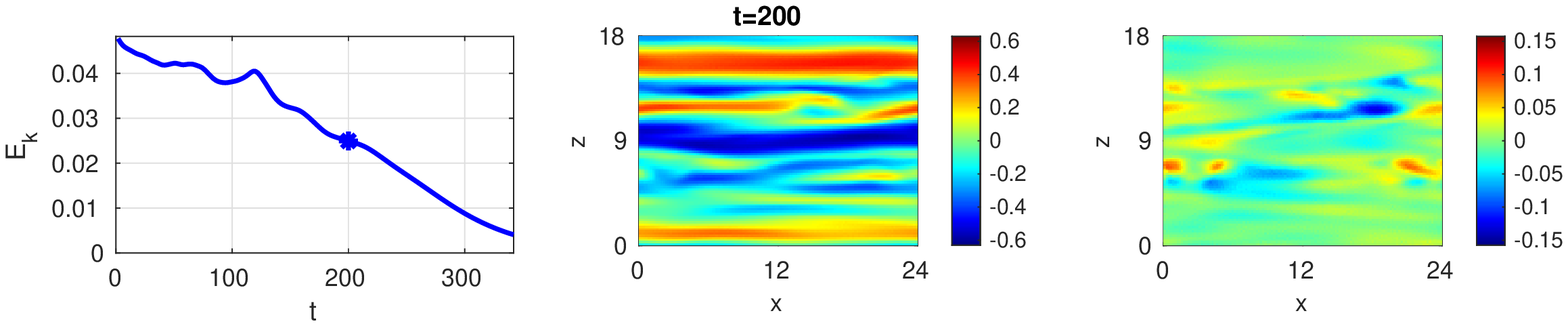}}
\centerline{\includegraphics[width=14cm,clip]{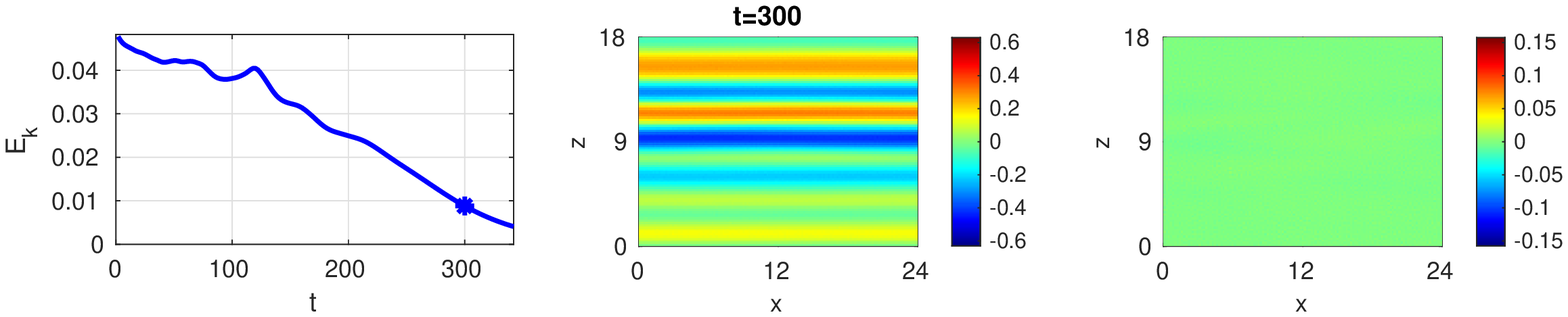}}
\caption{Time series of the kinetic energy (left panels) with a dot indicating the point in time of each line ($t=50$, $t=174$, $t=200$, $t=400$),
alongside colour levels of the streamwise velocity (central panels) and spanwise  velocity (right panels), at corresponding times, in the midplane $y=0$
during a collapse trajectory in a domain of size $L_x\times L_z=24\times 18$ computed by AMS.}
\label{coll24_18}
\end{figure}

\begin{figure}
\centerline{\includegraphics[width=14cm,clip]{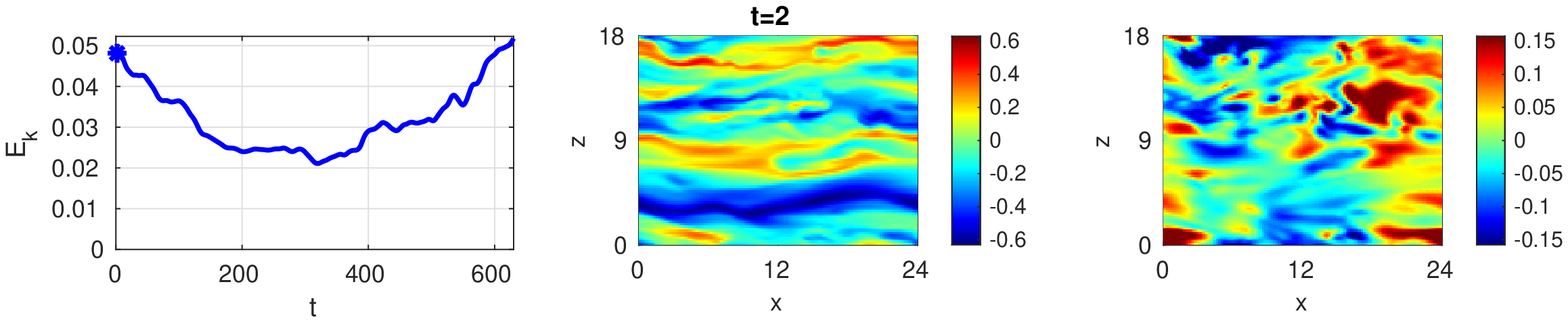}}
\centerline{\includegraphics[width=14cm,clip]{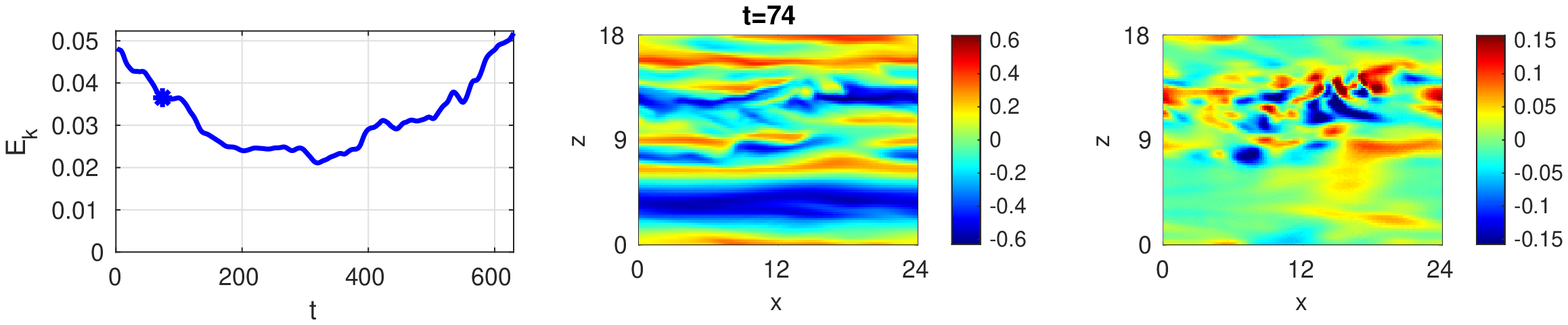}}
\centerline{\includegraphics[width=14cm,clip]{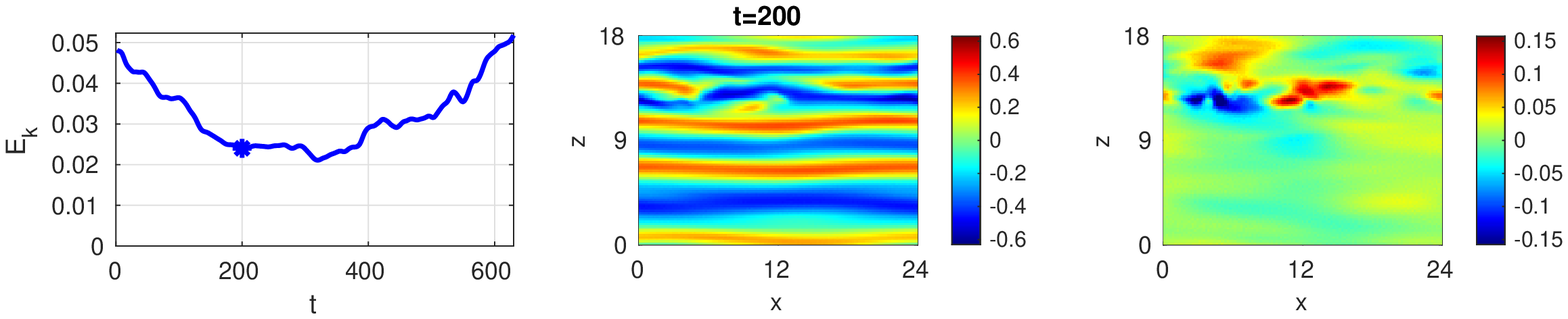}}
\centerline{\includegraphics[width=14cm,clip]{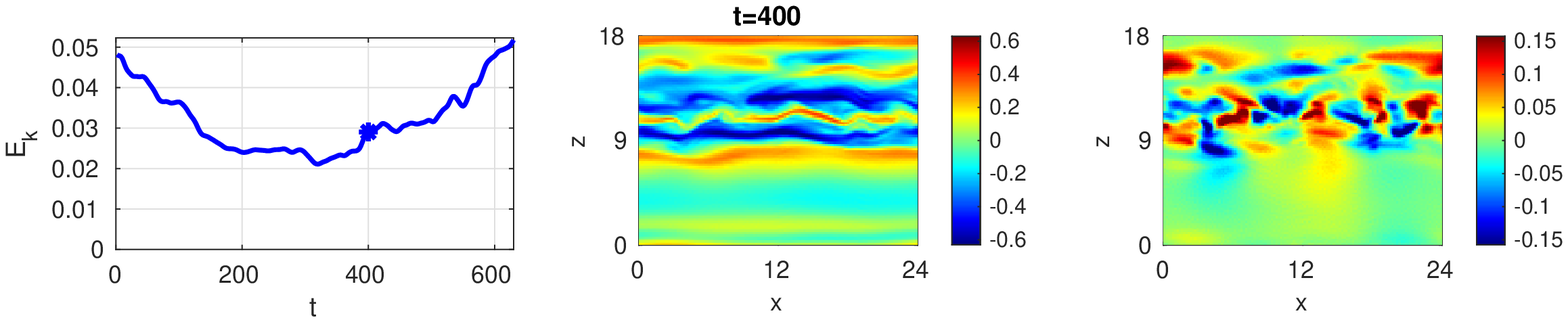}}
\centerline{\includegraphics[width=14cm,clip]{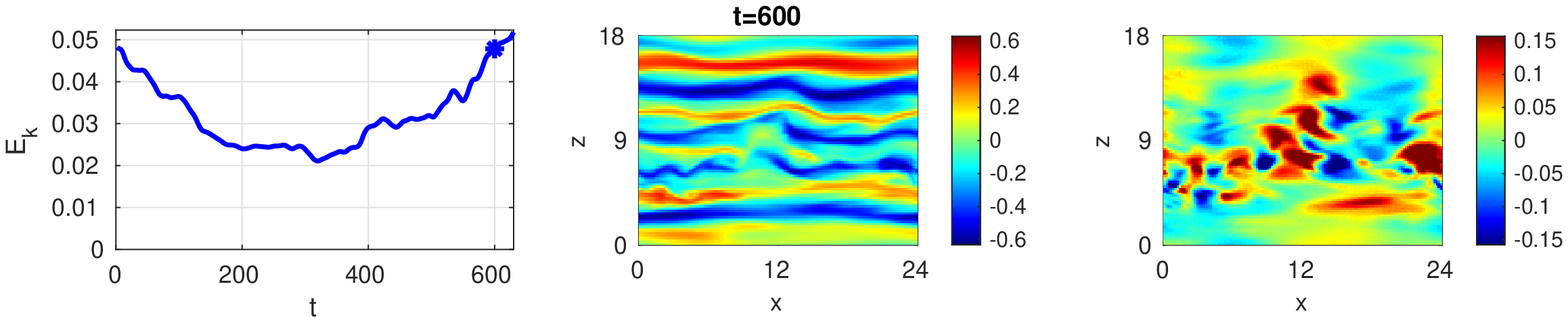}}
\caption{Time series of the kinetic energy (left panels) with a dot indicating the point in time of each line ($t=2$, $t=74$, $t=200$, $t=400$, $t=600$), alongside colour levels of the streamwise velocity (central panels) and spanwise velocity (right panels) in the midplane $y=0$ during a non-reactive trajectory (hole opening then closing), at corresponding times, in a domain of size $L_x\times L_z=24\times 18$ computed by AMS.}
\label{bounce24_18}
\end{figure}

We display colour levels of the streamwise and spanwise components of the velocity field in a midplane at successive instants during a collapse trajectory in figure~\ref{coll24_18} and during one of the non-reactive trajectory remaining at the last stage of AMS in figure~\ref{bounce24_18}.
We will focus on the two fields $(u_x,u_z)$ because of the physics of wall flows and experimental constraints:
\begin{itemize}
\item Indeed, on the one hand, it has been shown in models and observed in some experiments and simulations that there was a distinct behaviour of velocity components.
One finds velocity streaks contained mostly in $u_x$ as well as streamwise vortices, visible in the streamwise vorticity field $\omega_x=\partial_yu_{z}-\partial_zu_{y}$, but that can be observed through the component $u_z$ (they are the main contribution to this velocity component) \citep{hamilton1995regeneration}.
Moreover it has been very recently observed in models, simulations and experiments that decay of turbulence was much faster in streamwise vortices (or its proxies) than in velocity streaks  \citep{rolland_pre18,liu2020anisotropic,gome2020statistical}.
We will endeavour to  discuss this observation in specifically computed collapse trajectories.
\item On the other hand, most experimental observations are performed by PIV using a plane parallel to the wall,
thus capturing $u_x$ and $u_z$ in a $x-z$ plane near the midgap \citep{liu2020anisotropic,de2020transient}.
\end{itemize}

We first visualise the velocity fields as turbulence collapses. The reactive trajectory starts from uniform buffer layer turbulence in the whole domain (Fig.~\ref{coll24_18}, $t=50$).
The kinetic energy is of course close to the conditional average. As time goes on ($t=174$), the kinetic energy decreases, and observation indicates
that the spanwise component of the velocity field is much less intense than in the initial condition. Moreover, the largest values of $u_z$ are spatially localised
(intense small spatial scales for $9\lesssim z\lesssim 14$, less intense, largest spatial scales for $0\lesssim z\lesssim 9$).
The streamwise velocity field has not yet decayed in amplitude, however it is streamwise invariant where $u_z$ is almost $0$.
As the flow laminarises ($t=200$), the spanwise velocity field further decays while the streamwise velocity field becomes even more streamwise invariant. The amplitude of $u_x$ remains comparable to what was found in the initial condition $|u_x|\lesssim 0.6$.
This finally leads to a situation where $u_z$ is negligible and only streamwise velocity tubes are left in the flow ($t=300$).
These tubes then undergo viscous decay.

We can comparably examine a non-reactive trajectory, that is to say a realisation of the dynamics that undergoes a large enough excursion of kinetic energy to be retained at the last stage of AMS computations,
but that still goes back to a fully turbulent flow (Fig.~\ref{bounce24_18}).
As with all other trajectories, this one starts with turbulence in the whole domain (Fig.~\ref{bounce24_18}, $t=2$).
We again observe that $u_z$ decays faster than $u_x$ ($t=74$). In that case the spanwise localisation of the decay is not disputable. Along with the decay of $u_z$, the streamwise velocity component becomes streamwise invariant ($t=200$).
The kinetic energy fluctuates for some time around a plateau ($150\lesssim t\lesssim 350$), then increases again. The values taken by $u_z$ are getting more intense over an area of the domain that increases ($t=400$).
Note that the streamwise velocity field had decayed locally during the plateau (for $0\lesssim z\lesssim 7$).
We observe an asymmetry between the decay and the reinvasion processes. The component $u_z$ retracts to smaller areas than $u_x$ during the decay: the active-quiescent interface (typically planes $z=\text{cst}$) is not the same for $u_x$ and $u_z$. A strict definition of these interfaces will be proposed in section~\ref{hole} and figure~\ref{suicont}.
Then, both $u_z$ and a streamwise dependent $u_x$ then fully reinvades the domain ($t=600$), in that case the active-quiescent interface is the same for both components. We thus observe a concomitant restart of both components of the self sustaining process of turbulence.

\begin{figure}
\centerline{
\includegraphics[width=0.3\textwidth]{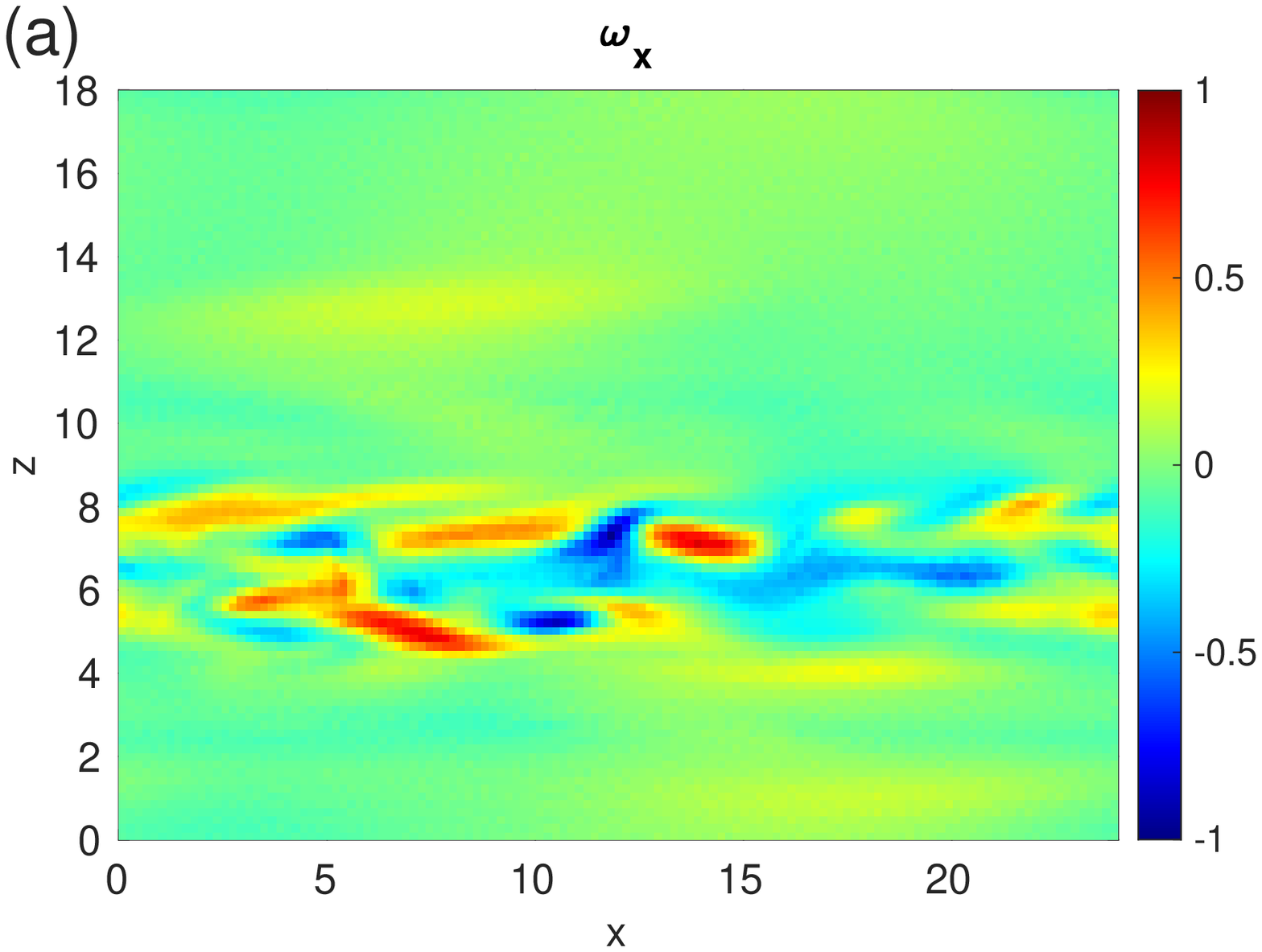}
\includegraphics[width=0.3\textwidth]{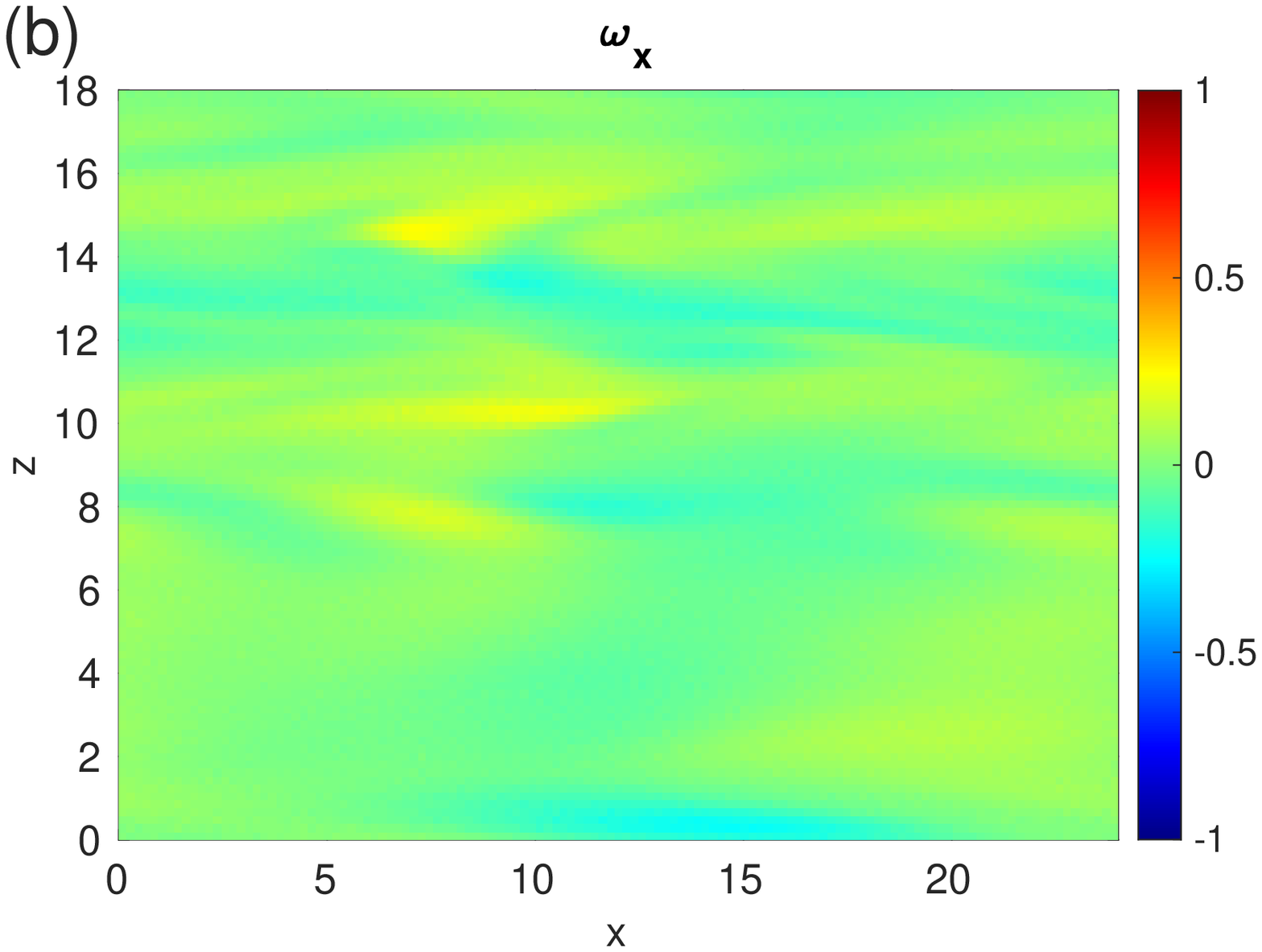}
\includegraphics[width=0.3\textwidth]{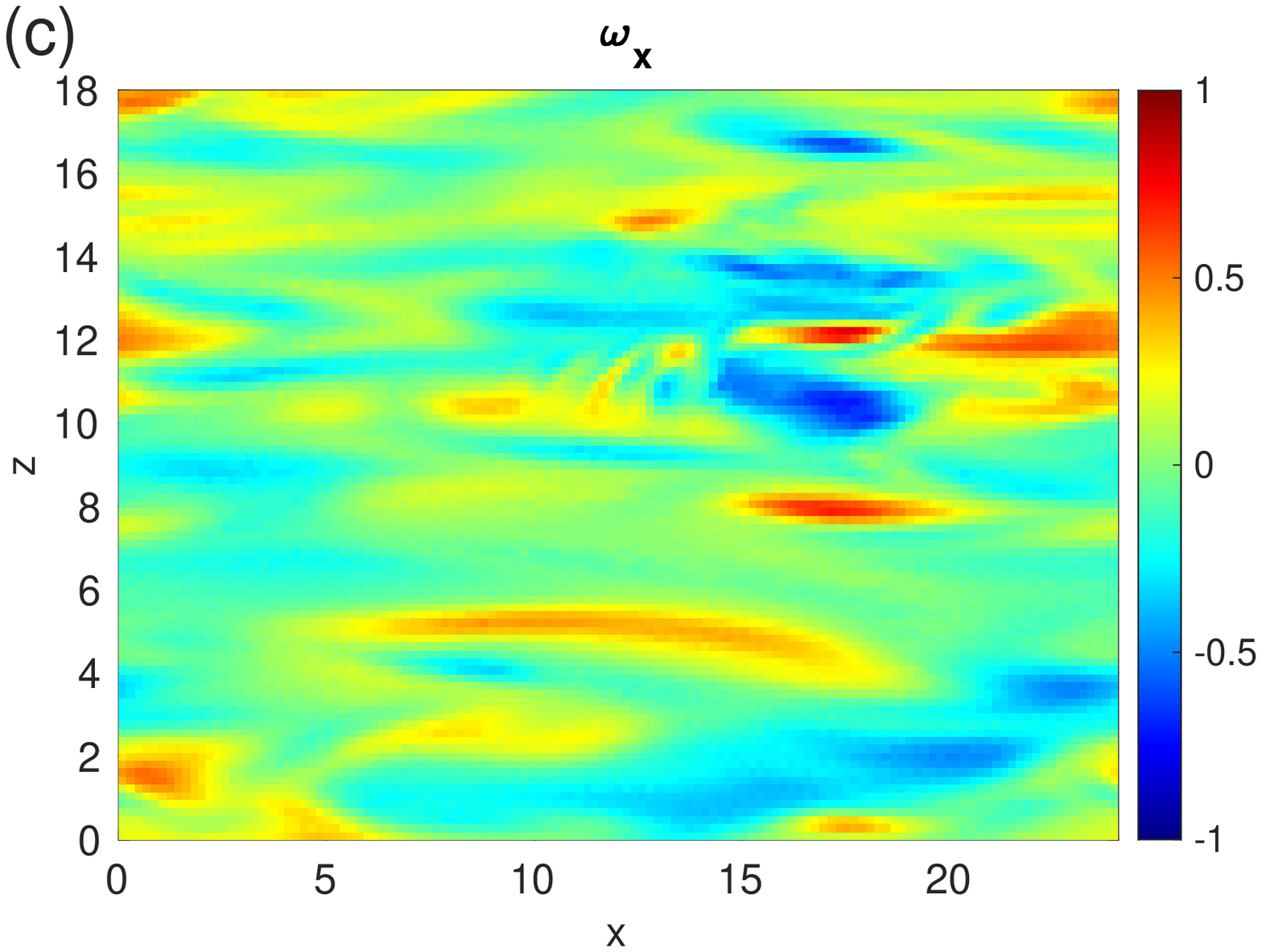}
}
\caption{Streamwise vorticity in the midplane $y=0$ in the system of size $L_x\times L_z=24\times 18$ for last states at the last branching stage of several AMS computations. (a) Run estimating $\hat{\alpha}=0.045$, $\hat{T}=1.5\cdot 10^4$, (b) run estimating $\hat{\alpha}=0.086$, $\hat{T}=1.1\cdot 10^4$, (c) run estimating $\hat{\alpha}=0.022$, $\hat{T}=3.7\cdot 10^4$.}
\label{last24_18}
\end{figure}

We can complement this view by observing the streamwise vorticity field $\omega_{x{\rm , last}}=\partial_yu_{z{\rm , last}}-\partial_zu_{y{\rm , last}}$ leading to the largest value of the reaction coordinate in the last suppressed state in the last stage of AMS (Fig.~\ref{last24_18}).
In this system of relatively moderate size, even if one always has an impression that turbulence collapses through the formation of a laminar hole, the structure of this turning point state varies from one run to another.
One can observe clearly localised streamwise vortices (Fig.~\ref{last24_18} (a)), an almost entirely quiescent flow (Fig.~\ref{last24_18} (b)) or rather active streamwise vortices all over the domain (Fig.~\ref{last24_18} (c)).
One of the reasons we have fairly different $\omega_{x{\rm , last}}$ is that they correspond to different instants on similar trajectories that do display a hole opening.
This is also caused by the fact that the size of the domain is not large enough for the reactive trajectories to be concentrated around a typical trajectory.

In any case, we can have a first view of the path followed during the collapse of turbulence.
Of all the manner for the cyclic self sustaining process of wall turbulence to fail (streaks first, vortices first, or concomitant decay of streaks and vorticies), the observed scenario is that of a first decay of streamwise vortices followed by a decay of the velocity streaks.
No more energy is extracted from the base flow so that the velocity streaks, turned into tubes, slowly decay.
They do not undergo any new streaks instability that would refuel the streamwise vortices \citep{marquillie2011instability,jimenez1991minimal}.
We will examine all the reactive trajectory to show that far from being some picked visualisations, this description is statistically significant.
We observed spanwise localisation of this failure of the self sustaining process of turbulence.
From the observations and monitoring of turbulent fraction, one can note that there is more variability in this respect.
We will examine a larger system in section~\ref{hole} to check whether this observation is disputable or not.

\begin{figure}
\centerline{\includegraphics[width=7cm]{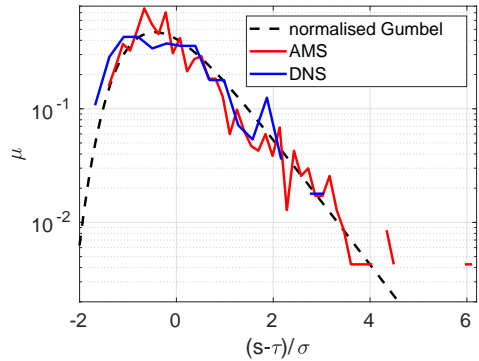}}
\caption{Normalised distribution of duration of collapse trajectories (in logarithmic scale) sampled by AMS ($1590$ samples), by DNS ($ 189$ samples) and compared to a normalised Gumbel distribution, in a system of size $L_x\times L_z=24\times 18$ at Reynolds number $R=370$.}
\label{duree_collapse}
\end{figure}

\subsubsection{Comparison between trajectories computed by AMS and by DNS}\label{comp_trajs}

We then start our comparison of the collapse trajectories computed by AMS and DNS by describing the duration of collapse trajectories, that is to say the time $s$ elapsed between the instant where the flow crossed the hypersurface $\mathcal{C}$ and the instant where the reaction coordinate reaches $\Phi=1$.
We have obtained 1590 trajectories from AMS and $189$ trajectories form DNS. We firstly compute the sample mean $\langle\tau\rangle_o$ and the sample variance $\sigma_\tau$ over AMS realisations of the trajectory durations, we  obtain $\tau_{AMS}=413\pm3$, $\tau_{DNS}=410\pm 7$ and $\sigma_{\tau,AMS}=105$ and $\sigma_{\tau,DNS}=91$.
As stated in section~\ref{sgood}, the error bars on $\tau$ are given by $\sigma_\tau$ divided by the square root of the number of samples\footnote{It is safe enough to assume each value is decorrelated enough from the large majority of the others to assert that the central limit theorem can be applied to the sampled mean: the most general formulation only require short length correlations.
There may be some degree of correlations between durations of reactive trajectories that arise from the same genealogy, but each genealogy is small enough compared to the sample size.}. We can see that both are within a few percent of one another and within the $66\%$ confidence interval of each other.
This shows that the reactive trajectories computed by AMS and DNS have almost the same duration, and thus that AMS has very little bias in the way it computes the collapse trajectories.
We then compare the sampled distributions of trajectory durations. In figure~\ref{duree_collapse}, we display the distribution of normalised durations $\tilde{s}=\frac{s-\tau}{\sigma_\tau}$ of collapse trajectories computed by means of AMS and of DNS. Both are very similar and seem to be very close to a Gumbel distribution. This  distribution of duration has been reported to be very close to a Gumbel in many instances of stochastically driven systems \citep{rolland_pre18,rolland2016computing,rolland2015statistical}.
This was often related to a simple structure of the reactive trajectory that went through a saddle of the deterministic part of the system. It is actually demonstrated that reactive trajectories durations have this distribution in one dimensional stochastic systems \citep{cerou2013length}.
To our knowledge, this is possibly the first time that such a distribution shape has been shown in turbulence without stochastic injection of energy.
This suggests us that the manner in which turbulence collapses in systems of large enough size can be compared to reactive trajectories in stochastic systems, in that it quite possibly follows some effective dynamics agitated by background turbulence.
 While it is more difficult to separate the dynamics into a deterministic part and a stochastic part, this indicates that the reactive trajectories can follow a rather simple path.

\begin{figure}
\centerline{
\textbf{(a)}
\includegraphics[width=0.3\textwidth]{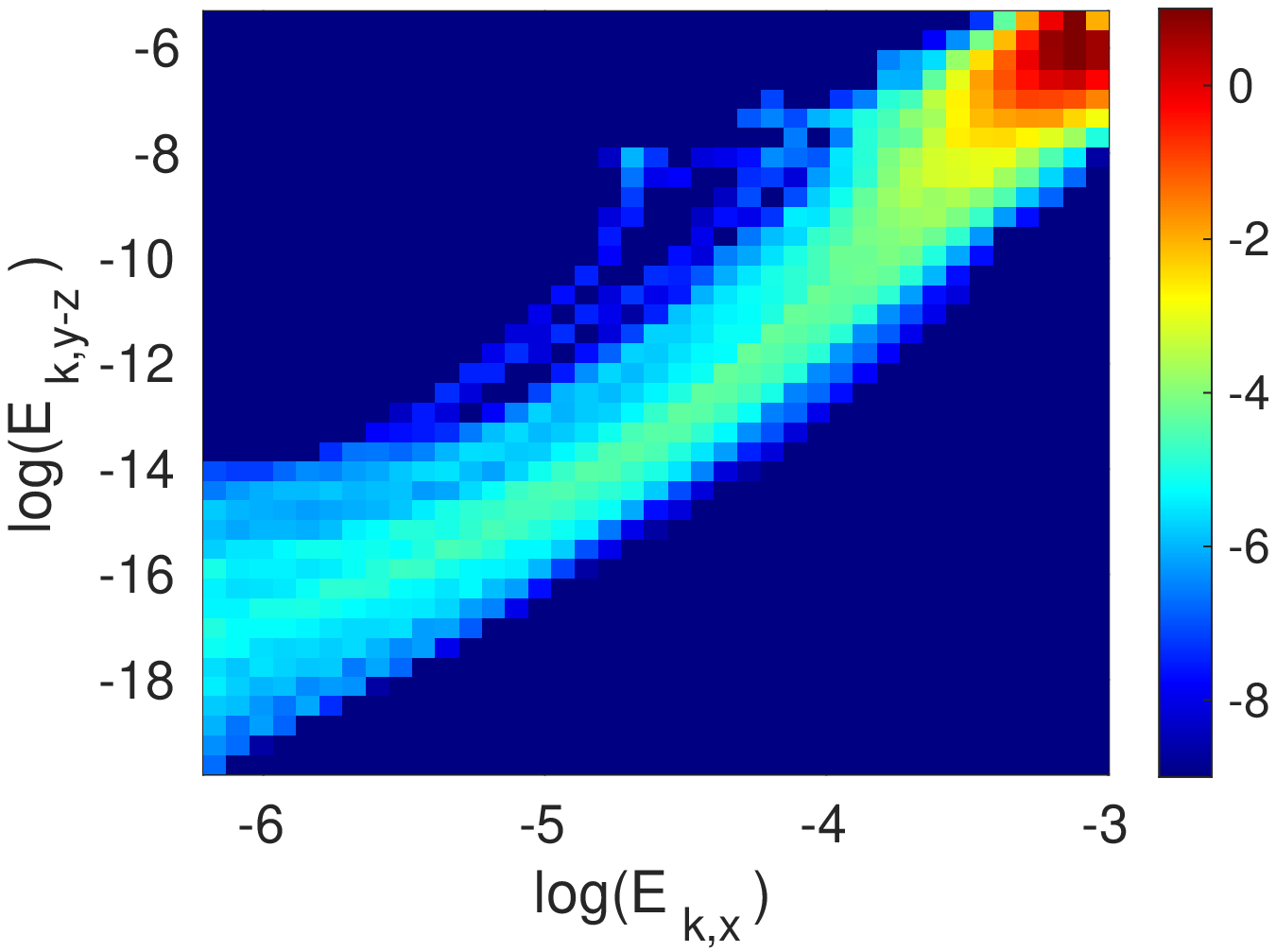}
\textbf{(b)}
\includegraphics[width=0.3\textwidth]{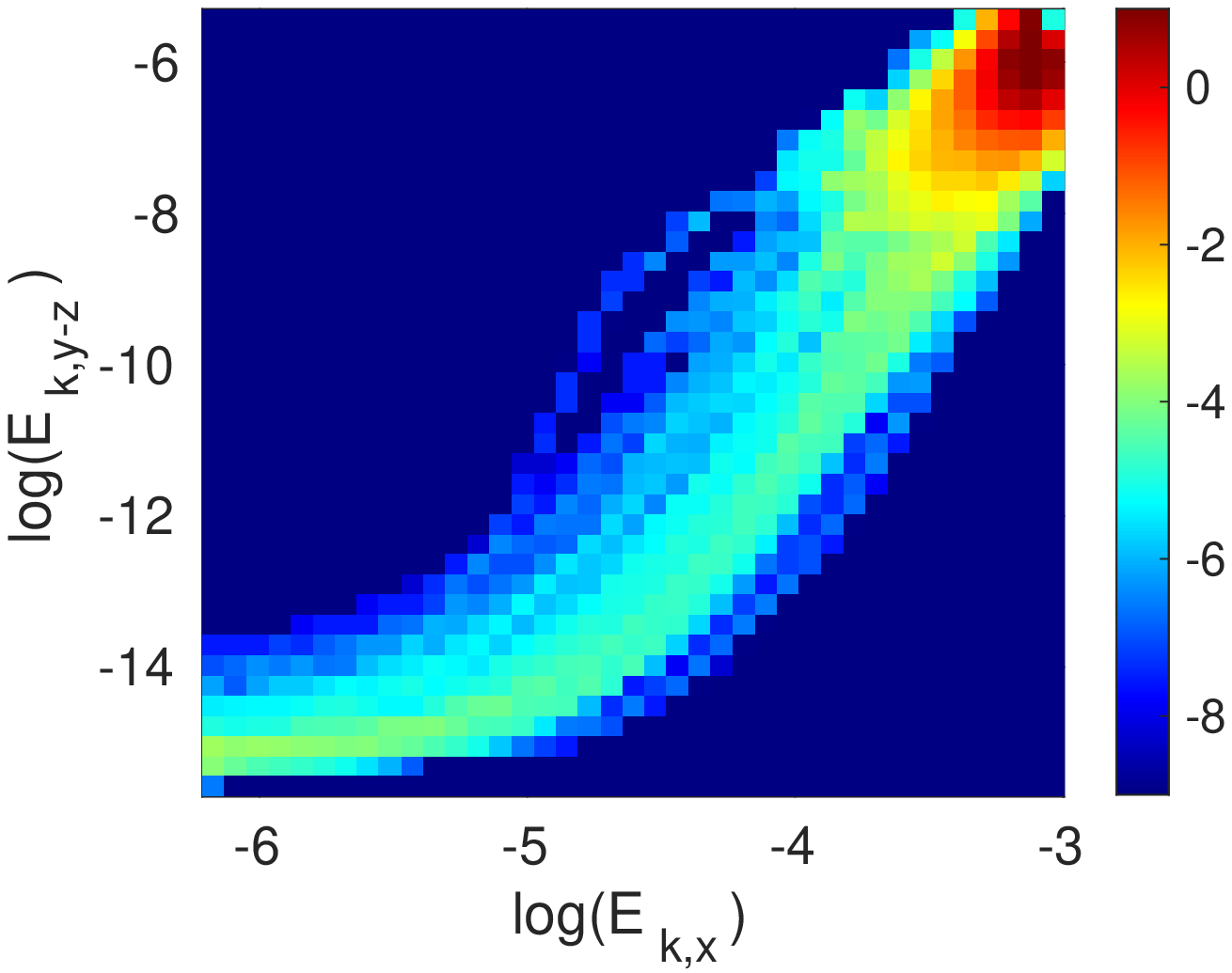}
\textbf{(c)}
\includegraphics[width=0.3\textwidth]{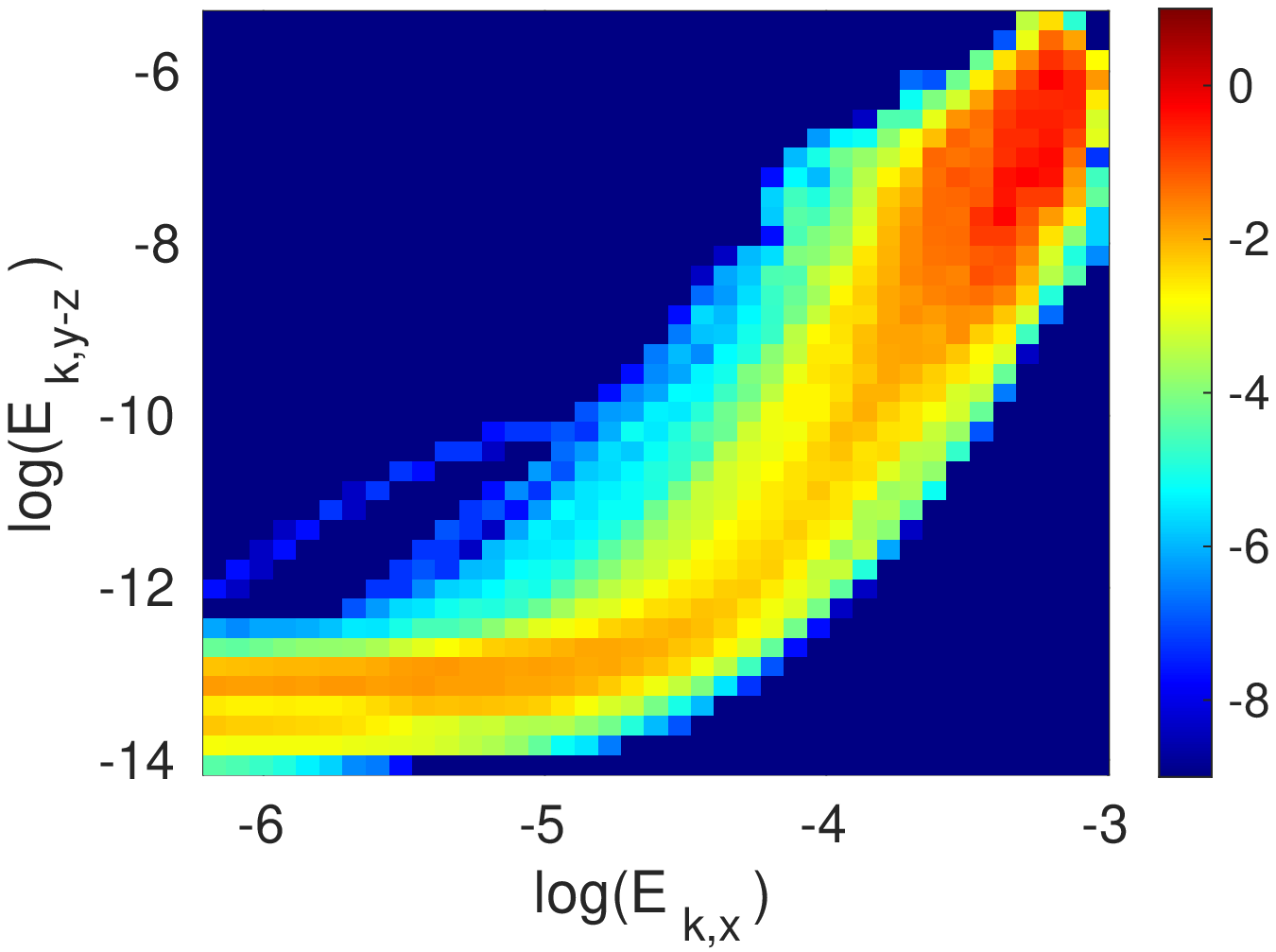}
}
\centerline{
\includegraphics[width=0.3\textwidth]{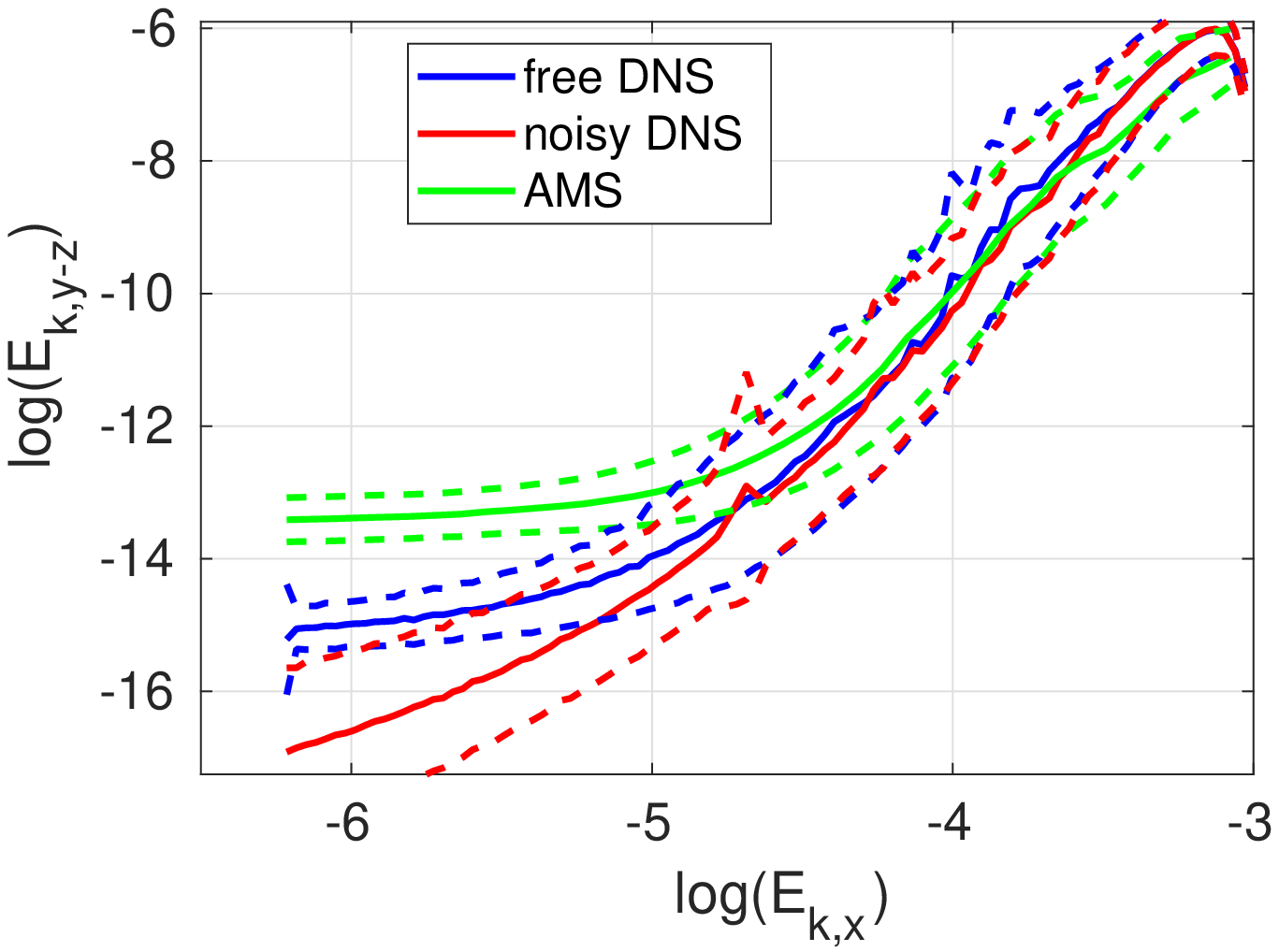}
}
\caption{Probability density functions in the $(E_{k,x},E_{k,y-z})$ plane, conditioned on being in a collapse trajectory, for (a) Trajectories from free DNS, (b) trajectories from noisy DNS, (c) 1138 trajectories from AMS. (d) Conditional Average trajectories and their variances.}
\label{trajs_24_18_370}
\end{figure}

In order to investigate to what extent the dynamics could follow a stochastic system-type transition path (see \citep{metzner2006illustration}) when turbulence collapses, we can then examine the trajectories themselves, computed by mean of AMS and DNS. Since we obtained mostly time series from direct numerical simulations, we will compare the path followed in the $(E_{k,x},E_{k,y-z})$ plane by the reactive trajectories (Fig~\ref{trajs_24_18_370}). We produce a visualisation in a small dimension space comparable to figure~6 of~\citep{prl_jet}. Therefore,
in order to examine the most probable path in this plane and how far trajectories depart from these, we construct the probability density functions $\rho_e$ in the $(E_{k,x},E_{k,y-z})$ plane using solely the collapse trajectories
computed in free Direct Numerical Simulations (Fig.~\ref{trajs_24_18_370} (a)), in noisy direct numerical simulations (Fig.~\ref{trajs_24_18_370} (b)) and AMS computations (Fig.~\ref{trajs_24_18_370} (c)).
We first note that the trajectories computed in all three cases have very similar beginnings: we observe in all three cases a decrease of $E_{k,y-z}$ much faster than that of $E_{k,x}$.
While $E_{k,x}$ is divided by $5$ (from approximately $0.05=\exp(-3)$ to $0.01\simeq \exp(-4.5)$), $E_{k,y-z}$ is divided by $400$ (from approximately $0.0025\simeq \exp(-6)$ to $5\cdot 10^{-6}\simeq \exp(-12)$).
This difference in decay rate quantifies what is observed in the velocity field: the streamwise vortices (whose amplitudes mostly contribute to $u_y$ and $u_z$) decay before the velocity streaks (whose amplitudes mostly contribute to $u_x$).
We note that as $E_{k,y-z}$ becomes smaller, the three PDF deviate in shape. In the case of free DNS (Fig.~\ref{trajs_24_18_370} (a)), both $E_{k,x}$ and $E_{k,y-z}$ keep decreasing, albeit at a smaller rate for $E_{k,y-z}$.
However, in situations where some noise in added to the flow (noisy DNS (Fig.~\ref{trajs_24_18_370} (b)) or AMS computations (Fig.~\ref{trajs_24_18_370} (c))), one way or the other, the kinetic energy contained in the wall normal and spanwise components $E_{k,y-z}$ reaches a neighbourhood of a minimal value while $E_{k,x}$ keeps decreasing (the now streamwise invariant streaks keep decaying in amplitude).
This minimum value is the consequence of the forcing exerted on the flow: $u_y$ and $u_z$ respond almost linearly to this forcing and fluctuate around these small values. In the case of AMS, this is the partially numerical effect of what is left in the spectrum of the white perturbations placed at branching.
The value around which the components fluctuate is a function of the noise variance and spectrum shape.
It has been checked in an AMS study of the build up of turbulence that the amount of energy given to the flow by this noise is small enough so that the probability turbulence restarts from this forcing is negligible. The differences of amplitude along the paths in theses PDF are (among other things) the consequence of the number of bins used, the number of sampled trajectories and the time spent by trajectories near low values of kinetic energy.

We can compare the trajectories more precisely by computing the conditional average $\langle \log(E_{k,y-z})\rangle$ as a function of $ \log(E_{k,x})$. Indeed, if we define the probability $\rho_x$ of having the value $\log(E_{k,x})$ during the reactive trajectories
\begin{equation}
\rho_x(\log(E_{k,x}))=\int \rho_e(\log(E_{k,x}),\log(e_{k,y-z}))\,{\rm d}\log(e_{k,y-z})\,,
\end{equation}
we have the conditional probability $\rho_c$ of observing $\log(E_{k,y-z})=\log(e_{k,y-z})$ if one has $\log(E_{k,x})$
 \begin{equation}
 \rho_c\left(\log(e_{k,y-z})\big|\log(E_{k,x})\right)=\frac{\rho_e(\log(E_{k,x}),\log(e_{k,y-z}))}{\rho_x(\log(E_{k,x}))}\,.
 \end{equation} and the conditional average is
\begin{equation}
\langle \log(E_{k,y-z})\rangle (\log(E_{k,x}))=\int \rho_c\left(\log(e_{k,y-z})\big|\log(E_{k,x})\right)\log(e_{k,y-z})\,{\rm d}\log(e_{k,y-z})\,.
\end{equation}
We can also compute the conditional variance
\begin{align}\notag
\sigma_{y-z}(\log(E_{k,x}))=
\\ \sqrt{\int \left(\rho_c(\log(e_{k,y-z}))|\log(E_{k,x})\log(e_{k,y-z})^2\right)\,{\rm d}\log(e_{k,y-z})-\langle \log(E_{k,y-z})\rangle^2 (\log(E_{k,x}))}\,.
\end{align}
We display the average $\langle \log(E_{k,y-z})\rangle (\log(E_{k,x}))$ in full line and the average plus/minus the variance $\langle \log(E_{k,y-z})\rangle (\log(E_{k,x}))\pm \sigma_{y-z}(\log(E_{k,x}))$ in dashed line for all three types of reactive trajectories in figure~\ref{trajs_24_18_370} (d).
This quantitatively confirms the observation of a faster decay of the streamwise vortices, and thus of $E_{k,y-z}$, than of the streaks, and thus of $E_{k,x}$.
This also confirms that using the AMS leads to the same collapse trajectories as using DNS, at a much smaller cost, up to the effect of the noise added at branching on $E_{k,y-z}$ when it is very small.
We can finally note that the trajectories seem to be concentrated around this average. Note however that the constant variance of $\log(E_{k,y-z})$ actually corresponds to a variance of $E_{k,y-z}$ which decreases exponentially as the collapse of turbulence goes on.
There is a larger variance of $E_{k,y-z}$ among the initial conditions and thus in the amplitude and distribution of streamwise vortices in the initial conditions and stronger fluctuations in the beginning of the trajectory.
On a final note, we state that in order to rigorously obtain instantons in stochastic systems, one has to take a limit of the variance of the added noise going to $0$.
In the case of our turbulent flows, this means either identifying an effective noise (which is a difficult task) or displaying large deviations in quantities like the probability density functions, the probability of collapse or the mean first passage time.
Quantitatively speaking, this amounts to identifying a small parameter $\epsilon$, proportional to the square of the suspected noise variance such that $\lim_{\epsilon\rightarrow 0}\epsilon\log(\alpha)$ or $\lim_{\epsilon\rightarrow 0}\epsilon\log(T)$ are independent on $\epsilon$ (these two limits are often the opposite of one another) \citep{touchette2009large}.
As it happens, such scalings have been displayed in probability density functions of kinetic energy in simulations when the sizes $L_x$, $L_z$ were increased so that domains contained up to 6 wavelengths of the laminar turbulent bands, with $\epsilon_{DNS}=\frac{1}{L_xL_z}$ \citep{rolland2015mechanical}.
In that case, the rate functions of the PDF of kinetic energy showed the tails for $E_k<\left\langle E_k\right\rangle$, describing the fluctuations of kinetic energy toward low values and eventually toward laminar flows.
This type of scaling was also shown in mean first passage times before collapse of laminar-turbulent coexistence in models of pipe flow where the length $L$ of the system was increased so that it contained up to ten puffs, with $\epsilon=\frac{1}{L}$,  \citep{rolland_pre18}.
One thus observes large deviations in the large system size limit, and in that limit one should expect to observe narrower and narrower concentration of trajectories,
and an increase of the mean first passage time before collapse which is exponential with the length of the system.
Note that models indicate that nothing in laminar-turbulent coexistence prevents the collapse of turbulence:
if the Reynolds number is increased slightly above the threshold of transition,
this only requires  the concomitant collapse of all the puffs.
This condition is at the origin of the exponential dependence of the lifetime of turbulence on size.
It may very well be that the very same thing happens in actual pipe flows or Couette flows,
where one would require the rare, but not impossible, concomitant collapse of all the puffs or of the bands of the flow.
This synchronisation could be at the origin of an exponential scaling in size.
Note that in that respect, there is a similar phenomenology in our rectangular boxes and larger domains:
in both case, the collapse of turbulence on a streamwise long corridor is required,
and in both cases, that collapse can be countered by contamination from the sides of the hole.
This would indicate that laminar turbulent coexistent is strictly permanent only in domains of infinite size.

Given these considerations, we can conclude that in order to observe more concentration of trajectories, and more generally large deviations of the probabilities and rate of probability of collapse, it may very well be necessary to observe said collapses in larger and larger domains.
For this purpose increasing the streamwise size $L_x$ may be more efficient than increasing the spanwise size $L_z$.
This may be a consequence of the specific mechanism of hole formation, leading to the formation of two active-quiescent fronts that move on.
Such situations have already been observed in theoretical physics systems \citep{rolland2016computing}. The rare event is the formation of the hole and of the fronts themselves.
Once said hole is created, the probability that it will entirely open is small and decreases with size, but this decrease is slower than exponential.


We finally note that even if each AMS run has some collapse trajectories in its first iterations (approximately $3.8\%$ of the trajectories generated in the first stage),
and that it will generate other collapse trajectories by branching on these collapse trajectories, it generates even more variability than this through mutation/selection on other initial conditions, with the help of anticipated branching.
This means that each run generates distinct, independent reactive trajectories, in particular trajectories with distinct initial conditions.
These trajectories are even more distinct in their later stages.
The amount of variability generated can be estimated by counting the number of genealogies, that is to say the number of group of trajectories that share the same starting point.
Each group will display trajectories decorrelated from those of another genealogy. On average, we generate 24 genealogies per AMS run with $120$ clones.
This value is obtained by averaging over all available AMS runs the number of distinct initial conditions in computed reactive trajectories.
See \citep{ragone2020rare} for a view of genealogies of trajectories with a different algorithm applied to climate.


\subsection{Mean first passage time and probability of collapse}\label{mfpt}

We now compare the mean first passage time before collapse $T$ and probability $\alpha$ of collapse computed by means of direct numerical simulations and AMS.
While the mean first passage time before collapse is the most physical quantity, the probability of collapse nevertheless contains a lot of information on the precision of the computations and the rarity of the event.
Looking into its statistics is always a good way of validating the AMS computations.
This is all the more true because this is the one quantity for which we have mathematical results of convergence \citep{rolland_CG07,brehier2016unbiasedness,cerou2019asymptotic}.
Using the methods presented in sections~\ref{AMS} and~\ref{dns_couette}, we obtain the averages with the 66\% confidence intervals $\langle \alpha\rangle_o=\alpha_{AMS}=0.042\pm 0.003$ with a variance $\sigma_{\alpha, AMS}\simeq 0.016$ using AMS, and $\alpha_{DNS}=0.037\pm 0.002$ using direct numerical simulations.
This indicates that the estimation of the probability of collapse is precise, we have an overlap of both 66\% interval of confidence (as presented in section~\ref{sgood}). When we suppress a fixed number (strictly larger than one) of clones at each iteration, the ideal variance of the probability of collapse is given by $\sigma_{id}^2=\frac{\alpha^2}{N}\left( \langle \kappa\rangle_o\frac{N_c}{N-N_c}+\frac{1-\langle r\rangle_o}{\langle r\rangle_o}\right)$ (where $\langle \kappa\rangle_o$ is the average number of iterations of AMS and $\langle r\rangle_o$ is the average proportion of reactive trajectories computed at the last iteration) \citep{rolland_CG07}. This is the minimum of variance of the estimator of $\alpha$. It is obtained when the ideal reaction coordinate, the committor  is used \citep{brehier2016unbiasedness,cerou2019asymptotic}. When this minimal variance is reached, AMS computations are performed with a minimal error. In practice this gives a value of the ideal variance $\sigma_{AMS,id}\simeq0.007$.
The variance on the estimate of the probability is twice the ideal variance. This indicates that the quality of the computation is acceptable, but not perfect, quite possibly due to imperfections in the reaction coordinate we used.

\begin{figure}
\centerline{\includegraphics[width=7cm]{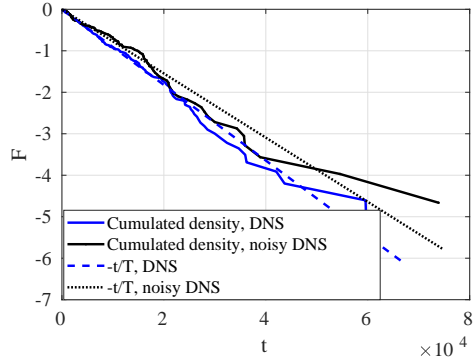}}
\caption{Logarithm of the cumulated distribution of waiting times in noiseless and noisy DNS for $L_x\times L_z=24\times 18$, $R=370$. We add the two linear functions $-t/\langle T\rangle$ for comparison.}
\label{fTDNS}
\end{figure}

\begin{figure}
\centerline{\includegraphics[width=7cm]{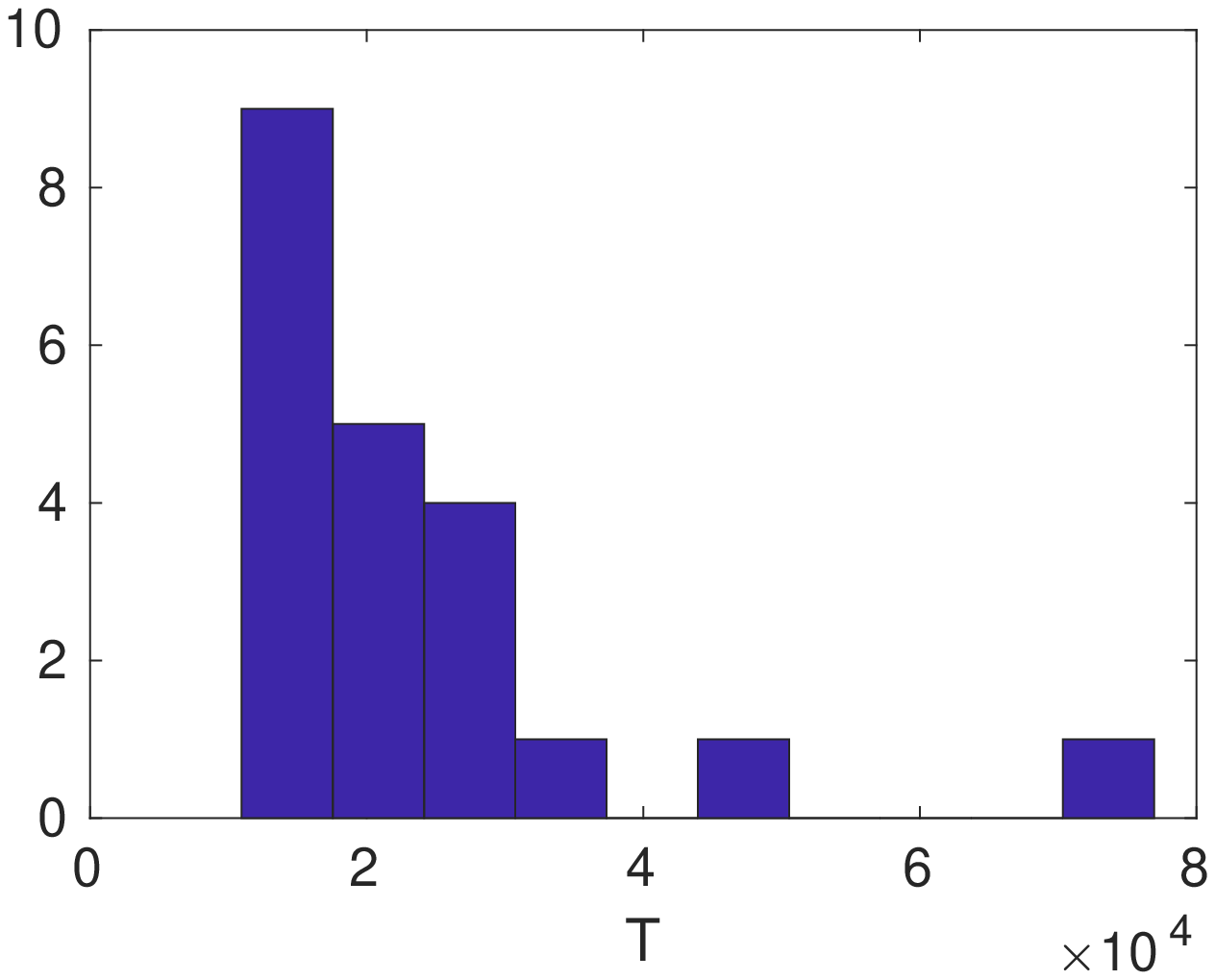}\includegraphics[width=7cm]{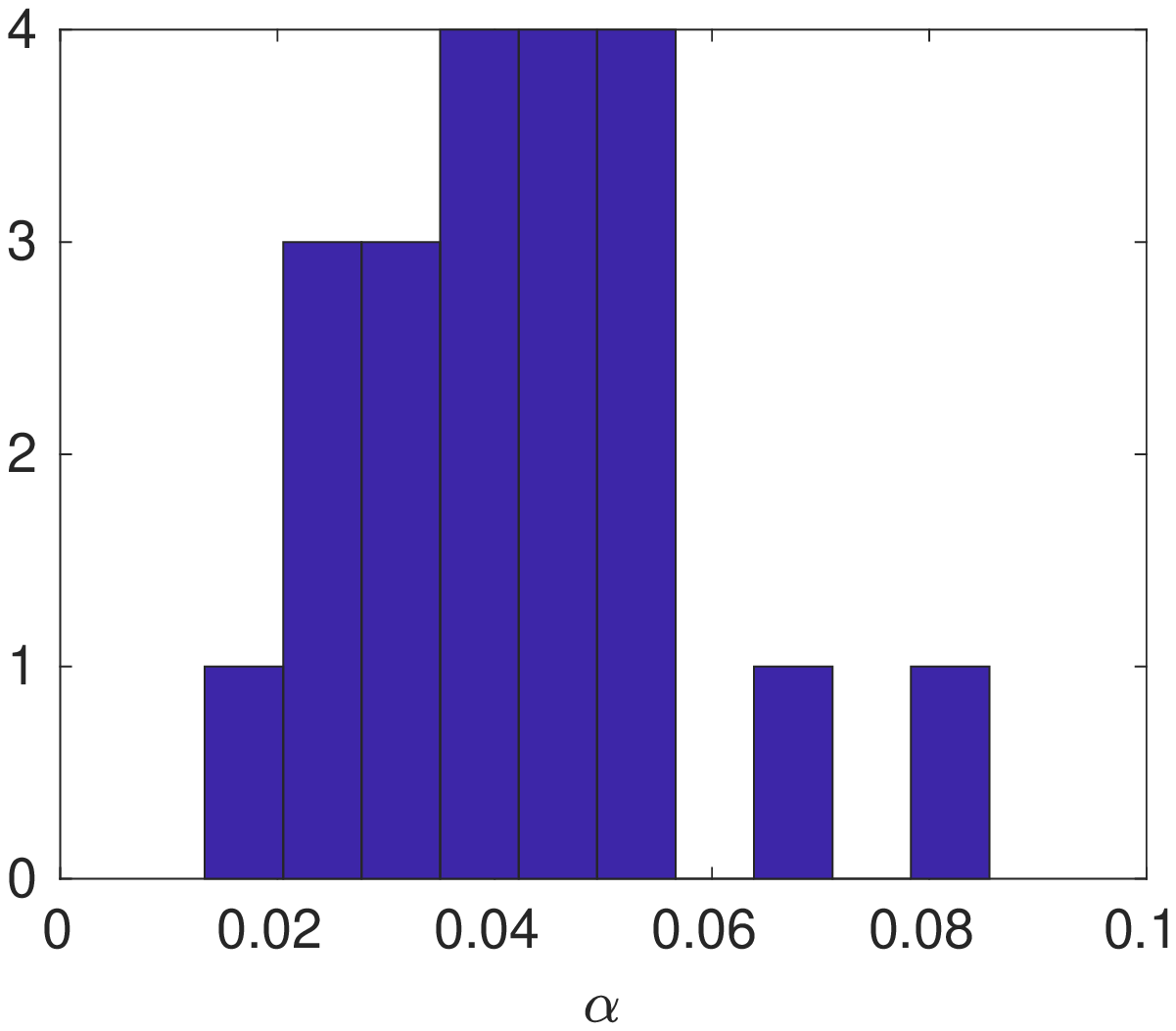}}
\caption{Histograms of $T$ (a) and $\alpha$ (b)  sampled by AMS in the system of size $L_x\times L_z=24\times 18$, at Reynolds number $R=370$.}
\label{hist}
\end{figure}

We then compare the mean first passage times before collapse.
By means of DNS, we sample the cumulated density of waiting times before collapse in simulations performed with and without additional noise, in order to estimate the mean first passage time as well as test the effect of noise on it.
We can see that cumulated densities $F(t)=\int_{t}^{\infty}f(\zeta)\,{\rm d}\zeta$, with $f(\zeta)$ the PDF of passage times, computed with and without noise decrease linearly in logarithmic scale (Fig.~\ref{fTDNS}). This is thus entirely consistent with an exponential distribution of passage times.
Both cumulated density functions have very similar slopes, which is itself almost equal to the estimated mean first passage time.
We estimate the mean first passage time before collapse in DNS by averaging over all sampled durations, we obtain $T_{DNS}=1.10\cdot 10^4\pm 8\cdot 10^2$ without noise and $T_{DNS,noisy}=1.3\cdot 10^4\pm 10^3$ with additional noise, where the error bars correspond to the 66\% confidence interval.
In both cases, the variance is equal to the average and the median is equal to $\sqrt{2}T$, further confirming that we have an exponential distribution.
While the addition of noise may increase the mean first passage time, it does not increase it dramatically, so that it is acceptable to use an additive small noise in simulations and AMS to help the separation of trajectories.
In turn, we estimate $T_{AMS}=2.4\cdot 10^4\pm 3\cdot 10^3$. Again, we used the 66\% interval of confidence as error bars. We note that $T_{AMS}$ over estimates the mean first passage time.
The confidence intervals do not overlap, which indicates that this overestimate is most likely an effect of a bias in the computation of $T_{AMS}$.
In order to understand why the estimate of the mean first passage time is not as good as what is obtained for other quantities, we examine the histograms of mean first passage times (Fig.~\ref{hist} (a)) and
probability of collapse (Fig.~\ref{hist} (b)) computed by AMS.
While the histogram of $\alpha$ is symmetric, that of $T$ is skewed. We note that the estimate of $T$ is polluted by a handful of very high values, originating from fairly small values of $\alpha$, which have a much smaller effect on the average over realisations $\alpha_{AMS}$.
On top of the improvement brought by a better reaction coordinate, the computation of the mean first passage time may very well be improved in AMS computations using more clones.

One can finally compare the relative cost of trajectory computations with AMS and DNS. Strictly speaking, the average physical duration of simulation to obtain one reactive trajectory by DNS is of order $\mathcal{O}(10^4)$, meanwhile the simulation time by means of AMS is of order $\mathcal{O}(10^3)$. Such a factor $10$ of acceleration of computation already seems interesting. This means that even in this case where the study of the system by DNS seems affordable, using AMS already leads to a substantial acceleration of computations.
Of course, with the use of checkpoints to store the trajectories during AMS computations, the reactive trajectories and non-reactive trajectories can be recorded without unnecessary use of RAM and disk-space.
Indeed, in many cases, multistability (or more precisely collapse of turbulence) is studied using long records of time series, where most of the information has to be eventually discarded.
Using such records to compute the velocity fields during collapse would be very inefficient.
Velocity fields presenting collapse are scarcely presented in DNS or experimental studies.
This does not have to be a consequence of the numerical procedure, as one can argue that only records spanning the last one thousand time units or so of the velocity have to be kept.
This requires careful memory management and has very seldom been explicitly presented in the literature to the author's knowledge.

\section{A very rare collapse in a domain of size $L_x\times L_z=36\times 27$: spatial organisation}\label{hole}

The observation of collapse of turbulence in the domain of size $L_x\times L_z=24\times 18$ indicated that, locally, collapse occurred by a more rapid failure of streamwise vortices than velocity streaks.
In that domain, however, it was not entirely clear whether this collapse was always a local process, leading to a streamwise laminar hole that then extended in the spanwise direction, or whether this could happen in a synchronous manner everywhere. The question is open as both behaviours have been observed in a finite number of decays forced by decrease of the Reynolds number to very low value \citep{manneville2011decay,de2020transient}.
We can therefore ask the question of whether we would always see both types of collapse when the domain size is increased and the simulation is closer to what can be observed in experiments.
For this reason we will consider the collapse of turbulence in a domain of size  $L_x\times L_z=36\times 27$ at Reynolds number $R=377$ (see Fig.~\ref{rolland_figint} for an illustration of the domain size).
This will also be the occasion to test anticipated AMS in a situation where the mean first passage time before collapse is much larger than in section~\ref{mfpt}.
In that case direct numerical simulations are not affordable and there will be no collapse trajectory computed at stage $0$ of AMS.
This will indicate how expensive an AMS computation is in that situation and whether anticipated AMS manages to create collapse trajectories and ensures that there is some variability between these trajectories.

For the AMS computations, we use $E_{\rm turb}=0.05$, $\Delta E=0.047$ in $\Phi_E$ (Eq.~(\ref{ephie})). We set the hypersurface $\mathcal{C}$ as the set of velocity fields such that $\Phi=0.07$. The initial conditions within $\mathcal{A}$ are prepared as stated in section~\ref{init}.
We use the parameter $\xi=0.15$, with the converging $\Phi_b$ (Eq.~(\ref{conv}), \S~\ref{rolland_sappa}) for the anticipation of branching.
In each computation, we use $60$ clones and suppressed 16 clones at each iterations. On 8 threads, this gives reasonable load balancing for the simulations.
Due to the rarity of collapse of turbulence for these control parameters, there is no collapse trajectory in the very first stage of AMS computation: the method selects them naturally.
A downside is that there is less variability between the trajectories computed in the same AMS run, in their first 100 time units.
There is of course variability from one AMS run to the other.
We note that the larger the domain is, the more decorrelated fluctuations are adding up in the flow and the less need there is for anticipation of branching.
With these parameters, we estimate $\alpha=8\cdot 10^{-5} \pm 7\cdot 10^{-5}$ and $T=8\cdot 10^{7}\pm 7\cdot 10^{7}$, where the error bars are given by the 66\% interval of confidence.
We have to simulate the flow for approximately $5000$ time units per trajectory obtained.
Even if one takes into account the fact that mean first passage times are somewhat overestimated with a finite number of clones, this still represents an acceleration of computations by a factor of more than one thousand.

\begin{figure}
\centerline{\includegraphics[width=7cm]{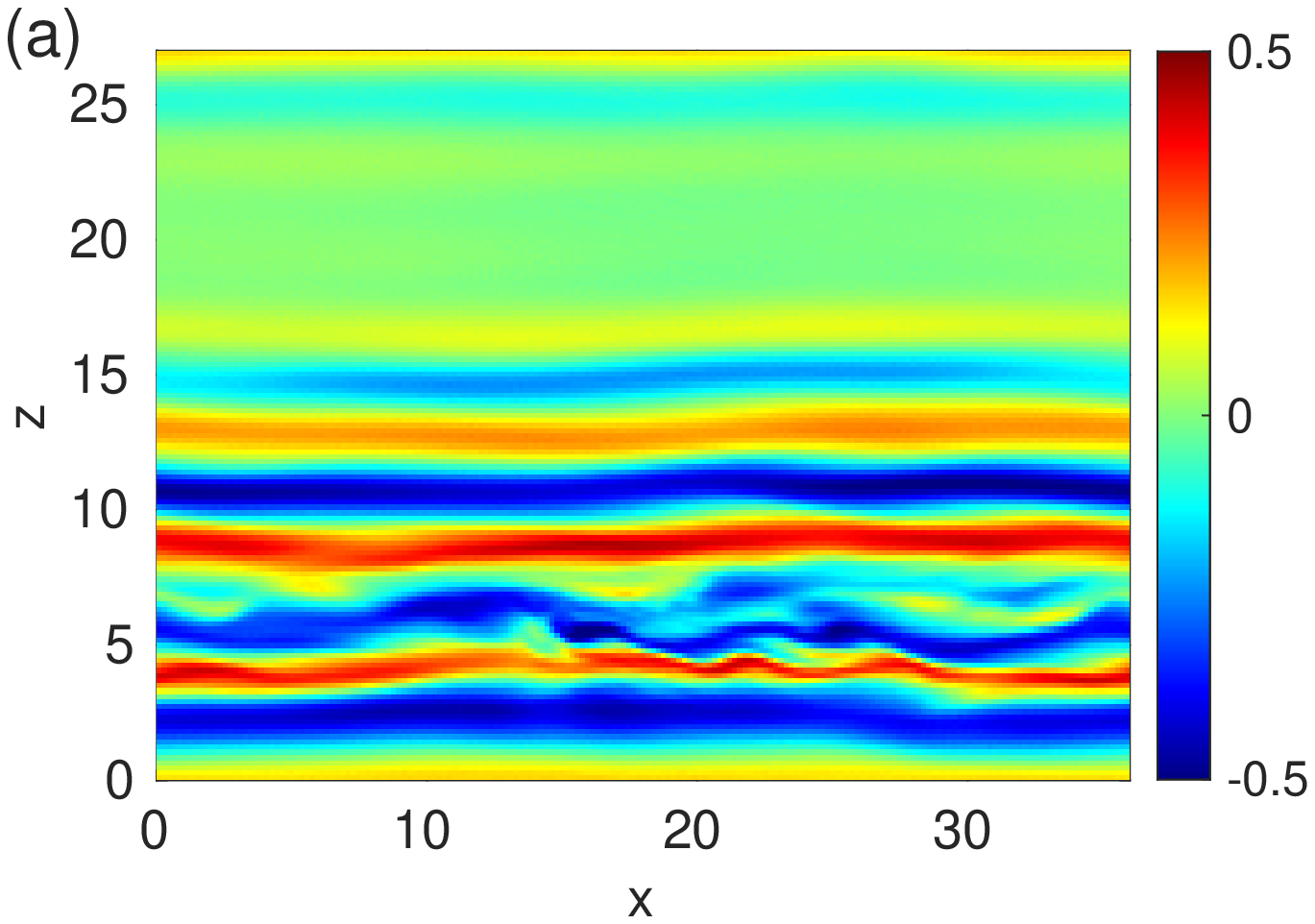}\includegraphics[width=7cm]{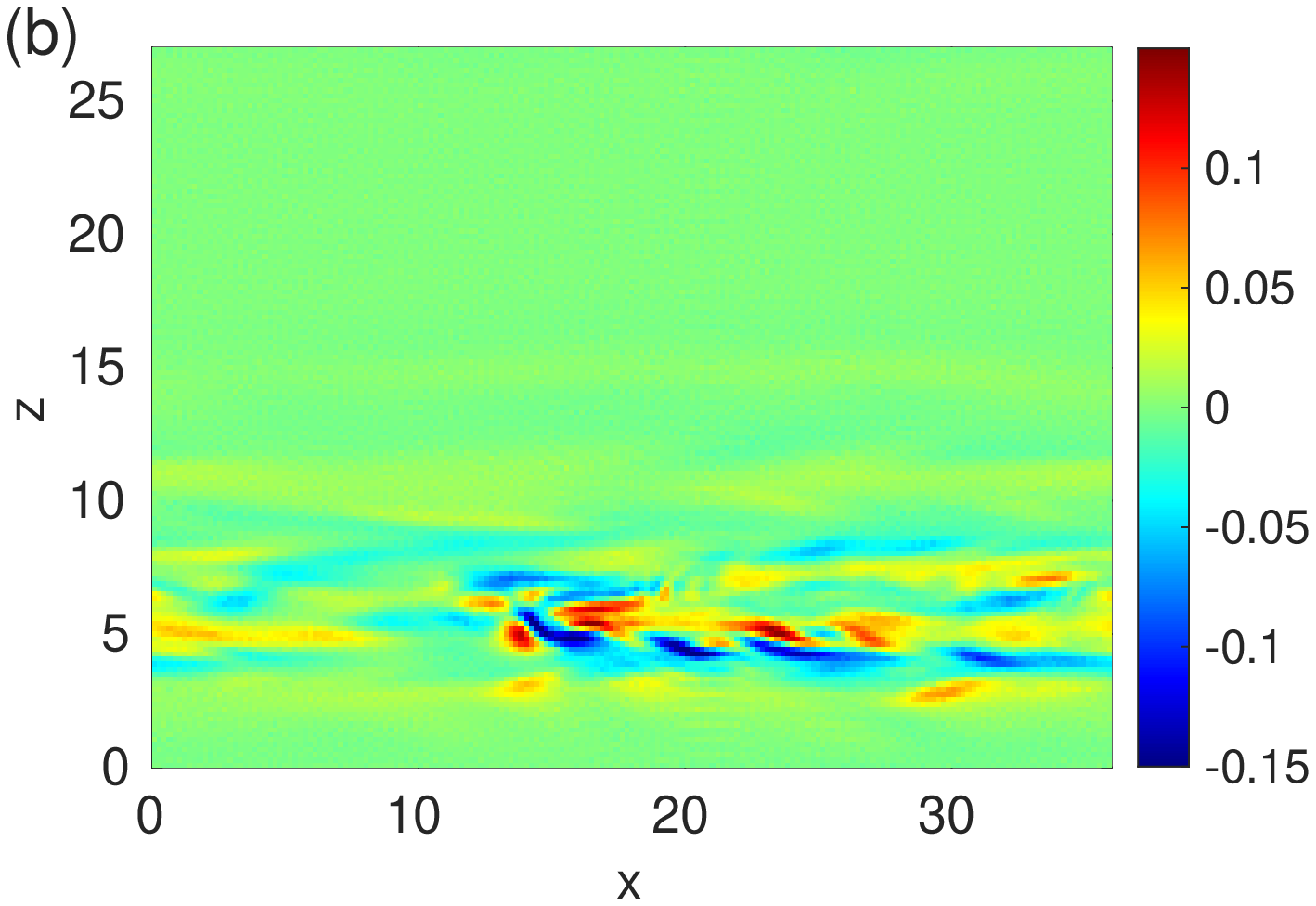}}
\centerline{\includegraphics[width=7cm]{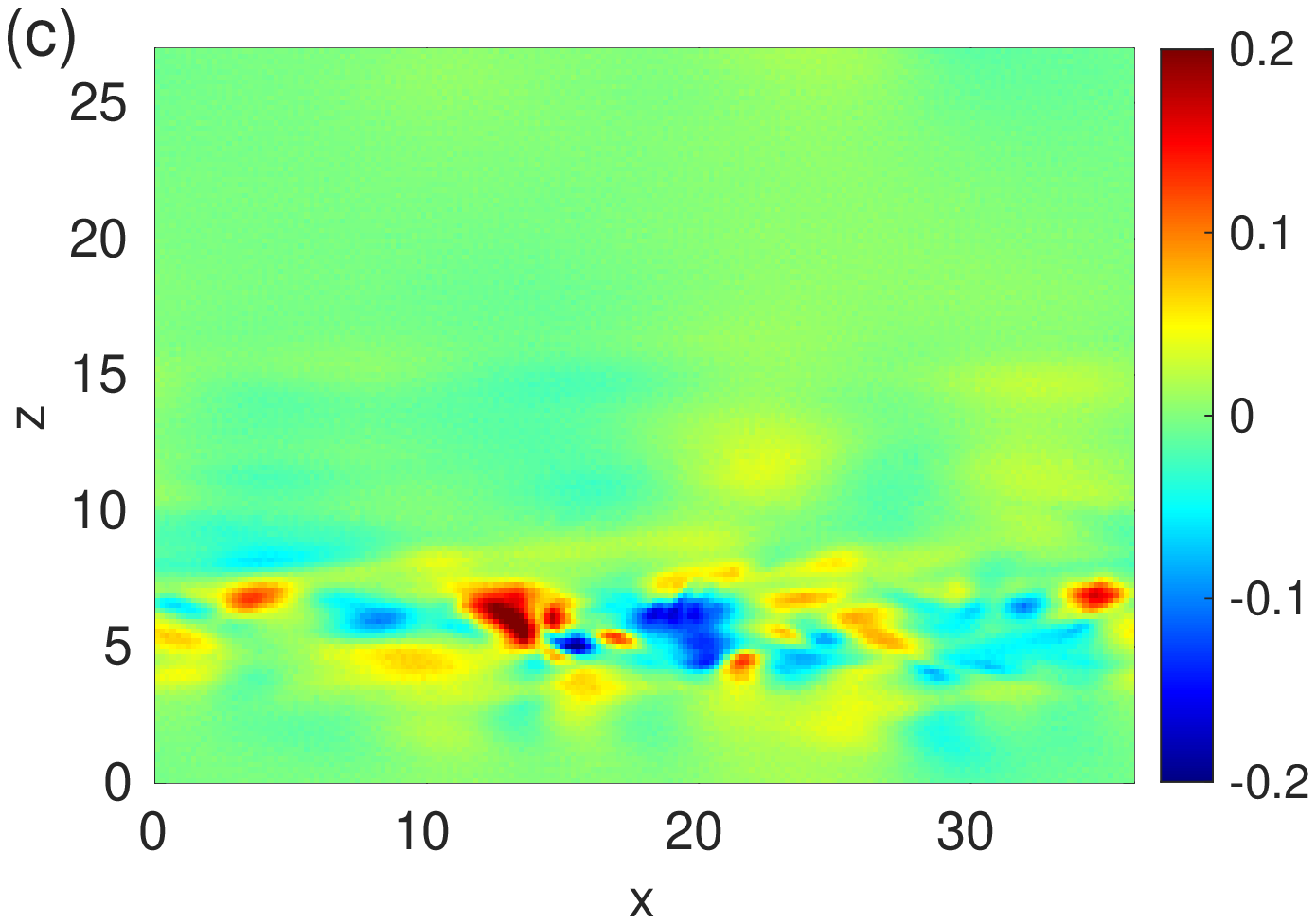}\includegraphics[width=7cm,clip]{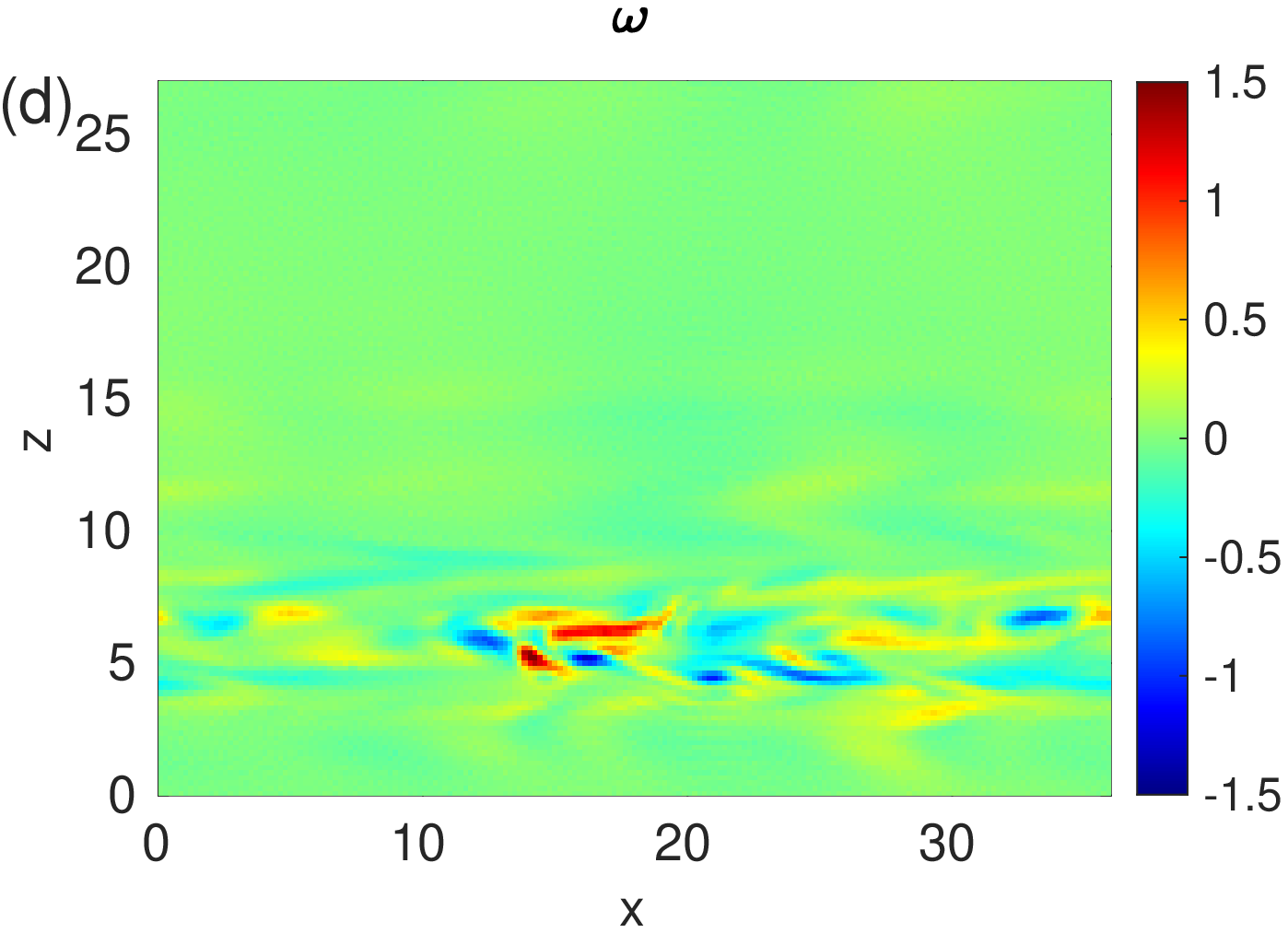}}
\centerline{\includegraphics[width=7cm]{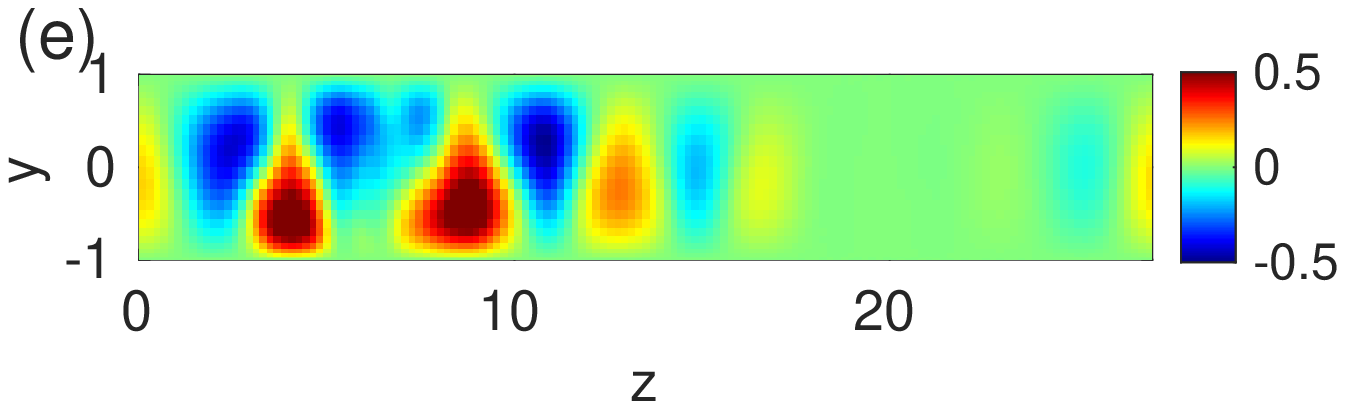}}
\caption{Colour levels of the streamwise (a), wall normal (b) and spanwise (c) components of the velocity field, in the $y=0$, $\vec e_x\times \vec e_z$ plane for the last stage at the last branching step in an AMS computation of collapse trajectories at $R=377$ in a domain of size $L_x\times L_z=36\times$, along with (d) the streamwise vorticity. (e) Colour levels of streamwise averaged streamwise velocity as a function of $y$ and $z$ for the same state.}
\label{col}
\end{figure}

We display the last state at the last branching stage $\mathbf{u}_{\rm last},\omega_{x{\rm , last}}$ from one of the AMS computations in figure~\ref{col}.
This state (which is not steady in any way) displays  a laminar hole localised in $z$ which occupies the whole streamwise length of the domain. For the streamwise component of the flow (Fig.~\ref{col} a,e), this hole is rather narrow and is situated for $17\lesssim z\lesssim 24$.
It is flanked by streamwise invariant streaks for $8\lesssim z\lesssim 17$ and $24\lesssim 27$, $0\lesssim 3$ (due to periodic boundary conditions in $z$).
In the wall normal component (Fig.~\ref{col} (b)), the spanwise component (Fig.~\ref{col} (c)) and subsequently the streamwise vorticity (Fig.~\ref{col} (d)), the laminar hole is actually much wider.
These components are away from zero only for $3\lesssim z \lesssim 8$, that is to say where the streamwise component displays fluctuations.
This is the only area left in the flow where the self sustaining process of turbulence is still active. This shift in the active-quiescent interface between the streaks and streamwise vortices flow components highlights again the scenario of collapse where the vortices disappear before the streaks.
Unlike in the domain of size $L_x\times L_z=24\times 18$, all the last state at the last branching stage that have been computed display such an opening of hole.
This has also been observed at other Reynolds numbers ($R=351$, $357$ and $370$, not shown here). The observation of this state confirms that for this regime of parameters, the scenario of collapse of turbulence is far more univocal. As we will see in the observation of collapse and partial collapse trajectories, there is again a local failure of streamwise vortices that disappear in a streamwise long hole leaving streamwise invariant streaks that decay viscously. These holes then extend in the streamwise direction.

\begin{figure}
\centerline{\includegraphics[width=14cm,clip]{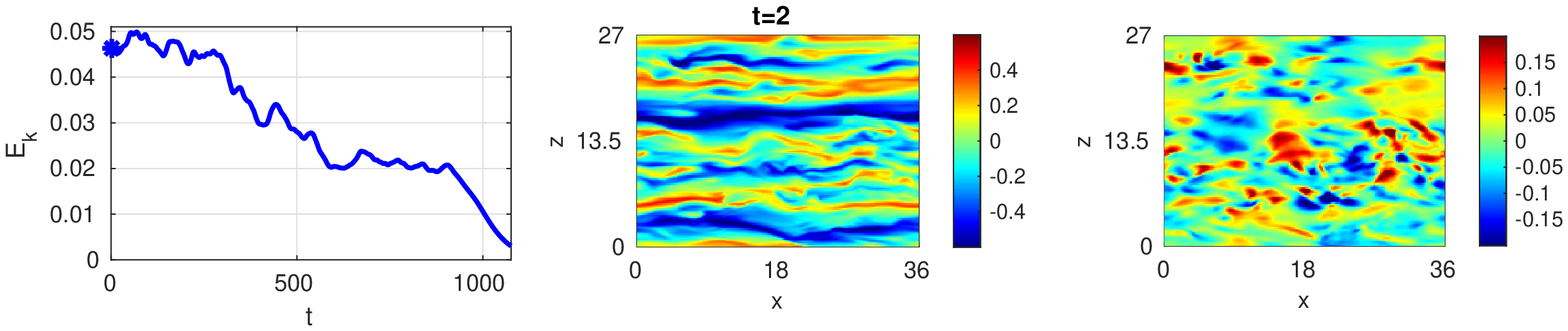}}
\centerline{\includegraphics[width=14cm,clip]{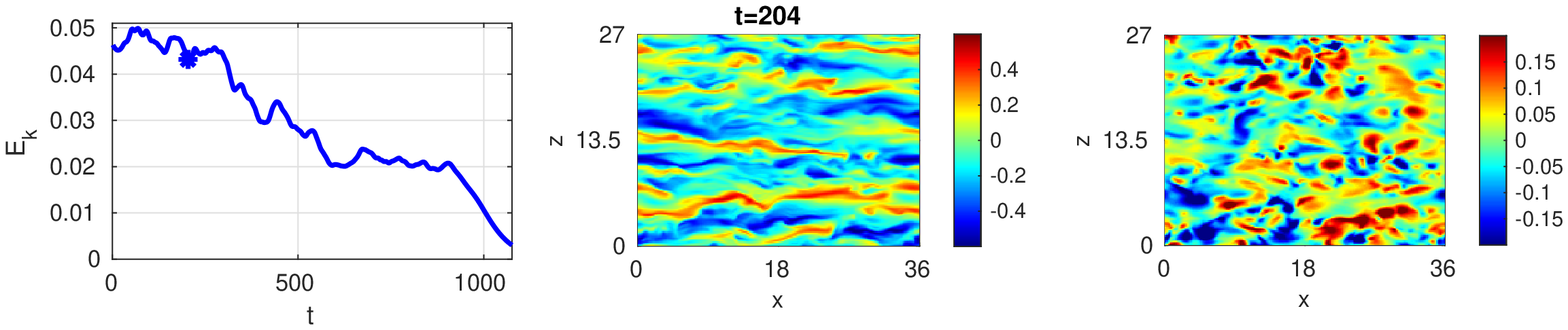}}
\centerline{\includegraphics[width=14cm,clip]{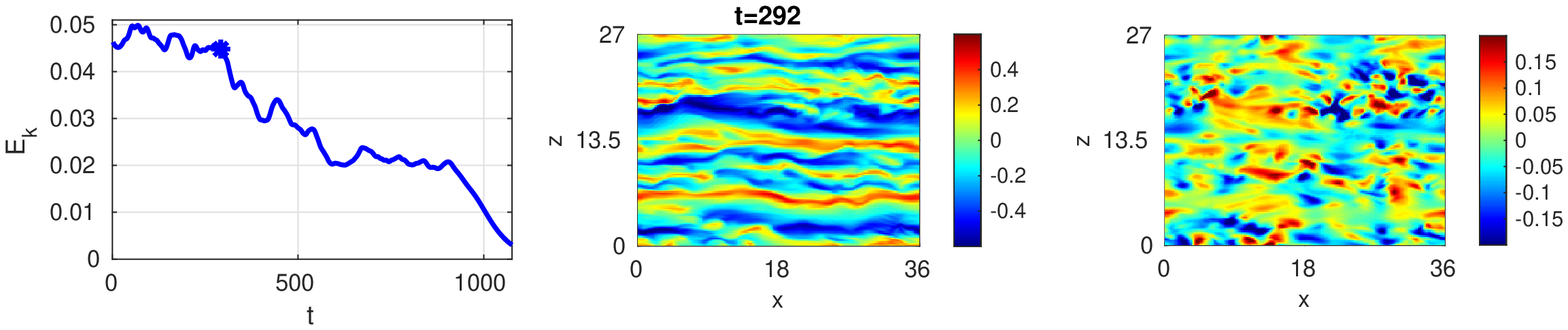}}
\centerline{\includegraphics[width=14cm,clip]{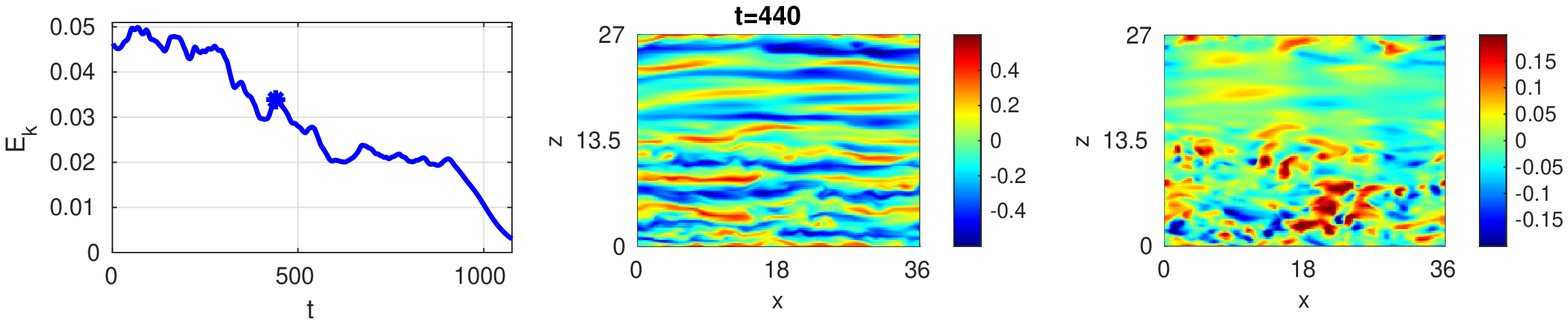}}
\centerline{\includegraphics[width=14cm,clip]{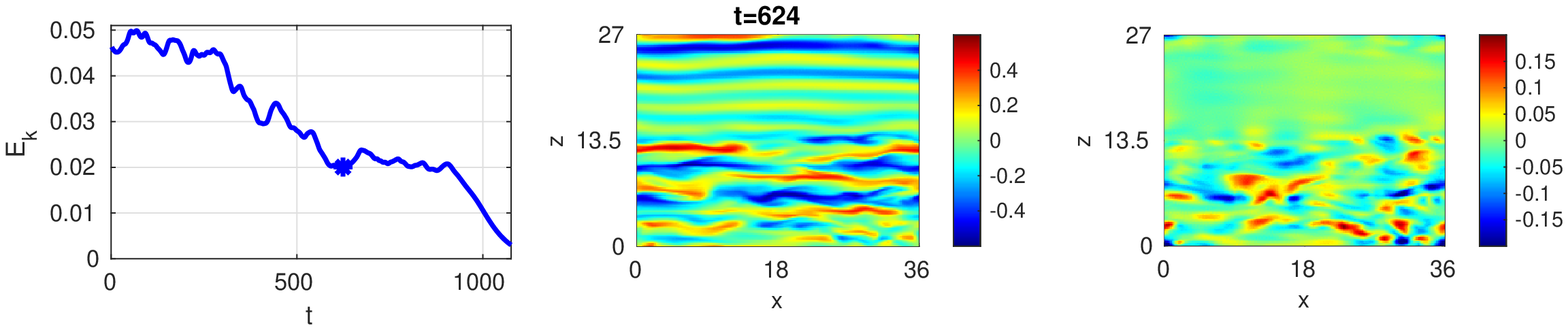}}
\centerline{\includegraphics[width=14cm,clip]{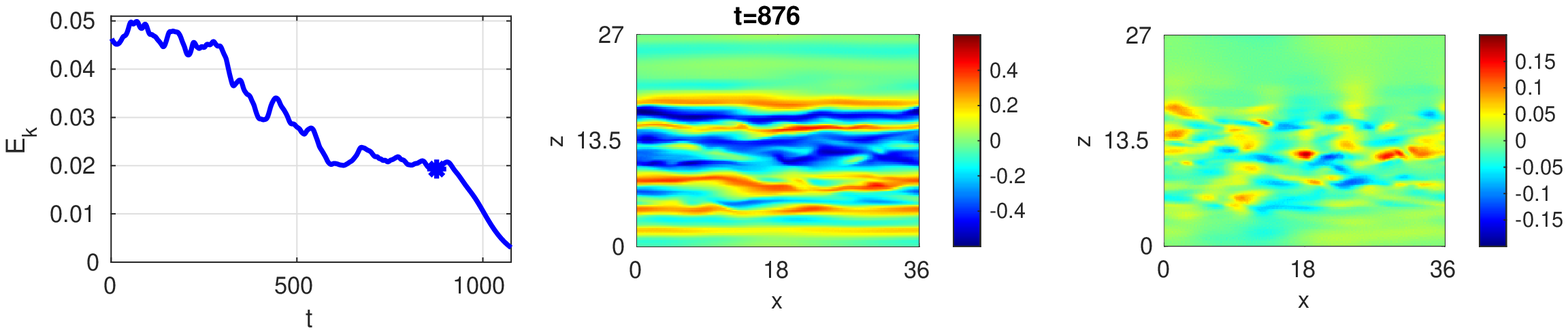}}
\centerline{\includegraphics[width=14cm,clip]{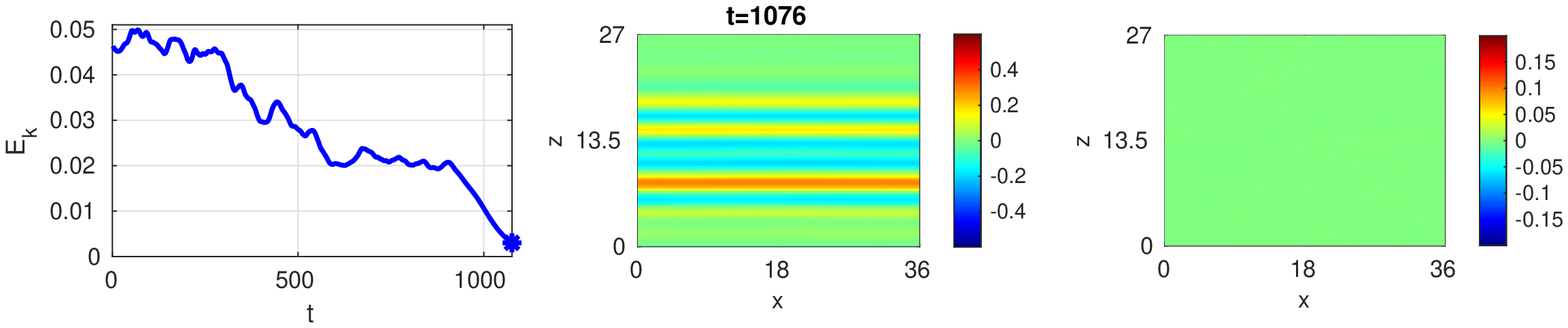}}
\caption{Following a collapse trajectory in a domain of size $L_x\times L_z=36\times 27$, $R=377$. The left panels show a time series of kinetic energy, with the dot indicating the instant at which the colour levels of streamwise velocity in the $y=0$ plane (middle panels) and spanwise velocity (right panels) at corresponding times in the $y=0$ plane are shown. \emph{A movie of this collapse trajectory is provided in the supplementary material}.}
\label{36_27_377_coll}
\end{figure}

We display a typical collapse trajectory obtained using AMS in figure~\ref{36_27_377_coll}. We follow the decay of the kinetic energy (left panels) along with the streamwise velocity and spanwise velocity in the midplane (central and right panels).
The trajectory starts from typical uniform wall turbulence at $t=2$, where the streamwise velocity is organised in low speed and high speed streaks, and the spanwise velocity varies at much shorter lengthscales, as would be expected from its participation in spanwise vortices.
After some slight decay in amplitude of the spanwise velocity component from $t=2$ to $t=292$, a laminar hole forms in $u_z$ from $t=292$ to $t=440$ as seen in the decay of kinetic energy. In that case, this hole is located in $14\lesssim z\lesssim 25$ for all $x$.
As to $u_x$, the velocity streaks are still present at that location, albeit almost streamwise invariant. At $t=624$, this hole has extended in the spanwise direction. The velocity streaks are now streamwise invariant for $z\gtrsim 13$ and decay in amplitude in this part of the flow.
Meanwhile the spanwise component of the flow has fallen to zero at the center of the laminar hole. As time moves forward, at $t=876$, $u_x$ totally falls to zero for $27\gtrsim z\gtrsim19$, one can notice that the interface between the active and quiescent domain has moved in the spanwise direction from $z\simeq 16$ to $z\simeq 19$.
Observation indicates that since the active-quiescent interface is the same for $u_z$ and $u_x$ in that case.
After this event, one nevertheless can note that at $t=824$ the amplitude of the spanwise velocity field has decreased in the turbulent domain indicating a failure of the streamwise vortices in the whole $3\lesssim z\lesssim 19$ area. The remaining streamwise vortices then completely decay, leaving only totally streamwise invariant streaks localised in $z$ at $t=1076$, that then viscously decay.

\begin{figure}
\centerline{\includegraphics[width=14cm,clip]{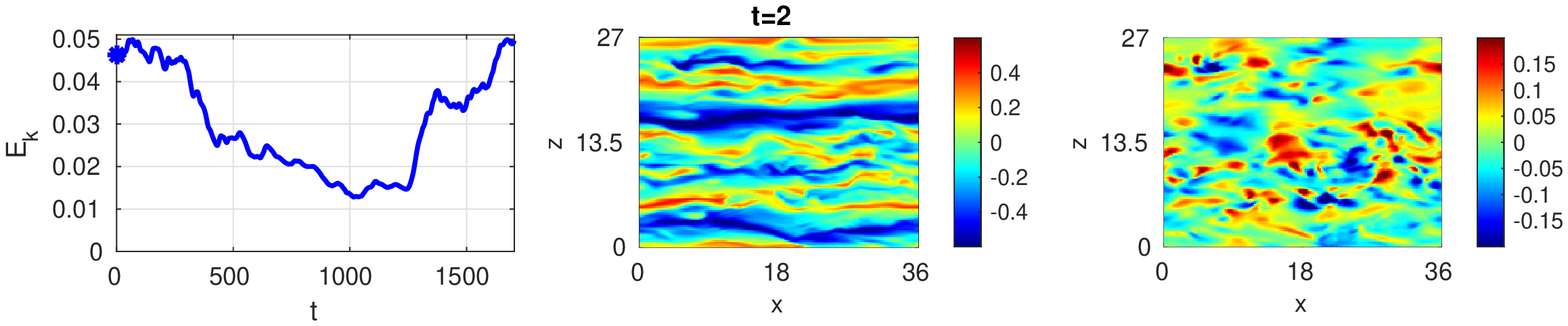}}
\centerline{\includegraphics[width=14cm,clip]{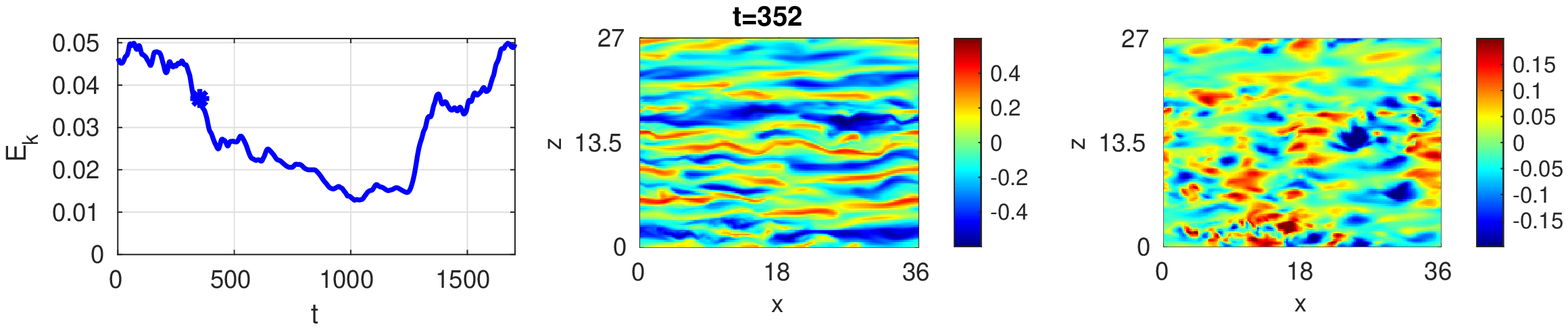}}
\centerline{\includegraphics[width=14cm,clip]{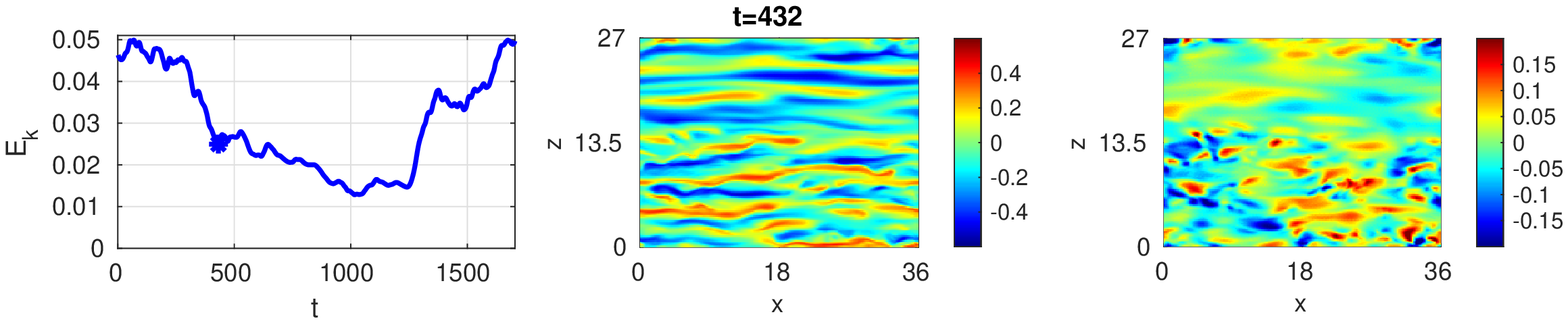}}
\centerline{\includegraphics[width=14cm,clip]{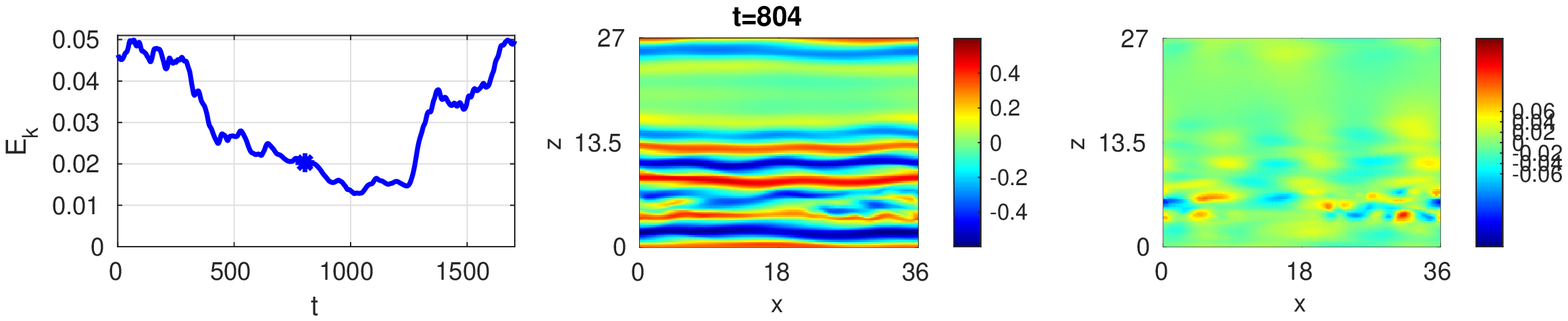}}
\centerline{\includegraphics[width=14cm,clip]{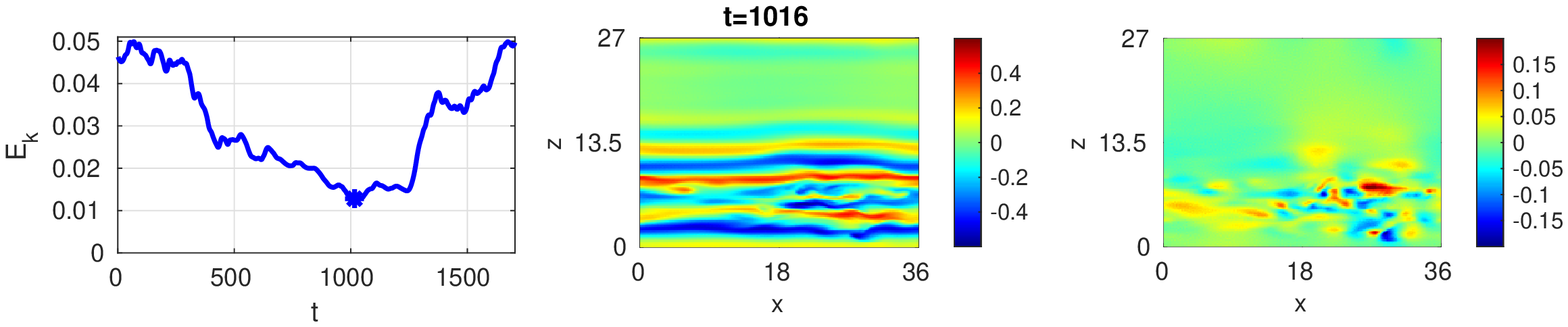}}
\centerline{\includegraphics[width=14cm,clip]{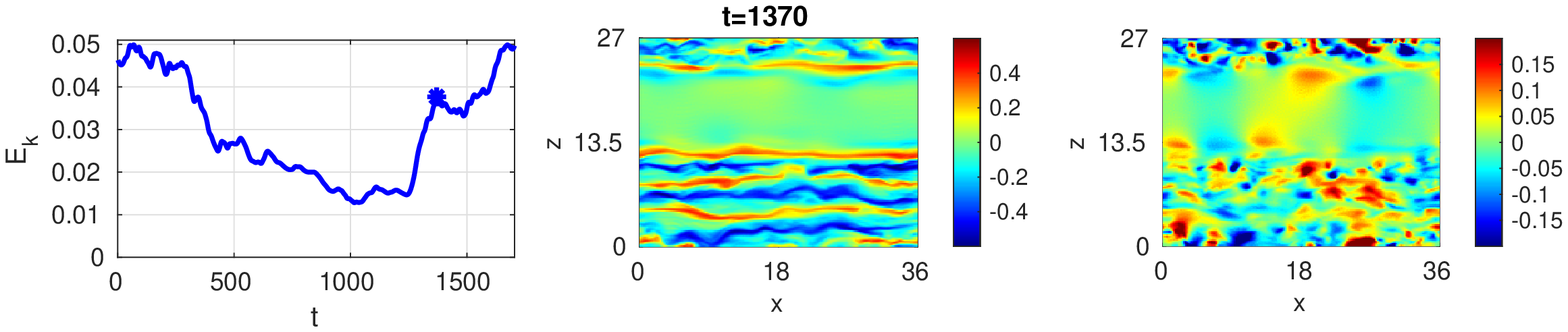}}
\centerline{\includegraphics[width=14cm,clip]{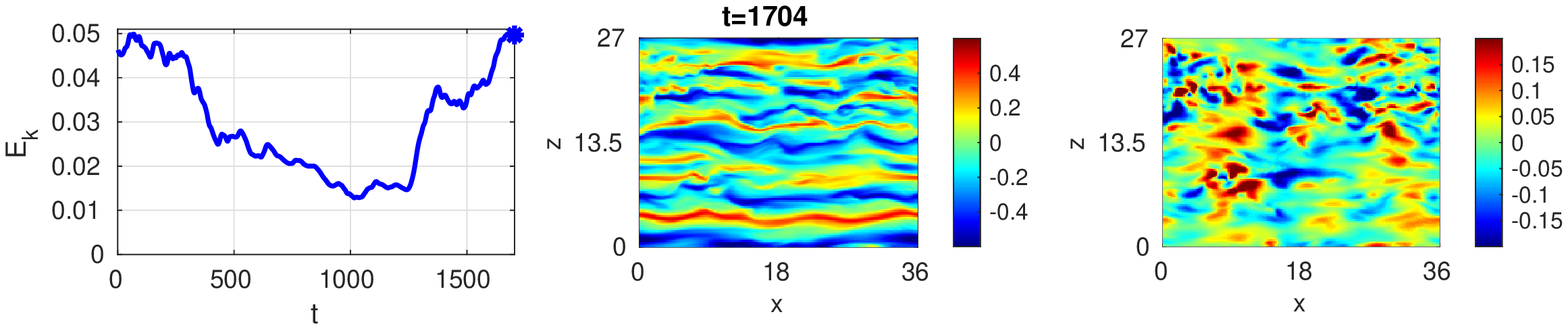}}
\caption{Following a non-reactive trajectory (hole opening then closing) in a domain of size $L_x\times L_z=36\times 27$, $R=377$. The left panels show a time series of kinetic energy, with the dot indicating the instant at which the colour levels of streamwise velocity in the $y=0$ plane (middle panels) and spanwise velocity (right panels) at corresponding times in the $y=0$ plane are shown. \emph{A movie of this non-reactive trajectory is provided in the supplementary material}.}
\label{36_27_377_fail}
\end{figure}

We display one of the $n< N_c$ non-reactive trajectories left at the last stage of an AMS computation in figure~\ref{36_27_377_fail}. This trajectory corresponds to an excursion toward laminar flow that evolves back to a fully turbulent flow.
We follow the excursion toward low values then increase of the kinetic energy (left panels) along with the streamwise velocity and spanwise velocity in the midplane (central and right panels).
This trajectory starts like the one of figure~\ref{36_27_377_coll}, by the opening of a laminar hole, first by decay of $u_z$ (for all $x$, for $20\lesssim z \lesssim 27$ at $t=352$), leaving almost streamwise invariant streaks at $t=432$.
Part of this similarity may be explained by the fact that the two trajectories share a common beginning, being branched from the same genealogy.
However they quickly separate, while displaying common features of laminar hole opening, and of simplification of the streak preceded by failure of the streamwise vortices.
At $t=806$, there is a laminar hole in $u_x$ for all $x$ and $17\lesssim z\lesssim 22$ and the vortices have almost decayed.
At $t=1076$ the streaks have further receded, however, the self sustaining process of turbulence has restarted in the middle of the turbulent domain, as seen by the increase of amplitude of $u_z$ in the $2\lesssim z\lesssim10$.
From this point on, turbulence reinvades the domain. We can note that at $t=1370$, the turbulent hole is closing, as seen in the surge of kinetic energy.
However the opening and the closing are different processes: now the active-quiescent interfaces are almost at the same position in $z$ for both $u_x$ and $u_z$.
In the end, turbulence reinvades the whole domain.

\begin{figure}
\centerline{\includegraphics[width=7cm]{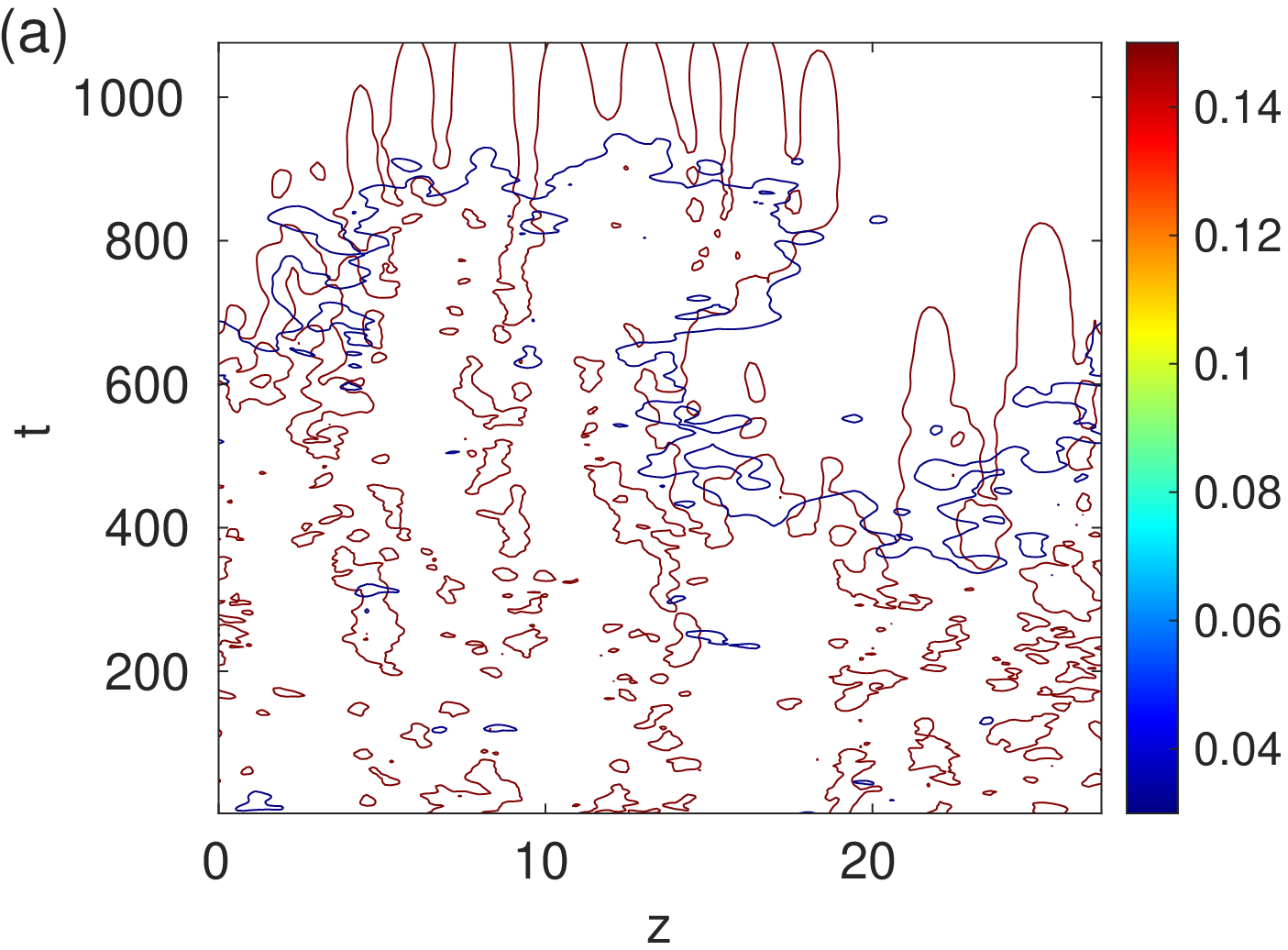}\includegraphics[width=7cm]{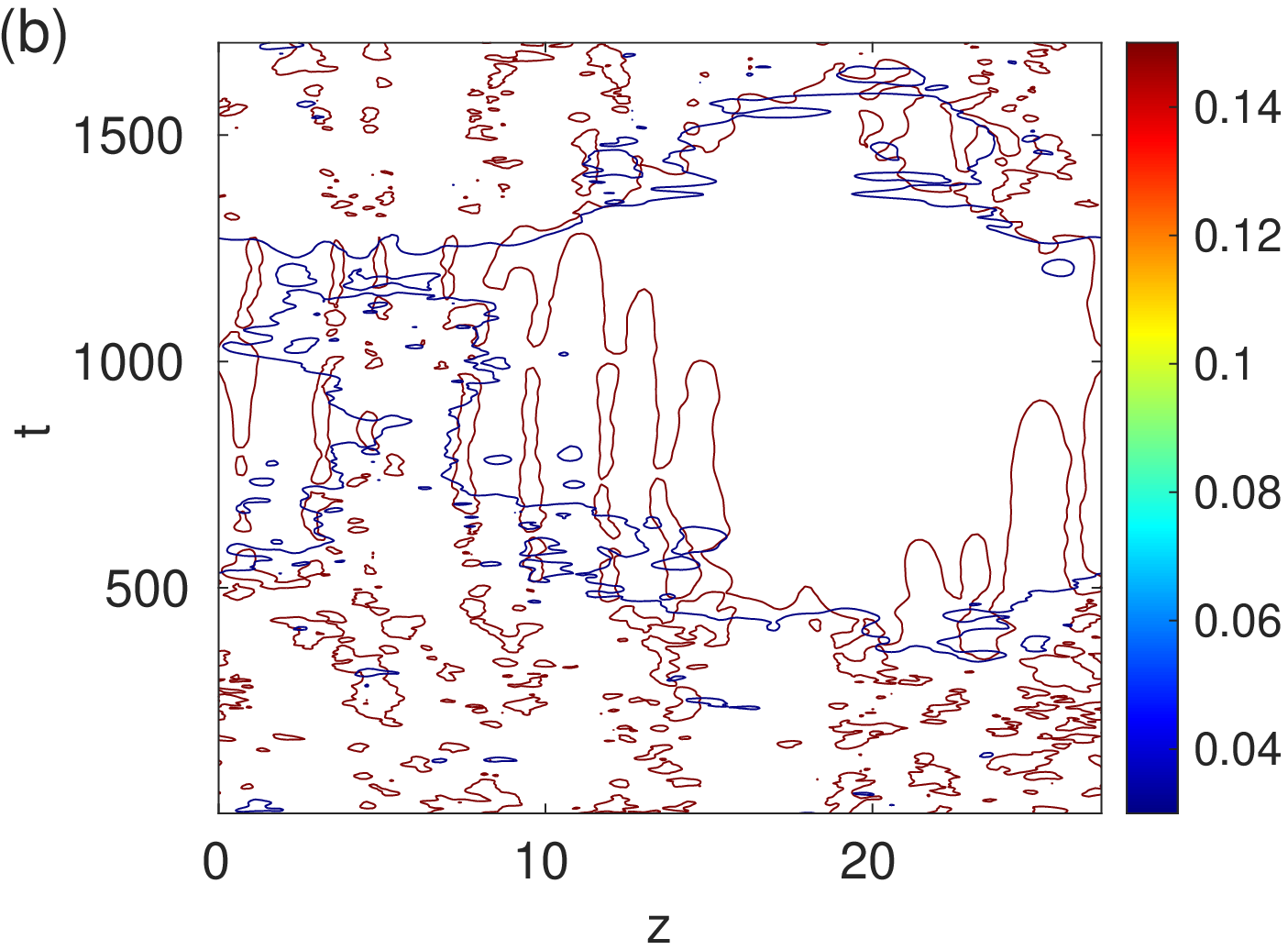}}
\caption{Contours of $\sqrt{\int_{x=0}^{36}u_z^2\,{\rm d}x-\left( \int_{x=0}^{36}u_z\,{\rm d}x\right)^2}=0.03$ and $\sqrt{\int_{x=0}^{36}u_x^2\,{\rm d}x}=0.15$ as a function of $z$ and $t$ for (a) the collapse trajectory of figure~\ref{36_27_377_coll} and (b) the non-reactive trajectory of figure~\ref{36_27_377_fail}.}
\label{suicont}
\end{figure}

We can observe the movement of the active-quiescent interfaces in more detail for the reactive  trajectory (Fig.~\ref{suicont} (a)) and the non-reactive trajectory (Fig.~\ref{suicont} (b)).
These figures use both the contours of $\sqrt{\int_{x=0}^{36}u_z^2\,{\rm d}x-\left( \int_{x=0}^{36}u_z\,{\rm d}x\right)^2}=0.03$ and $\sqrt{\int_{x=0}^{36}u_x^2\,{\rm d}x}=0.15$ which respectively indicate the spanwise location of active-quiescent interface for the streamwise vortices and for the velocity streaks.
These can be compared to earlier visualisations (Fig.~\ref{36_27_377_coll} and Fig.~\ref{36_27_377_fail}), that justified the study of the purely spanwise location of the active quiescent interface in this type of domain.
In the case of the collapse trajectory, we can observe the survival of streaks where streamwise vortices have disappeared for $500\lesssim t \lesssim 800$ and $20\lesssim z\lesssim 27$ and for $900\lesssim t\lesssim 1100$ and $5\lesssim z\lesssim 20$ (Fig.~\ref{suicont} (a)).
While this is partially caused by the method of detection of the interface (see Fig.~\ref{36_27_377_coll} $t=624$) and while the scenario is not at one hundred percent univocal, the situation where streamwise vortices survive where streaks have decayed is never observed in the reactive and non-reactive trajectories.
Similar observations can be made in the case of the non-reactive trajectory (Fig.~ \ref{suicont} (b)). We have an apparent collapse of both streaks and vortices at $t=400$ and for $15\lesssim z\lesssim 20$. For $500\lesssim t\lesssim 800$, only the  active-quiescent interface of the vortices recedes.
For $800\lesssim 1200$, we can see some reduction of the surface occupied by the streaks. For $t\gtrsim 1200$, we can see that the surface occupied by the vortices increases again until the active-quiescent interface for the streaks and the vortices is again the same and they both reinvade the domain.

\section{Conclusion}\label{concl}

In this work we have presented one of the first applications of a rare event sampling method to study multistability in a turbulent flow not forced stochastically.
Namely, Adaptive Multilevel Splitting was used with anticipated branching to compute and study turbulence collapse trajectories in transitional plane Couette flow.
A large number of collapse trajectories could thus be computed using a dramatically reduced amount of computational time. In some of the cases considered here, simulations can be more than one thousand times faster.
Calculations may be even more accelerated for rarer events as the computational cost of AMS typically evolves like the logarithm of the computational time of DNS.
Moreover, the collapse trajectories can be straightforwardly recorded in full details using the checkpoints created during AMS computations.
The reactive trajectories in phase space, statistics of reactive trajectories durations, velocity fields during reactive and non-reactive trajectories as well as estimates of the probability of turbulence collapse
and mean first passage time before turbulence collapse (with their respective error bars) are then available at low cost using this approach.

A first step in this work has been to ensure the validity of the result of AMS computations. This has been done in a system of size $L_x\times L_z=24\times 18$ at Reynolds number $R=370$.
Computation of reactive trajectories and mean first passage time are amenable in this system, so that rare event computations could be compared to a reference.
It could thus be shown that reactive trajectories (trajectories in phase space, statistics of durations) are computed precisely.
Some effects of the low noise added on the flow for the separation of trajectories were identified in the very last stages of reactive trajectories, when the flow is finally laminar.
The bias added to the collapse computations is small enough so that the use of this additional noise is entirely acceptable.
Similarly, the probability of collapse computed using AMS was shown to be in the same 66\% interval of confidence as the one computed by DNS.
The computed mean first passage times are  overestimated. Histograms of the computed probabilities of collapse and mean first passage time before collapse indicate that this overestimate is caused by a handful of overly large estimates of $T$ related to underestimates of $\alpha$. This is most likely an effect of the number of clones used in this computation. Other estimators of mean first passage time could be used to improve the computation \citep{lestang2020numerical}, which have their own drawbacks for trajectory computations, such as constraints on trajectory durations.

A large part of this work was of course devoted to the study of the collapse trajectories properties. In the system of size $L_x\times L_z=24\times 18$,
the visualisation of collapse trajectories showed that the laminarisation could be sketched in two stages.
The velocity components $u_y$ and $u_z$, and as a consequence, streamwise vortices, decayed first, so that the flow is driven out of the self sustaining process of turbulence \citep{waleffe1997self}.
As a consequence, streamwise velocity has less dependence on $x$, and then slowly decays without displaying streaks instability \citep{marquillie2011instability,kawahara2003linear}.
Examination of the velocity fields shows that this process can happen locally, leading to a streamwise long laminar hole in $u_y$, $u_z$ and then extends in $z$
 : streamwise vortices decay everywhere in this small domain.
Viewed in the quadrant $E_{k,x}\ge 0,E_{k,y-z}\ge 0$, these visualisations all correspond to trajectories concentrated along the same path.
While this path is more continuous than the sketch given here, it still displays a first stage where $E_{k,y-z}$ decays much faster than $E_{k,x}$, then a second stage where $E_{k,y-z}$ is small (around the noise level) and $E_{k,x}$ decays in turn.
Such concentration of trajectories are commonplace in stochastic models which display rare transitions when the noise variance goes to zero.
While the separation of the dynamics of an actually turbulent flow into a deterministic part and a noise part is
very difficult to make, and that the flow displays transient chaos when the domain is very small \citep{eckhardt2007turbulence}, this is a first indication that the collapse of turbulence in Couette or Poiseuille flows of increasing size is best understood as a stochastic system. Given the concentration of trajectories, one may be tempted to draw a comparison to Instantons computed in stochastic systems \citep{prl_jet,grafke2019numerical}. In order to give a meaning to such a description, one would need a low noise limit. This can possibly be the case in the large size limits as observed in models \citep{rolland_pre18} and in the large deviations of the kinetic energy \citep{rolland2015mechanical}.
The statistics of reactive trajectories durations are another feature of the trajectories which are very close to their stochastic counterparts.
Both DNS and AMS computations show that the normalised distribution of durations is very close to a Gumbel distribution.
In stochastic multistable systems it is demonstrated in one dimensional systems that this is the distribution of durations \citep{cerou2013length}.
In extended systems, it is observed that this distribution is found when the multistability occurs when the system transits through a single curved saddle of its deterministic part \citep{rolland2016computing,rolland_pre18}.
This would place the collapse of turbulence in plane Couette flow in systems of increasing sizes in this category.

The second part of the study of trajectories concerned the question of the localisation of the laminar holes.
For this matter the size of the studied system has been increased, from $L_x\times L_z=24 \times 18$ to $36\times 27$. Increasing the domain size led to the computation of a single type of reactive trajectories.
They all display the local opening of a spanwise localised streamwise elongated hole, first in streamwise vorticity, then in streamwise velocity as this hole extends in the spanwise direction. Laminar--turbulent front movements have been presented by \citep{duguet2011stochastic}.
There has been some observations of similar hole opening, as well as global collapse in numerical simulations and in experiments \citep{manneville2011decay,de2020transient}. However, there has been so far no general way to ascertain the weight of each scenario in turbulence collapse for given values of control parameters.
Using a method like AMS and sampling a large number of trajectories at a comparably low cost can therefore help give some statistical weight to each scenario of collapse.
Moreover, this suppressed the need to perform a quench or a more continuous decrease of the Reynolds number in order to observe said collapses.
This reopens the way for the study of collapse of non isolated turbulent puffs and spots, which had been considered experimentally \citep{bottin1998statistical} and more recently in models.
The dynamics of couette flow, where turbulence can extend in two direction, are even richer than in the case of the collapse of laminar-turbulence coexistence in models of the canonical pipe flow puff \citep{rolland_pre18}.

Another interesting point of the use of a method like AMS is the ability to identify a turning point in the dynamics of collapse \citep{simonnet2020multistability}.
This is done by recording the state leading to the largest reaction coordinate in the suppressed trajectories of the last stage of AMS. In stochastically forced systems, this state corresponds to an effective saddle between the two metastable states, and can alternatively be computed through a dichotomy procedure (see \citep{willis2009turbulent,schneider2007turbulence}), or even by computing the saddle point of the deterministic point of the system in systems forced by an additive noise.
This turning point gives us some information on the mechanism of multistability and in our case on collapse of turbulence.
It displayed localised streaks and streamwise vortices in the largest system, which backs the scenario of collapse through hole formation.
In the case of collapse of turbulence in plane Couette flow, these states did not necessarily make the best starting points for edge tracking.
Edge tracking has already been performed with the results of collapse of turbulence in pipe flow \citep{de2012edge}.
Note that edge tracking can be performed using the result of AMS computation: this has been done in the study of build up of turbulence under some additive noise and should be presented in a later text.

While the use of AMS to compute the reactive trajectories has proven successful, there is still some room for improvement, in particular for the estimation of the mean first passage time before collapse.
With a given set up, one can always bring improvement on the estimate by increasing the number of clones used to compute the trajectories (and thus their numerical cost).
This has been repeatedly measured in stochastically driven systems \citep{rolland2015statistical} and demonstrated for such systems asymptotically \citep{brehier2016unbiasedness,cerou2019asymptotic}.
However, one could wish to perform these improvements at little additional numerical cost.
Said studies have also shown that much improvement  can be brought by improving the reaction coordinate used in computations \citep{rolland2015statistical,brehier2016unbiasedness,rolland2016computing,cerou2019asymptotic}.
This has been so far performed by hand with trial and error by integrating physical properties of the system in the reaction coordinate
in order to mimic the committor function: the probability of reach the arrival state before the departure state from any given state \citep{onsager1938initial,brehier2016unbiasedness,rolland2015statistical,cerou2019asymptotic}.
Indeed it has been demonstrated that it was the ideal reaction coordinate: the one that lead to the best estimates \citep{cerou2019asymptotic}.
However, this committor function is in a way the answer to the question we ask when we study multistability.
Systematic methods to approximate the committor and improve the reaction coordinate are being validated in few degrees of freedom systems.
They could be directly applied to transitional turbulence. Indeed, the requirement for these methods to function are met: trajectories,
reactive or not, are reasonably well estimated.
These approaches may provide another estimate of the mean first passage time.
This could improve the estimates of $T$ provided in this text and avoid the accumulation of error that occur when equation~(\ref{EQT}) is used.

Finally, the computation of reactive trajectories in transitional wall turbulence opens the way for similar computations in high Reynolds number turbulence. Systematic computation of reactive trajectories in systems of aerodynamical interest \citep{kim1988investigation}, or geophysical interest are more than conceivable \citep{herbert2020atmospheric}.

\section*{Acknowledgements}

The author thanks F. Bouchet, C.-E. Br\'ehier, T. Lestang and E. Simonnet as well as T. Liu and B. Semin for interesting discussions.
The author thanks the help of the PSMN platform of ENS de Lyon, where most of the computation of this work have been performed. The author finally thanks the hospitality of Institut PPrime and ENSMA and their computational resources, where this work has been started, as well as CICADA (now AZZURA, centre de calcul interactif de l'Universit\'e de Nice). The author also thanks Felipe Alves Portela for help proofreading the manuscript.
The author also acknowledges the comments of  the anonymous referees that helped improve the manuscript.

\appendix
\section{Anticipated AMS}\label{rolland_sappa}

\begin{figure}
\centerline{\includegraphics[width=15cm]{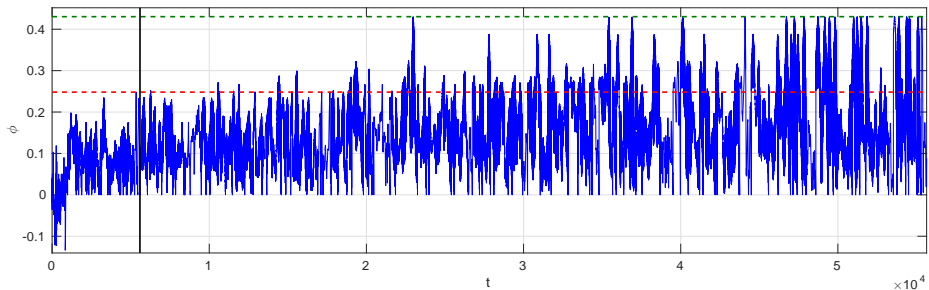}}
\caption{Time series of $\Phi$ during an AMS calculation with the  $24\times 16$ system leading to extinction (the laminar state corresponds to $\phi\ge 1$). The black line indicates the end of the stage 0 of AMS and the red line the maximal $\phi=0.248\pm 0.001$ reached during this stage (``contained in the initial conditions''). The green line indicate the maximum $\phi=0.430\pm0.001$ reached during the whole AMS calculation.}
\label{ext_24_16}
\end{figure}

The various anticipated branching strategies proposed in section~\ref{AMS} help increasing the variability amongst trajectories.
More importantly this approach helps avoiding so called extinction.
Indeed, when using AMS, we wish to avoid a situation where no trajectories are able to go further than $\Phi_{\rm ext}<\Phi(\partial\mathcal{B})$.
In that case, no matter how many additional steps are performed, we have at stage $k$ $\Phi_{\max,k}=\max_i\max_t(\Phi_i(t))=\Phi_{\rm ext}$ and reactive trajectories cannot be computed.
We illustrate this situation in figure~\ref{ext_24_16} where we display the time series of the reaction coordinate $\Phi$ for all trajectories successively computed by AMS in a run where no anticipation is performed. The reaction coordinate used here is based on the asymmetry of the flow (see Eq.~\ref{eqasym}). The figure was obtained in a sequential run of AMS where all the time series of the reaction coordinate for all the trajectories simulated were successively saved. We draw a distinction between the trajectories computed in the initial step of the algorithm (times before the black vertical line)
and those computed during the mutation selection process (times after the black line). During the initial free run stage of AMS (``stage 0''), the maximum of reaction coordinate over all trajectories is $\Phi_{\max,0}=0.248\pm 0.001$.
Note that this maximum of $\Phi$ on a trajectory corresponds to a turning point where $\Phi (t)$ will decrease with an extremely high probability even if slightly perturbed. In the first steps of the classical AMS computation, the levels of reaction coordinate at which trajectories are branched are rather below $\Phi_{\max,0}$, so that variability among trajectories is initially created
and the maximum of $\Phi$ eventually reaches $\Phi_{\max,k}=0.430\pm0.001$ after $k$ stages. Extinction occurs because branching are performed at higher and higher levels of $\Phi$ which are closer and closer to the value $\Phi_{\max,k}$.
Eventually all trajectories reach that value, all branching is performed at $\Phi_{\max,k}$ and even with small perturbation, all trajectories subsequently decay in $\Phi$. Closer examination shows that they all
differ from one another after some time. The issue here is not that trajectories do not separate from one another, but that they do so during a decay phase of $\Phi$ and thus cannot lead to a progress of the
reaction coordinate. This has also been seen in the study of turbulent wakes \citep{lestang2020numerical}. Anticipation of branching consists in branching trajectories before these turning points so that
they all have the time to separate and generate  new trajectories going further and further in $\Phi$.

\begin{figure}
\centerline{\includegraphics[width=18cm]{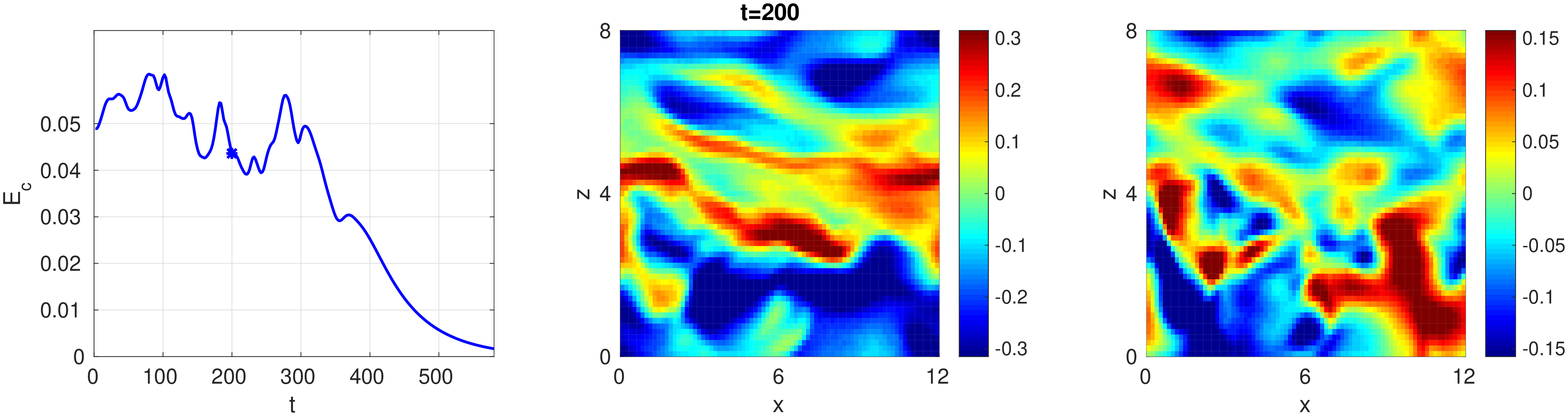}}
\centerline{\includegraphics[width=18cm]{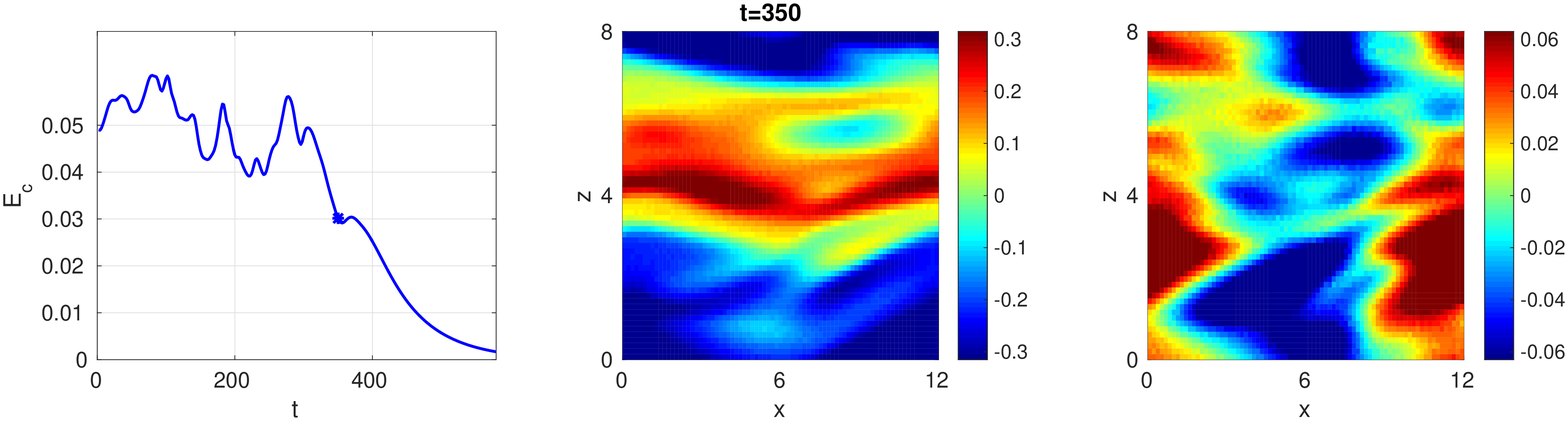}}
\centerline{\includegraphics[width=18cm]{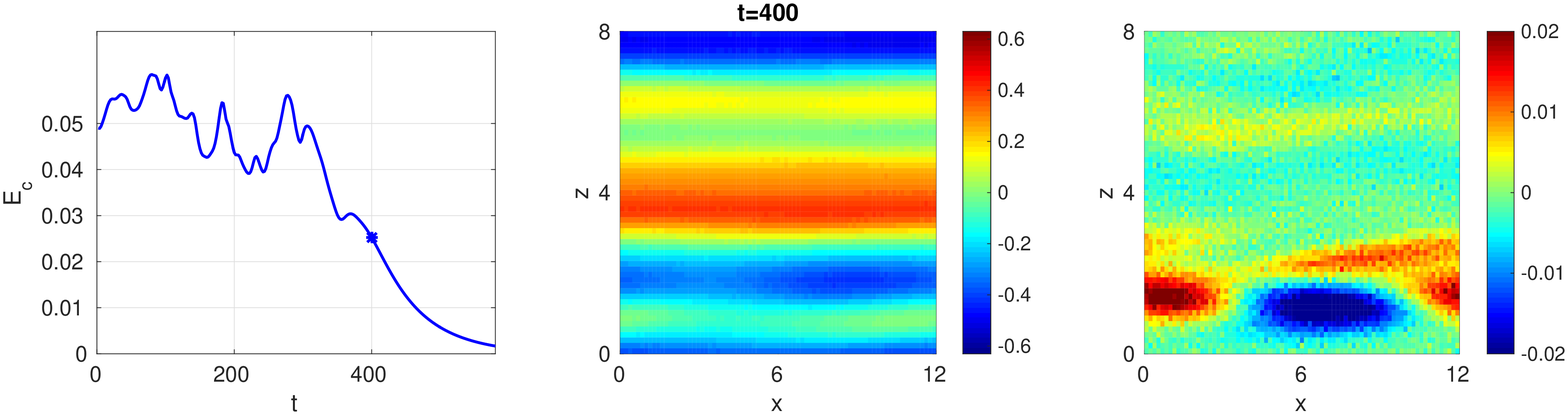}}
\centerline{\includegraphics[width=18cm]{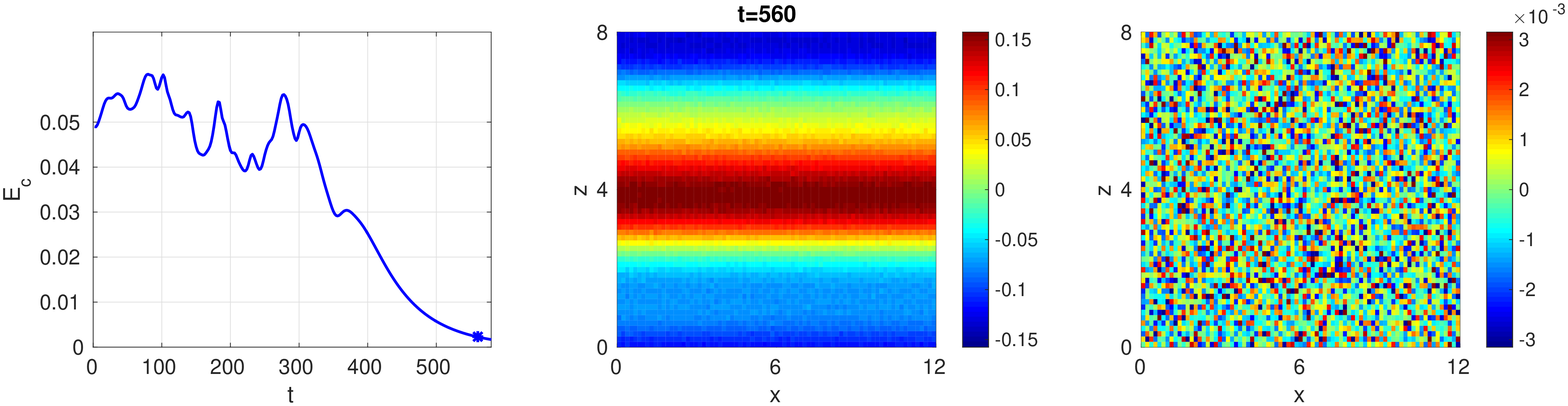}}
\caption{Collapse of turbulence in a very small system of size $L_x\times L_z=12\times 8$. Left column: time series of kinetic energy with dot indicating the instant in the simulation. middle column, colour levels of streamwise velocity in the horizontal midplane. Right column: colour levels of spanwise velocity in the horizontal midplane.}
\label{serie}
\end{figure}

The anticipation strategy which should be followed depends on the properties of the system. In the case of the collapse of turbulence in plane Couette flow, this depends on the size of the system.
The line is drawn between small systems (typically of size $L_x\lesssim 15$, and $L_z\lesssim 10$) and larger systems.
Small systems which have very few degrees of freedom, display at most four or five velocity streaks (high or low velocity) and a handful of streamwise vortices.
Their  behaviour can be described by the escape away from a chaotic repeller \citep{eckhardt2008dynamical}.
They necessarily collapse globally in space, this means that the larger $\Phi$ is, the more globally quiescent
the flow becomes: the mixing is less intense and there is little separation of trajectories.
This is shown in figure~\ref{serie} in a small system of size $L_x\times L_z=12\times 8$. The flow required some reorganisation before collapse was possible (part of the trajectory for $0\le t\le 300$), and at most two low speed and two high speed streaks are present.
When the collapse takes place (between $t=350$ and $t=400$) $u_z$ collapses almost uniformly. Larger systems have more and more degrees of freedom, and contain more velocity streaks and vortices.
An important feature
is that they collapse through the formation of a laminar hole. This means  that for a large portion of the reactive trajectory some area of the flow remains turbulent. This turbulent area can serve as some sort of heat bath which
fuels the trajectory with perturbations and thus can greatly help the separation of trajectories. This means that in small systems, branching must indeed be performed on trajectories having greater excursions,
but on their initial stages, where the flow is still agitated and thus will separate the trajectories. In that case, we use a so called saturated anticipation
\begin{equation}\Phi_{b,sat}=\xi\left(1-\exp\left(-\frac{\Phi_{N_c}}{\xi}\right)\right)\label{sat}\end{equation}
this function is shown with $\xi=0.25$ in figure~\ref{rolland_figrare} (c), with the ``saturated'' label. If we follow that strategy in the case of larger systems, it will require too many tries for branched trajectories
 to reach $\Phi_{N_c}$. A more efficient strategy is to anticipate the branching with a small shift, this can for instance be done with a function that slowly reaches its asymptote, the line of slope $1$,
 \begin{equation}
\Phi_{b, conv}=\frac{\Phi_{N_c}\left(1+\tanh\left( \frac{\Phi_{N_c}-\xi}{\xi}\right) \right)}{2}\,,\label{conv}
\end{equation}
This function is displayed in figure ~\ref{rolland_figrare} (c), with the ``converging'' label.

On a final note, we state that the strong necessity for anticipated branching may come from the imperfection of the reaction coordinate. Indeed, we have so far used a reaction coordinate constructed with a formula
under simple physical arguments. For instance, in the case of the kinetic energy, we considered that $E$ decreases as the flow goes toward the laminar state. In the case of large systems, it
may very well be that the committor, the function which associates the probability of reaching the final state to (in our case) an instantaneous velocity field, may not be so far from an affine function of $E$.
In the case of very small systems, which display a behaviour close to temporal chaos, it is likely that the committor has very large variation on isosurfaces of constant $E$, in particular when one moves slightly away from turbulence. This is caused by the complex structure of the laminar-turbulent boundary in small scale systems \citep{schmiegel1997fractal}.
Both situations are relevant to collapse of turbulence in shear flows in general. Indeed, pipe flows will most of the time display a localised turbulent puff which keeps a small number of degrees of freedom as the length of the pipe is
increased. Meanwhile spatially extended channel flows such as Couette or Poiseuille flow will display localised turbulence which will have more and more degrees of freedom as the domain size is increased.
The option of anticipated branching should remain on the table as there is no certainty that methods to estimate better reaction coordinate could work without prior runs of AMS.

\bibliographystyle{apalike}
\bibliography{ref_coll}

\end{document}